\documentclass[journal,onecolumn]{IEEEtran}
\ifCLASSINFOpdf
\else
\fi
\usepackage{amssymb,amsfonts,amsmath,amscd,dsfont,mathrsfs}
\usepackage[T1]{fontenc}
\usepackage{graphicx,float,psfrag,epsfig}
\usepackage{wrapfig}
\usepackage{epstopdf,color}
\usepackage{hyperref,bbm}
\usepackage{balance}
\usepackage{caption}
\usepackage{mathtools}

\usepackage{algorithm}
\usepackage[noend]{algpseudocode}
\DeclareMathAlphabet{\mathpzc}{OT1}{pzc}{m}{it}

\makeatletter
\def\thmhead@plain#1#2#3{%
  \thmname{#1}\thmnumber{\@ifnotempty{#1}{ }\@upn{#2}}%
  \thmnote{ {\the\thm@notefont#3}}}
\let\thmhead\thmhead@plain
\makeatother
\newtheorem{propo}{Proposition}[section]
\newtheorem{lemma}[propo]{Lemma}

\newtheorem{corollary}[propo]{Corollary}
\newtheorem{theorem}[propo]{Theorem}

\newcommand{\reals}{{\mathbb R}}
\newcommand{\prob}{\mathbb P}
\newcommand{\E}{\mathbb E}
\newcommand{\V}{V}
\newcommand{\G}{G}

\newcommand{\hv}{\hat{v}}

\def\ind{\mathbb{I}}

\def\hv{{\hat v}}

\def\q{{q}}
\def\cG{{\cal G}}

\renewcommand{\algorithmicrequire}{\textbf{Input:}}
\renewcommand{\algorithmicensure}{\textbf{Output:}}

\begin{document}
%
\title{Hiding the Rumor Source}
%
%
%

\author{Giulia~Fanti,~\IEEEmembership{Member,~IEEE,}
        Peter~Kairouz,~\IEEEmembership{Member,~IEEE,}
        Sewoong~Oh,~\IEEEmembership{Member,~IEEE,}
        ~Kannan~Ramchandran,~\IEEEmembership{Fellow,~IEEE}
        and Pramod~Viswanath,~\IEEEmembership{Fellow,~IEEE,}
\thanks{G. Fanti, P. Kairouz, S. Oh, and P. Viswanath are with the University of Illinois at Urbana-Champaign, Urbana, IL 61801 USA e-mail: \{fanti,kairouz2,swoh,pramodv\}@illinois.edu .}
\thanks{ K. Ramchandran is with the Department
of Electrical and Computer Engineering, University of California, Berkeley,
CA, 30332 USA e-mail: kannanr@eecs.berkeley.edu .}
\thanks{This work was presented in part at ACM Sigmetrics 2015 and 2016.}%
}

\maketitle

\begin{abstract}
Anonymous social media platforms like Secret, Yik Yak, and Whisper have emerged as important tools for sharing ideas without the fear of judgment. 
Such anonymous platforms are also important in nations under authoritarian rule, where freedom of expression and the personal safety of message authors may depend on anonymity. Whether for fear of judgment or retribution, it is sometimes crucial to hide the identities of users who post sensitive messages. In this paper, we consider a global adversary who wishes to identify the author of a message; it observes either a snapshot of the spread of a message at a certain time, sampled timestamp metadata, or both. Recent advances in rumor source detection show that existing messaging protocols are vulnerable against such an adversary. We introduce a novel messaging protocol, which we call adaptive diffusion, and show that under the snapshot adversarial model, adaptive diffusion spreads content fast and achieves perfect obfuscation of the source when the underlying contact network is an infinite regular tree. That is, all users with the message are nearly equally likely to have been the origin of the message. 
When the contact network is an irregular tree, we characterize the probability of maximum likelihood detection by proving a concentration result over Galton-Watson trees.
Experiments on a sampled Facebook network demonstrate that adaptive diffusion effectively hides the location of the source even when the graph is finite, irregular and has cycles.
\end{abstract}


%
\IEEEpeerreviewmaketitle

\section{Introduction}
\label{sec:intro}



Microblogging platforms are central to the fabric of the present Internet; popular examples include Twitter and Facebook. 
In such platforms, users propagate short messages (texts, images, videos) over a contact graph, which represents a social network in most cases.
Message forwarding  often occurs through built-in mechanisms that rely on user input, such as clicking ``like" or ``share" on a particular post.
Brevity of message, fluidity of user interface, and trusted party communication combine to make these microblogging platforms a major communication mode of modern times.

However, the popularity of microblogging services also makes them a prime target for invasive user monitoring by employers, service providers, or government agencies.
This monitoring typically exploits \emph{metadata}: non-content data that characterizes content, like timestamps.
Metadata can often be as sensitive as data itself \cite{narayanan2009anonymizing,greschbach2012devil};
this reality was publicized by Michael Hayden, former Director of the CIA, with his observation that ``We kill people based on metadata" \cite{hayden}.

The alarming privacy implications of these platforms has spurred the growth of {\em anonymous microblogging} platforms, like Whisper \cite{whisper}, Yik Yak \cite{yikyak}, and the now-defunct Secret \cite{secret}.
These platforms enable users to share messages with their friends without revealing authorship metadata.

Existing anonymous messaging services
store both messages and authorship information on centralized servers, which makes them vulnerable to
government subpoenas, hacking, or direct company access.
An alternative solution would be to store this information in a distributed fashion;
each node would know only its own friends, and message authorship information would never be transmitted to any party.
Distributed systems are  more robust to monitoring due to lack of central points of failure.
However, even under distributed architectures,
simple anonymous messaging protocols (such as those used by commercial anonymous microblogging apps) are still vulnerable against an adversary with side information, as proved in recent advances in rumor source detection \cite{SZ11a,PTV12}.
In this work, we study a basic building block of the messaging protocol
 that would underpin truly anonymous microblogging platform 
-- {\em how to anonymously broadcast a single message on a contact network}, even in the face of a strong deanonymizing adversary with access to metadata.
Specifically, we focus on
anonymous microblogging
built atop an underlying \emph{social network}, such as a network of phone contacts or Facebook friends.

\vspace{0.1in}
\noindent {\bf Adversaries.}
We consider three adversarial models, which capture different approaches to collecting metadata. 
In each case, the underlying contact network is modeled as a graph that is known to the adversary.
Beyond that, the adversary could proceed in a few different ways.

The adversary might use side channels to infer whether a node is infected, i.e., whether it received the message.
If an adversary collects only infection metadata for all network users, we call it a {\em snapshot} adversary. 
This could represent a state-level adversary that attends a Twitter-organized protest; it implicitly learns who received the protest advertisement by observing which individuals are physically present, but not the associated metadata.
The snapshot adversary is well-studied in the literature, primarily in the related problem of \emph{source identification} \cite{SZ11a,WDZT14,PVF12,FC12,luo2013identify}. 
We focus primarily on the snapshot adversarial model in this paper.

Alternatively, the adversary might explicitly corrupt some fraction of nodes by bribery or coercion; these corrupted {\em spy nodes} could pass along metadata like message timestamps and relay IDs. If an adversary only collects information from spies, we call it a {\em spy-based} adversary. A spy-based adversary could represent a government agency participating in social media to study users, for instance.
The adversary's reach may be limited by factors like account creation, contact network structure \cite{danezis2009sybilinfer}, or the cost of corrupting participants.
This adversarial model is discussed in detail in \cite{fkorv2016}, but we include the relevant theoretical results in this paper for the sake of completeness. 

Finally, an adversary could combine the spy-based and snapshot adversarial models by using both forms of metadata.
If an adversary uses spies and a snapshot, we call it a {\em spy+snapshot} adversary. 
This adversarial model allows us to study the 
capabilities of both snapshot and spy metadata types, combining the results on snapshot adversary capabilities derived here with those of spy adversary capabilities derived in  
in \cite{fkorv2016}.


\vspace{0.1in}
\noindent {\bf Spreading models.}
In social networks,
messages are typically propagated based on users' approval, which is expressed via liking, sharing or retweeting. This mechanism, which enables social filtering and reduces spam, has inherent  random delays associated with each user's time of impression and decision to ``like" the message (or not). Standard models of rumor spreading in networks  explicitly model such random delays via a {\em diffusion} process: messages are spread independently over different edges with a fixed probability of spreading (discrete time model) or an exponential spreading time (continuous time model).
The designer can partially control the spreading rate
by introducing artificial delays on top of the usual random delays due to users' approval of the messages. 

We model this physical setup as a discrete-time system.
At time $t=0$, a single user $v^*\in \V$ starts to spread a message over the contact network $\G=(\V,E)$ where
users and contacts are represented by nodes and edges, respectively.
Upon receiving the message, nodes approve it immediately.  
The assumption that all nodes are willing to approve and pass the message is common in rumor spreading analysis \cite{SZ11a,SZ11b,ZY13}. 
However, by assuming message approval is immediate, we abstract away the natural random delays typically modeled by diffusion.
At the following timestep, 
the protocol decides which neighbors will receive the message, and
how much propagation delay to introduce.
Given this control, the system designer wishes to design a spreading protocol that
makes message source inference difficult. 

Specifically, after $T$ timesteps, let $\V_T\subseteq \V$, $G_T$, and $N_T \triangleq |\V_T|$ denote the set of infected nodes, the subgraph of $G$ containing only $V_T$, and the number of infected nodes, respectively.
at a given time $T$, the adversary uses all available metadata to estimate the source. We assume no prior knowledge of the source, so 
the adversary computes a maximum-likelihood (ML) estimate of the source $\hv_{\rm ML}$. 
We desire a spreading protocol that minimizes the probability of detection $P_D=\prob(\hv_{\rm ML}=v^*)$.

\emph{Current state-of-the-art:}
Diffusion is commonly used to model epidemic propagation over contact networks. While simplistic (it ignores factors like individual user preferences), diffusion is a commonly-studied and useful model  due to its simplicity and first-order approximation of actual propagation dynamics.
Critically, it captures the \emph{symmetric} spreading used by most social media platforms.

However, diffusion has been shown to exhibit poor anonymity properties; under the adversarial models we consider, the source can be identified reliably \cite{SZ11a,PTV12}. 
We therefore seek a different spreading model with strong anonymity guarantees.
 We wish to achieve the following performance metrics:
\begin{itemize}
	\item [$(a)$] We say a protocol has an {\em order-optimal rate of spread} if the
	expected time for the message to reach $n$ nodes scales linearly compared to the time required by the fastest spreading protocol.
	\item [$(b)$] We say a protocol achieves a {\em perfect obfuscation} if the probability of source detection for the maximum likelihood estimator is order-optimal. The definition of optimality differs for different adversarial models, so we define this metric separately for each adversarial model. 
\end{itemize}

\begin{figure}[tb]
	\vspace{-0.1cm}
	\begin{center}
	\includegraphics[width=.5\textwidth]{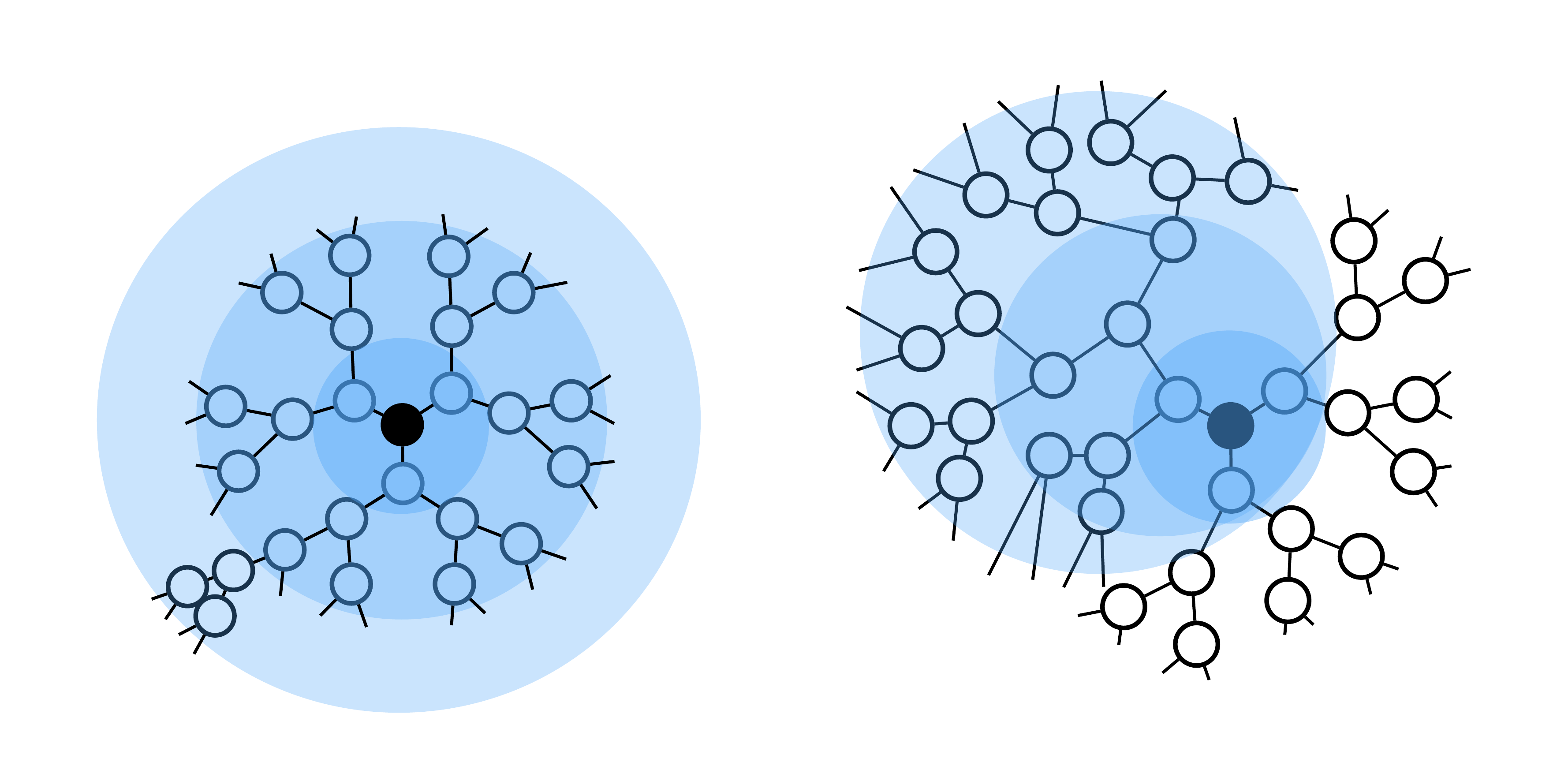}
	\end{center}
	\vspace{-.2cm}
	\caption{Illustration of a spread of infection when spreading immediately (left) and
	under adaptive diffusion (right). }
	\label{fig:intro}
\end{figure}

\vspace{0.1in}
\noindent{\bf Contributions.}
We introduce {\em adaptive diffusion}, a novel messaging protocol with provable author anonymity guarantees
against all of the discussed adversarial models. 
Whereas diffusion spreads the message symmetrically in all directions, 
adaptive diffusion breaks that symmetry (Figure \ref{fig:intro}). This has different implications for different adversarial models, but it consistently yields stronger anonymity guarantees than diffusion. 
Adaptive diffusion is also inherently distributed and spreads messages fast, i.e.,
the time it takes adaptive diffusion to reach $n$ users is at most twice the time it takes the
fastest spreading scheme which immediately passes the message to all its neighbors.

We prove that over $d$-regular trees, adaptive diffusion provides perfect obfuscation of the source 
 under the snapshot adversarial model. 
That is, 
the likelihood of an infected user being the source of the infection is equal among all infected users. 
We derive exact expressions for the probability of detection, and show that this expression is optimal for the snapshot adversary by providing a matching 
 fundamental lower bound. 

In practice, the contact networks are not regular infinite trees. 
For a general class of graphs which can be finite, irregular and have cycles, 
we provide results of numerical experiments on real-world social networks and synthetic networks 
showing that the protocol hides the source at nearly the best possible level of obfuscation under 
the snapshot adversarial model.
The same is true for spy-based adversaries; such  simulation results for that adversarial model are discussed in \cite{fkorv2016}.
Further, for a specific family of random {\em irregular} infinite trees, known as Galton-Watson trees, 
we characterize the probability of detection under adaptive diffusion and a snapshot adversary. 
In the process, we prove a strong concentration for the extreme paths 
in the Galton-Watson tree that consists of nodes with large degrees, which might be of independent interest. 

Finally, we characterize the probability of maximum likelihood detection of adaptive diffusion in various edge cases,
such as when the adversary can take \emph{multiple snapshots}, 
and when the underlying graph contains regular cycles, as in an infinite lattice graph.






\vspace{0.1in}
\noindent{\bf Related work.}
Anonymous communication has been a popular research topic for decades.
For instance, anonymous {point-to-point} communication
allows a sender to communicate with a receiver without the receiver learning the sender's identity.
A great deal of work has emerged in this area, including Tor \cite{tor}, Freenet \cite{freenet}, Free Haven \cite{freeHavenProject}, and Tarzan \cite{tarzan}.
In contrast to this body of work, we address the problem of anonymously \emph{broadcasting} a message over
an underlying contact network (e.g., a network of Facebook friendships or phone contacts). 

Anonymous broadcast communication has been most studied in context of the dining cryptographers' (DC) problem.
We diverge from the literature on this topic \cite{chaum88,corrigan2010dissent,goel2003herbivore,golle2004dining,von2003k} in approach and formulation.
We consider statistical spreading models rather than cryptographic encodings,
accommodate  computationally unbounded adversaries,  and consider domain-specific contact networks
 rather than a fully connected communication network.

Recently, Riposte addressed a similar problem of anonymously writing to a public message board \cite{corrigan2014riposte}. 
It uses techniques from private information retrieval to 
store multiple, corrupted copies of messages on distributed servers.
This corruption is designed so that no subset of colluding servers can determine the author. 
However, Riposte places no restrictions  on communication with the servers, thereby facilitating spam.  
Differences in the communication model and adversarial model prevent Riposte from effectively solving our problem of interest.


Within the realm of statistical message spreading models, the problem of detecting the origin of an epidemic or the source of a rumor
has been studied under the {\em diffusion} model.
Recent advances in \cite{SZ11a,SZ11b,WDZT14,PVF12,FC12,luo2013identify,ZY13,Austin1,Austin2,Austin3,Austin4,khim2015confidence} 
show that
it is possible to identify the source within a few hops with high probability.
Consider an adversary who has access to the underlying {\em contact network} of friendship links
and the snapshot of infected nodes at a certain time. The  problem of
locating a rumor source,   first posed in \cite{SZ11a}, naturally corresponds to
  {graph-centrality}-based inference algorithms: for a continuous time model,  \cite{SZ11a,SZ11b} used the
{rumor centrality} measure to correctly identify the source after time $T$ (with  probability converging to a positive number for large $d$-regular and random trees, and with probability proportional to $1/\sqrt{T}$ for lines).
The probability of identifying the source increases even further when multiple infections
from the same source are observed \cite{WDZT14}.
With multiple sources of infections,
spectral methods have been proposed for
estimating the number of sources and the set of source nodes in \cite{PVF12,FC12}.
When infected nodes are allowed to recover as in the susceptible-infected-recovered (SIR) model,
{Jordan centrality} was proposed in \cite{luo2013identify,ZY13} to estimate the source.
In \cite{ZY13}, it is shown that the Jordan center is still within a bounded hop distance from the
true source with high probability, independent of the number of infected nodes.

When the adversary collects timestamps (and other metadata) from spy nodes,
standard diffusion reveals the location of the source \cite{PTV12,ZY13,LMOZ13}.
However, ML estimation is known to be NP-hard \cite{ZCY14},
and analyzing the probability of detection is also challenging.
the source can be effectively identified.


In summary, under natural, diffusion-based message spreading---as seen in almost every content-sharing platform today---an 
adversary with some side information can identify the rumor source with high confidence. 
We overcome this vulnerability by asking the {reverse question}: can we {design} messaging protocols that spread fast while protecting the anonymity of the source?
Related challenges include $(a)$ identifying the best algorithm that the adversary might use to infer the location of the source;
$(b)$ providing analytical guarantees for the proposed spreading model; and
$(c)$ identifying the fundamental limit on what any spreading model can achieve. 
We address all of these challenges for snapshot adversarial model (Section \ref{sec:snapshot}), 
spy-based adversarial model (Section \ref{sec:spies}), and finally the spy+snapshot model (Section \ref{sec:spy_snapshot}). 
In this paper, our primarily focus is on the snapshot adversarial model; 
 the spy-based and spy+snapshot adversaries are discussed in detail in \cite{fkorv2016}.

Our work fits into a larger ecosystem that enables anonymous messaging; we implicitly assume that the ecosystem is healthy.
For instance, we assume that nodes communicate securely in a distributed fashion, but anonymity-preserving, peer-to-peer (P2P) address lookup is still an active research area \cite{borisov2014dp5}, as is privacy-preserving distributed data storage in P2P systems \cite{jawad2009protecting}.
We do not consider adversaries that operate below the application layer (e.g., by monitoring the network or even physical layer) \cite{winter2012great,fbi}.
Lower-level solutions may be more appropriate against such an opponent, harnessing factors like physical proximity of users \cite{firechat}.
In that space, physical layer security and privacy attacks pose a very real threat, as has been documented extensively in prior work \cite{bauer2009physical,danev2010attacks,karlof2003secure}.

\vspace{0.1in}
\noindent{\bf Organization.}
The remainder of this paper is organized as follows:
To begin, we introduce the general adaptive diffusion protocol in Section \ref{sec:adaptive}.
In Section \ref{sec:snapshot}, we describe how to specialize adaptive diffusion under a snapshot adversarial model. 
In Section \ref{sec:spies}, we describe how to apply adaptive diffusion under a spy-based adversarial model. 
Combining the key insights of these two approaches,
we introduce results from the spy+snapshot adversarial model in Section \ref{sec:spy_snapshot}. 
For each adversarial model, we first describe the precise version of adaptive diffusion that applies to infinite $d$-regular trees, 
and show that it achieves perfect obfuscation of the source. 
We then provide 
extensions to irregular trees. 
We conclude by presenting simulated results over real graphs: finite, irregular, and containing cycles. 
In Section \ref{sec:Polya}, we make a connection between adaptive diffusion on a line and P\'olya's urn processes. 
This connection, while interesting in itself, provides a novel analysis technique for precisely capturing the price of control packets that are 
passed along with the messages in order to coordinate the spread of messages as per adaptive diffusion.



\section{Adaptive diffusion} \label{sec:adaptive}
In this section, we describe adaptive diffusion in its most general form, 
and leave for later sections the specific choice of parameters involved. 
For the purpose of introducing adaptive diffusion, we specifically on an infinite $d$-regular tree as the underlying contact network. 

\begin{figure}[b!]
    \centering
  \includegraphics[scale=0.5]{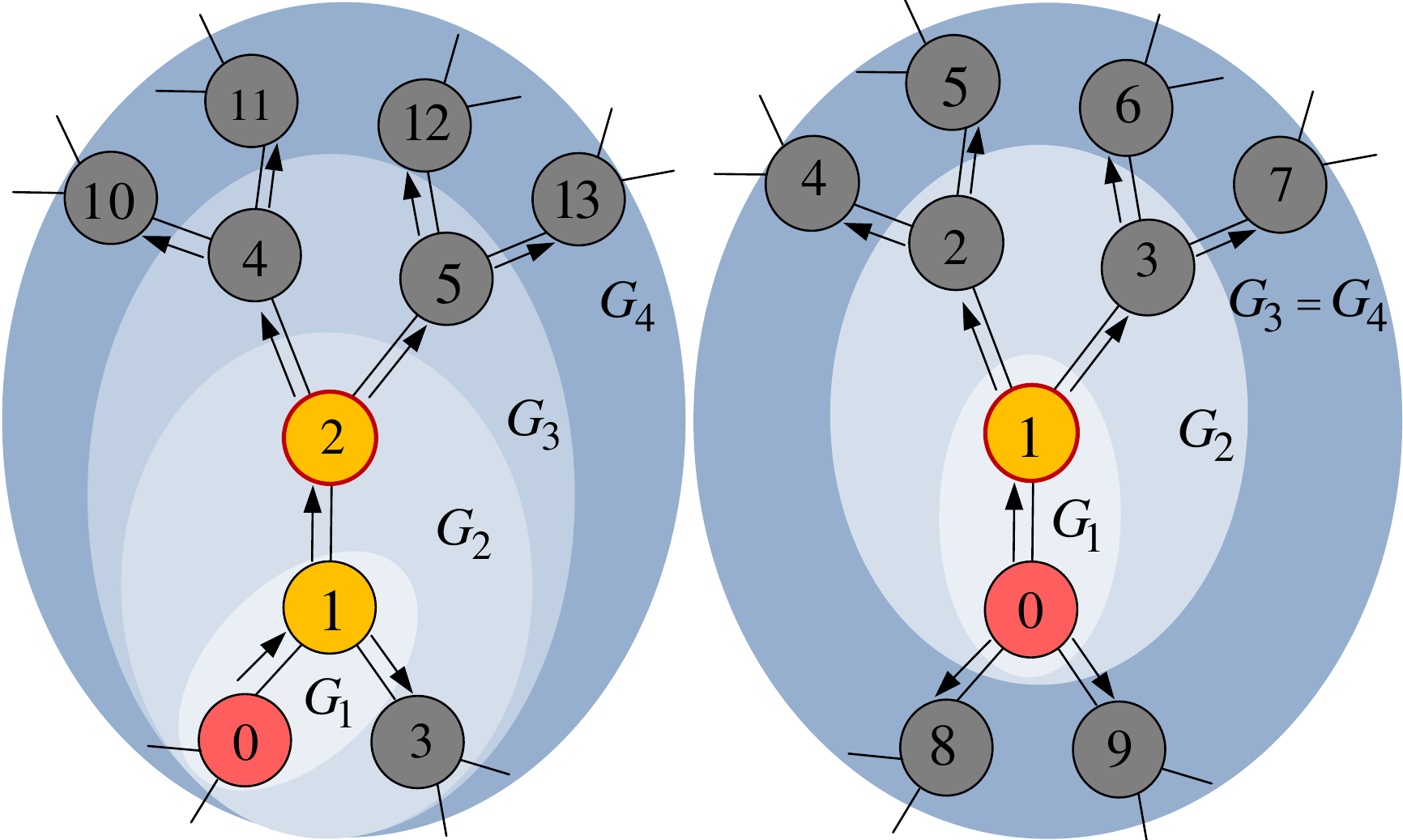}
  \caption{Adaptive diffusion over regular trees. Yellow nodes indicate the set of virtual sources (past and present), and for $T=4$, the virtual source node is outlined in red.}
  \label{fig:graph}
\end{figure}

We step through the intuition of the adaptive diffusion spreading model with an example, partially illustrated in Figure \ref{fig:graph}. 
The precise algorithm description is provided in Protocol \ref{alg:adp_diff}.
Adaptive diffusion ensures that the infected subgraph $G_t$ at any even timestep $t\in\{2,4,\ldots\}$ is
a balanced tree of depth $t/2$, i.e. the hop distance from any leaf to the root (or the center of the graph) is $t/2$.
We call the root node of $G_t$ the ``virtual source'' at time $t$, and denote it by $v_t$.
We use $v_0=v^*$ to denote the true source.
To keep the regular structure at even timesteps,
we use the odd timesteps to transition from one regular subtree $G_t$ to another one $G_{t+2}$ with depth incremented by one.

More concretely, the first three steps are always the same.
At time $t=0$, the rumor source $v^*$ selects, uniformly at random, one of its neighbors to be the virtual source $v_2$; at time $t=1$, $v^*$ passes the message to $v_2$.
Next at $t=2$, the new virtual source $v_2$ infects all its uninfected neighbors forming $G_2$ (see Figure \ref{fig:graph}).
Then node $v_2$ chooses to either keep the virtual source token 
or to pass it along.

If $v_2$ chooses to remain the virtual source i.e., $v_4=v_2$, it passes `infection messages' to all the leaf nodes in the infected subtree, telling each leaf to infect all its uninfected neighbors. Since the virtual source is not connected to the leaf nodes in the infected subtree, these infection messages get relayed by the interior nodes of the subtree. This leads to $N_t$ messages getting passed in total (we assume this happens instantaneously). These messages cause the rumor to spread symmetrically in all directions at $t=3$. At $t=4$,
no spreading occurs (Figure \ref{fig:graph}, right panel).

If $v_2$ does \emph{not} choose to remain the virtual source,
it passes the virtual source token to a randomly chosen neighbor $v_4$, excluding the previous virtual source (in this example, $v_0$).
Thus, if the virtual source moves, it moves away from the true source by one hop.
Once $v_4$ receives the virtual source token, it sends out infection messages. However, these messages do not get passed back in the direction of the previous virtual source. This causes the infection to spread asymmetrically over only one subtree of the infected graph ($G_3$ in Figure \ref{fig:graph}, left panel).
In the subsequent timestep ($t=4$), the virtual source remains fixed and passes the same infection messages again. After this second round of asymmetric spreading, the infected graph is once again symmetric about the virtual source $v_4$ ($G_4$ in Figure \ref{fig:graph}, left panel).

This process continues at each timestep: the virtual source $v_t$ chooses whether to keep or pass the virtual source token.
Conditioned on this decision, the infected subgraph grows deterministically as needed to ensure symmetry about the new virtual source, $v_{t+2}$.

\begin{algorithm}[ht!]
\caption{Adaptive Diffusion}
\label{alg:adp_diff}
\begin{algorithmic}[1]
\Require contact network $G=(V,E)$, source $v^*$, time $T$, degree $d$
\Ensure set of infected nodes $V_T$
\State $V_0 \gets \{v^*\}$, $h \gets 0$, $v_0 \gets v^*$
\State $v^*$ selects one of its neighbors $u$ at random
\State $V_1 \gets V_0 \cup \{u\}$, $h \gets 1$, $v_1 \gets u$
\State let $N(u)$ represent $u$'s neighbors
\State $V_2 \gets V_1 \cup N(u)\setminus \{v^*\}$, $v_2 \gets v_1$
\State $t \gets 3$
\For{$t \leq T$}
\State $v_{t-1}$ selects a random variable $X \sim U(0,1)$
\If{$X \leq \alpha_d(t-1, h)$}
\ForAll{$v \in N({v_{t-1}})$}
\State Infection Message({$G$,$v_{t-1}$,$v$,$G_t$})
\EndFor
\Else
\State $v_{t-1}$ randomly selects $u \in N({v_{t-1}})\setminus \{v_{t-2}\}$
\State $h \gets h + 1$
\State $v_{t} \gets u$
\ForAll{$v \in N({v_{t}})\setminus \left\{v_{t-1}\right\}$}
\State Infection Message({$G$,$v_{t}$,$v$,$V_t$})
\If{$t+1>T$}
\State \text{break}
\EndIf
\State Infection Message({$G$,$v_{t}$,$v$,$V_t$})
\EndFor
\EndIf
\State $t \gets t + 2$
\EndFor
\Procedure{Infection Message}{$G$,$u$,$v$,$V_t$}
\If{$v \in V_t$}
\ForAll{$w \in N({v})\setminus \left\{u\right\}$}
\State Infection Message({$G$,$v$,$w$,$G_t$})
\EndFor
\Else
\State $V_t \gets V_{t-2} \cup \{v\}$
\EndIf
\EndProcedure
\end{algorithmic}
\end{algorithm}

As we will see momentarily, adaptive diffusion uses varying amounts of control information to coordinate the spread of messages. 
In some adversarial models (snapshot), this control information does not hurt anonymity; in others (spy-based), it can be problematic. We therefore introduce different implementations of adaptive diffusion as needed, using different amounts of control information.
In each implementation, the resulting distribution of the random infection process is the same (if the same parameters 
 are used).

This random infection process can be defined as a time-inhomogeneous (time-dependent) Markov chain over the state defined by the location of the current virtual source $\{v_t\}_{t\in\{0,2,4,\ldots\}}$.
By the symmetry of the underlying contact network, which we assume is an infinite $d$-regular tree, and
the fact that the next virtual source is chosen uniformly at random among the neighbors of the current virtual source,
it is sufficient to consider a Markov chain over the hop distance between the true source $v^*$ and $v_t$, the virtual source at time $t$.
Therefore, we design a Markov chain over the state
\begin{eqnarray*}
	h_t &=& \delta_H(v^*,v_t)\;,
\end{eqnarray*}
for even $t$, where $\delta_H(v^*,v_t)$ denotes the hop distance between nodes $v^*$ (the true source) and $v_t$ (the virtual source).
Figure \ref{fig:graph} shows an example with $(h_2,h_4)= (1,2)$ on the left and
 $(h_2,h_4)=(1,1)$ on the right.

At every even timestep, the protocol randomly determines  whether
to keep the virtual source token ($h_{t+2} = h_t$) or to pass it ($h_{t+2}=h_t + 1$).
We specify the resulting time-inhomogeneous Markov chain over $\{h_t\}_{t\in\{2,4,6,\ldots\}}$ by choosing appropriate transition probabilities as a function of
 time $t$ and current state $h_t$.
 For even $t$, we denote this probability by
\begin{eqnarray}
	\alpha_d(t,h) &\triangleq& \prob\big(\,h_{t+2}=h_t | h_t=h\,\big) \;,
\end{eqnarray}
where the subscript $d$ denotes the degree of the underlying contact network.
In Figure \ref{fig:graph}, 
at $t=2$, the virtual source remains at the current node (right) with probability $\alpha_3(2,1)$,
or passes the virtual source to a neighbor with probability $1-\alpha_3(2,1)$ (left).
The parameters $\alpha_d(t,h)$ fully describe the transition probability of the Markov chain defined over $h_t \in \{1,2,\ldots,t/2\}$.
For example, if we choose $\alpha_d(t,h)=1$ for all $t$ and $h$, then the virtual source never moves for $t>1$. The message spreads almost symmetrically, so the source can be caught with high probability, much like diffusion.
If we instead choose $\alpha_d(t,h)=0$ for all $t$ and $h$, the virtual source \emph{always} moves. This ensures that the source is always at one of the leaves of the infected subgraph. We return to this special case when addressing spy-based adversaries in Section \ref{sec:spies}.

The real challenge, then, is choosing the parameters $\alpha_d(t,h)$, which fully specify the virtual source transition probabilities. These parameters can significantly alter the anonymity and spreading properties of adaptive diffusion. 
In this work, we explain how to choose this parameter $\alpha_d$ to achieve desired source obfuscation.

\section{ snapshot-based adversarial model} \label{sec:snapshot}

Under the snapshot adversarial model, an adversary observes the infected subgraph $G_T$ at a certain time $T$
and produces an estimate $\hat{v}$ of the source $v^*$ of the message. 
Since the adversary is assumed to not
have  any prior information on which node is likely to be the source,
we analyze the performance of the maximum likelihood estimator
\begin{equation}
\label{eq:maxlhest}
\hv_{\rm ML} = \arg\max_{v\in G_T} \prob(G_T|v).
\end{equation}
We show that adaptive diffusion with appropriate parameters can achieve {\em perfect obfuscation}, i.e. 
the probability of detection for the ML estimator when  $n$ nodes are infected is close to $1/n$: 
	\begin{eqnarray}
		\prob\big(\,\hv_{\rm ML}=v^* | N_T=n\,\big) &=& \frac{1}{n}  + o\Big(\frac{1}{n}\Big)\;.
	\end{eqnarray}
This is the best source obfuscation that can be achieved by any protocol, since there are only $n$ candidates for the source and 
they are all equally likely.

\subsection{Main Result (Snapshot Model)}
\label{sec:adap_diff}

In this section, we show that for appropriate choice of parameters $\alpha_d(t,h)$, we can achieve both fast spreading and perfect obfuscation over $d$-regular trees.
We start by giving baseline spreading rates for deterministic spreading and diffusion. 

Given a contact network of an infinite $d$-regular tree, $d>2$,
consider the following deterministic spreading protocol. At time $t=1$,
the source node infects all its neighbors. At $t \geq 2$,
the nodes at the boundary of the infection spread the message to their uninfected neighbors.
Thus,  the message spreads one hop in every direction at each timestep.
This approach is the fastest-possible spreading, infecting $N_T=1+d((d-1)^T-1)/(d-2)$ nodes at time $T$,
but the source is trivially identified as the center
of the infected subtree.
In this case, the infected subtree is a
balanced regular tree where all leaves are at equal depth from the source.

Now consider a random diffusion model. At each timestep, each uninfected neighbor of an infected node
is independently infected with probability $q$.
In this case, $\E[N_T] = 1+ qd((d-1)^T-1)/(d-2)$, and it was shown in \cite{SZ11a}
that the probability of correct detection for the maximum likelihood estimator of the rumor source
is $ \prob(\hv_{\rm ML}=v^*) \geq C_d $ for some positive constant $C_d$ that only depends on the degree $d$. Hence, the source is only hidden in a constant number of nodes close to the center, even when
the total number of infected nodes is arbitrarily large.

Now we consider the spreading and anonymity properties of adaptive diffusion. 
Let $p^{(t)} = [p^{(t)}_h]_{h\in\{1,\ldots,t/2\}}$
denote the distribution of the state of the Markov chain at time $t$, i.e.
$p^{(t)}_h = \prob(h_t = h)$.
The state transition can be represented as  the following $((t/2)+1)\times (t/2)$ dimensional column stochastic matrices:
\begin{eqnarray*}
	\label{eq:MC}
	\mathclap{p^{(t+2)} = \begin{bmatrix}
	\alpha_d(t,1) & & & \\
	1-\alpha_d(t,1) &\alpha_d(t,2) & & \\
	& 1-\alpha_d(t,2)& \ddots& \\
	& & \ddots& \alpha_d(t,t/2) \\
	& & &  1-\alpha_d(t,t/2) \\
	\end{bmatrix}p^{(t)}}\;.
\end{eqnarray*}

We treat $h_t$ as strictly positive, because at time $t=0$, when $h_0=0$, the virtual source is always passed. Thus, $h_t\geq1$ afterwards. 
At all even $t$, we desire $p^{(t)}$ to be
\begin{eqnarray}
	\label{eq:MCp}
	p^{(t)} = \frac{d-2}{(d-1)^{t/2}-1}\begin{bmatrix}
		1\\
		(d-1)\\
		\vdots\\
		(d-1)^{t/2-1}
	\end{bmatrix} \;\in\reals^{t/2} \;,
\end{eqnarray}
for $d>2$ and for $d=2$,
$p^{(t)} = (2/t) {\mathbf 1}_{t/2}$ where ${\mathbf 1}_{t/2}$ is all ones vector in $\reals^{t/2}$.
There are $d(d-1)^{h-1}$ nodes at distance $h$ from the virtual source,
and by symmetry all of them are equally likely to have been the source:
\begin{eqnarray}
	\prob(G_T| v^*, \delta_H(v^*,v_t)=h) &=& \frac{1}{d(d-1)^{h-1}} p^{(t)}_{h} \nonumber\\
		&=& \frac{d-2}{d((d-1)^{t/2}-1)} \;,\nonumber
\end{eqnarray}
for $d>2$, which is independent of $h$.
Hence, all the infected nodes (except for the virtual source) are equally likely to have been the source of the origin. This statement is made precise in Equation \eqref{eq:p_diff}.

Together with the desired probability distribution in Equation \eqref{eq:MCp},
this gives a recursion over $t$ and $h$ for computing the appropriate $\alpha_d(t,h)$'s.
After some algebra and an initial state $p^{(2)}=1$, we get that the following choice ensures the desired
Equation \eqref{eq:MCp}:
\begin{eqnarray}
\alpha_d(t,h) = \left\{
  \begin{array}{lr}
    \frac{(d-1)^{t/2-h+1}-1}{(d-1)^{t/2+1}-1} & ~\text{if}~ d > 2\\
    \frac{t-2h+2}{t+2} & ~\text{if}~ d=2
  \end{array}
\right.
\label{eq:alpha}
\end{eqnarray}
With this choice of parameters, we show that adaptive diffusion spreads fast,
infecting $N_t = O((d-1)^{t/2})$ nodes at time $t$ and
each of the nodes except for the virtual source is equally likely to have been the source.

\begin{theorem}
\label{thm:main}
Suppose the contact network is a $d$-regular tree with $d\geq2$,
and one node $v^*$ in $G$ starts to spread a message according to
Protocol  \ref{alg:adp_diff} at time $t=0$, with $\alpha_d(t,h)$ chosen according to Equation \ref{eq:alpha}.
At a certain time $T\geq0$ an adversary estimates the location of the source $v^*$
using the maximum likelihood estimator $\hv_{\rm ML}$.
The following properties hold for Protocol \ref{alg:adp_diff}:

\begin{itemize}
	\item [$(a)$] the number of infected nodes at time $T$ is
		\begin{eqnarray}
		N_{T}\geq \left\{
		  \begin{array}{lr}
			\frac{2(d-1)^{(T+1)/2}-d}{(d-2)}+1 & \text{if}~d>2\\
			T+1 & \text{if}~d=2
		\end{array}
		\right.
		\label{eq:n_diff}
		\end{eqnarray}
	\item [$(b)$] the probability of source detection for the maximum likelihood estimator at time $T$ is
	\begin{eqnarray}
		\prob\left(\hat{v}_{\rm ML}=v^*\right)  \leq \left\{
		  \begin{array}{lr}
		 \frac{d-2}{2(d-1)^{(T+1)/2}-d} &\text{if}~d>2\\
		(1/T) & \text{if}~d=2
		\label{eq:p_diff}
		\end{array}
		\right.
	\end{eqnarray}
	\item [$(c)$] the expected hop-distance between the true source $v^*$ and its estimate $\hat{v}_{\rm ML}$ under maximum likelihood estimation is lower bounded by
	\begin{eqnarray}
		\E[d(\hv_{\rm ML},v^*)] \geq
		\frac{d-1}{d}\frac{T}{2}.
		\label{eq:d_diff}
	\end{eqnarray}
\end{itemize}
\end{theorem}
(Proof in Section \ref{sec:proofs_snapshot_main})

Although this choice of parameters achieves perfect obfuscation, the spreading rate is slower than
the deterministic spreading model, which infects $O((d-1)^T)$ nodes at time $T$. 
However, this type of constant-factor loss in the spreading rate is inevitable: the only way to deviate from the deterministic spreading model is
to introduce appropriate delays.

In order to spread according to adaptive diffusion with the prescribed $\alpha_d(h,t)$, the system needs to know the 
degree $d$ of the underlying contact network. 
However, performance is insensitive to knowledge of $d$ for certain parameter settings, as shown in the following proposition. 
Specifically, one can choose $\alpha_d(h,t)=0$ for all $d$, $h$, and $t$ and still achieve performance comparable to the  optimal choice. 
The main idea is that there are as many nodes in the boundary of the snapshot (leaf nodes) as there are in the interior, so it is sufficient to hide among the leaves. 
One caveat is that if the underlying contact network is a line (i.e. $d=2$) then this approach fails since there are only two leaf nodes at any given time, and the probability of detection is trivially $1/2$.
\begin{propo}
\label{pro:tree}
Suppose that the underlying contact network $G$ is an infinite $d$-regular tree with $d>2$,
and one node $v^*$ in $G$ starts to spread a message  at time $t=0$ according to
Protocol  \ref{alg:adp_diff} with $\alpha_d(h,t)=0$ for all $d$, $h$, and $t$.
At a certain time $T\geq1$ an adversary estimates the location of the source $v^*$
using the maximum likelihood estimator $\hv_{\rm ML}$.
Then the following properties hold: 
\begin{itemize}
	\item[$(a)$] the number of infected nodes at time $T\geq1$ is at least
	\begin{equation}
	\label{eq:n_tree}
	N_T \;\geq \;
	\frac{(d-1)^{(T+1)/2}}{d-2}  \; ;
	\end{equation}
	
	\item[$(b)$] \vspace{-0.2cm} the probability of source detection for the maximum likelihood estimator at time $T$ is
	\begin{eqnarray}
\prob\big( \hv_{\rm ML} =v^* \big) &=& \frac{d-1}{2+(d-2)N_T} \; ; \text{ and }
\label{eq:pd_tree}
	\end{eqnarray}
	
	\item[$(c)$] \vspace{-0.2cm} the expected hop-distance between the true source $v^*$ and its estimate $\hat{v}$
	is lower bounded by
	\begin{eqnarray}
\E[\delta_H(v^*,\hv_{\rm ML})] &\geq& \frac{T}{2} \;.
		\label{eq:treedist}
	\end{eqnarray}
	\end{itemize}
\end{propo}

(Proof in Section \ref{sec:proofs_tree}).

\vspace{0.1in}
\noindent \textbf{Multiple snapshots.}
The results in Theorem \ref{thm:main} and Proposition \ref{pro:tree} hold for a single snapshot. 
However, an adversary could in principle take multiple snapshots of the same message's spread, at different points in time. 
We show that doing so increases the probability of detection at most by a logarithmic factor, compared to what it learns from the first snapshot (on average).
\begin{propo}
\label{pro:multiple_snap}
Suppose that the underlying contact network $G$ is an infinite $d$-regular tree with $d>2$,
and one node $v^*$ in $G$ starts to spread a message  at time $t=0$ according to
Protocol \ref{alg:adp_diff}, with $\alpha_d(t,h)$ chosen according to Equation \ref{eq:alpha}.
At a certain time $T\geq0$ an adversary observes a snapshot $G_T$ with $N_T$ nodes. 
In timesteps $\{T_1, T_2, \ldots, T_m\}$, where $T_i > T$ for all $i\in \{1,2,\ldots m\}$, 
the adversary again observes snapshots $G_{T_i}$. 
The adversary then estimates the location of the source $v^*$
using a maximum likelihood estimator $\hv_{\rm ML}$, based on knowledge of all observed snapshots.
Then the probability of source detection for the maximum likelihood estimator at time $T$ is upper bounded as follows:
	\begin{eqnarray}
\prob\big( \hv_{\rm ML} =v^* \big) &\leq & C\frac{\log_{d-1} N_T}{N_T-1} + o\left (\frac{\log_{d-1}N_T}{N_T} \right ) \; 
\label{eq:pd_tree_multiple}
	\end{eqnarray}
	where the constant $C$ depends only only on the tree degree $d$.
\end{propo}

(Proof in Section \ref{sec:proofs_multiple_snap}).

This result suggests that an adversary cannot learn much more than the information it learns from the first snapshot;
i.e., the probability of detection increases at most from $O(1/N_T)$ to $O(\log N_T/N_T)$.
Moving forward, we will assume that the snapshot adversary observes only one snapshot, at time $T$.

\subsection{Irregular Trees}\label{sec:irregular_trees}
In this section, we study adaptive diffusion on irregular trees, with potentially different degrees at the vertices. 
Although the degrees are irregular,
we still apply adaptive diffusion
with $\alpha_{d_0}(t,h)$'s chosen for a specific $d_0$ that might be mismatched with the graph due to degree irregularities.
There are a few challenges in this degree-mismatched adaptive diffusion.
First, finding the maximum likelihood estimate of the source is not immediate, due to degree irregularities.
Second, it is not clear \emph{a priori} which choice of $d_0$ is good.
We first show an efficient message-passing algorithm for computing the maximum likelihood source estimate.
Using this estimate, we illustrate through simulations how  adaptive diffusion
performs and show that the detection probability is not too sensitive to the choice of $d_0$ as long as $d_0$ is above a threshold that depends on the degree distribution.

Then, for the special choice of $d_0=\infty$, we precisely characterize the maximum likelihood probability of detection
and demonstrate that adaptive diffusion does not provide perfect obfuscation.
Doing so requires proving a concentration result for an extreme value defined over Galton-Watson branching processes, which may be of independent interest.
We use the associated analysis to propose a modification of adaptive diffusion called preferential-attachment adaptive diffusion (PAAD), 
which empirically improves the probability of detection over irregular trees, compared to standard adaptive diffusion. 

\vspace{0.1in}
\noindent{\bf Efficient ML estimation}. To keep the discussion simple, we assume that $T$ is even. The same approach can be naturally extended to  odd $T$. Since the spreading pattern in adaptive diffusion is entirely deterministic given the sequence of virtual sources at each timestep,
computing the likelihood $\prob(G_T|v^*=v)$ is equivalent to computing the probability of the virtual source moving from $v$ to $v_T$ over $T$ timesteps. On trees, there is only one path from $v$ to $v_T$ and since we do not allow the virtual source to ``backtrack", we only need to compute the probability of every virtual source sequence $(v_0, v_2, \ldots, v_T)$ that meets the constraint  $v_0=v$. Due to the Markov property exhibited by adaptive diffusion, we have
$ \prob(G_T| \{(v_t, h_t)\}_{t\in\{2,4,\ldots,T\}}) = \prod_{\substack{t<T-1\\t\text{ even}}} \prob(v_{t+2}|v_{t}, h_t)$, where $h_t =\delta_H(v_0,v_t)$.
For $t$ even, $\prob(v_{t+2}|v_{t}, h_t)=\alpha_d(t, h_t)$ if $v_t=v_{t+2}$ and $\frac{1 - \alpha_d(t,h_t)}{d_{v_{t}}-1}$ otherwise.
Here $d_{v_t}$ denotes the degree of node $v_t$ in $G$.
Given a virtual source trajectory $\mathcal P=(v_0, v_2, \ldots, v_T)$, let $\mathcal J_{\mathcal P}=(j_1, \ldots, j_{\delta_H(v_0,v_T)})$ denote the timesteps at which a new virtual source is introduced, with $1 \leq j_i \leq T$ . It always holds that $j_1=2$ because after $t=0$, the true source chooses a new virtual source and $v_2 \neq v_0$.
If the virtual source at $t=2$ were to keep the token exactly once after receiving it (so $v_2 = v_4$), then $j_2=6$, and so forth. To find the likelihood of a node being the true source, we sum over \emph{all} such trajectories
{\small
\begin{align}
\label{eq:likelihood}
&\prob(G_T|v_0) = \sum_{\mathcal J_{\mathcal P}: {\mathcal P}\in {\mathcal S(v_0,v_T,T)}} \underbrace{\frac{1}{d_{v_0}}\prod_{k=1}^{\delta_H(v_0,v_T)-1} \frac{1}{d_{v_{j_k}}-1}}_{A_{v_0}} \times \nonumber\\
 & \underbrace{\prod_{\substack{t< T\\t \text{~even}}} \left(\mathds 1_{\{t+2 \notin \mathcal J_{\mathcal P}\}} \alpha_d(t,h_t) + {\mathds 1}_{\{t+2 \in \mathcal J_{\mathcal P}\}} (1-\alpha_d(t,h_t))\right), }_{B_{v_0}}
\end{align}
}%
where $\mathds 1$ is the indicator function and \newline $S(v_0,v_T,T)=\{{\mathcal P}: {\mathcal P}=(v_0,v_2,\ldots,v_T)\text{ is a valid trajectory of the virtual source}\}$.
Intuitively, part $A_{v_0}$ of the above expression is the probability of choosing the set of virtual sources specified by $\mathcal P$, and part $B_{v_0}$ is the probability of keeping or passing the virtual source token at the specified timesteps. Equation \eqref{eq:likelihood} holds for both regular and irregular trees.
Since the path between two nodes in a tree is unique, and part $A_{v_0}$
is (approximately) the product of node degrees in that path,
$A_{v_0}$ is identical
for all trajectories $\mathcal P$.
Pulling $A_{v_0}$ out of the summation, we wish to compute the summation over all valid paths $\mathcal P$ of part $B_{v_0}$
(for ease of exposition, we will use $B_{v_0}$ to refer to this whole summation).
Although there are combinatorially many valid paths, we can simplify the formula in Equation \eqref{eq:likelihood}
for the particular choice of $\alpha_d(t,h)$'s defined in \eqref{eq:alpha}.

\begin{propo}\label{prop:irregular}
Suppose that the underlying contact network $\tilde G$ is an infinite tree with degree of each node larger than one.
One node $\tilde v^*$ in $\tilde G$ starts to spread a message  at time $t=0$ according to
Protocol  \ref{alg:adp_diff} with the choice of $d=d_0$.
At a certain even time $T\geq 0$,
the maximum likelihood estimate of $\tilde v^*$ given a snapshot of the infected subtree $\tilde G_T$ is
\begin{eqnarray}
	\arg \max_{v\in \tilde G_T\setminus \tilde v_T} &&  \frac{d_0}{d_v} 
\prod_{v' \in P(\tilde v_T,v)\setminus\{\tilde v_T,v\}} \frac{d_0-1}{ d_{v'}-1 }
		\;\label{eq:mismatch}
\end{eqnarray}
where $\tilde v_T$ is the (Jordan) center of the infected subtree $\tilde G_T$,
$P(\tilde v_T, v)$ is the unique shortest path from $\tilde v_T$ to $v$,
and $d_{v'}$ is the degree of node $v'$.
\end{propo}

\begin{figure}[h]
	    \centering
  \includegraphics[scale=0.64]{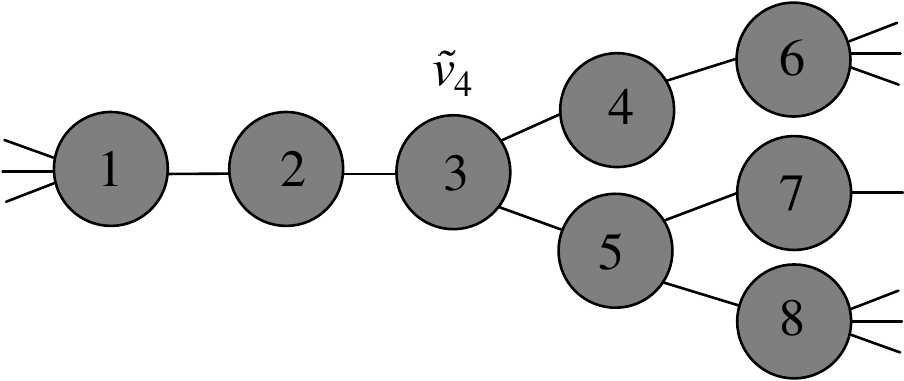}
  \caption{Irregular tree $\tilde G_4$ with virtual source $\tilde v_4$.}
  \label{fig:irregular}
\end{figure}

To understand this proposition, consider Figure \ref{fig:irregular}, which was spread using adaptive diffusion (Protocol \ref{alg:adp_diff}) with
 a choice of $d_0=2$.
  Then Equation \eqref{eq:mismatch} can be computed easily for each node, giving
  $[1/2,1,0,1,2/3,1/2,1/2,1/4]$ for nodes $[1,2,3,4,5,6,7,8]$, respectively. Hence, nodes 2 and 4 are most likely.
  Intuitively, nodes whose path to the center have small degrees are more likely.
  However, if we repeat this estimation assuming $d_0=4$, then
  Equation \eqref{eq:mismatch} gives $[3,2,0,2,4/3,3,3,3/2]$.
  In this case, nodes $1$, $6$, and $7$ are most likely.
  When $d_0$ is large, adaptive diffusion tends to place the source closer to the leaves of the infected subtree, so
  leaf nodes are more likely to have been the source.

\begin{IEEEproof}[Proof of Proposition \ref{prop:irregular}]
We first make two observations:
($a$) Over regular trees, $\prob(G_T|u)=\prob(G_T|w)$ for any $u\neq w \in G_T$, even if they are different distances from the virtual souce. ($b$) Part $B_{v_0}$ is identical for regular and irregular graphs, as long as the distance from the candidate source node to $v_T$ is the same in both, and the same $d_0$ is used to compute $\alpha_{d_0}(t,h)$.
That is, let $\tilde G_T$ denote an infected subtree over an \emph{irregular} tree network, with virtual source $\tilde v_T$, and $G_T$ will denote a \emph{regular} infected subtree with virtual source $v_T$. For candidate sources $\tilde v_0 \in \tilde G_T$ and $v_0\in G_T$, if $\delta_H(\tilde v_T, \tilde v_0) = \delta_H(v_T, v_0)=h$, then $B_{v_0} = B_{\tilde v_0}$. So to find the likelihood of $\tilde v_0\in \tilde G_T$, we can solve for $B_{\tilde v_0}$ using the likelihood of $v_0\in G_T$, and compute $A_{\tilde v_0}$ using the degree information of every node in the infected, irregular subgraph.


To solve for $B_{\tilde v_0}$,
note that over regular graphs, $A_{v}=1/(d_0 \, (d_0-1)^{\delta_H(v,v_T)-1})$, where $d_0$ is the degree of the regular graph.
If $G$ is a regular tree,
Equation \eqref{eq:likelihood} still applies.
Critically, for regular trees,  the $\alpha_{d_0}(t,h)$'s are designed such that
the likelihood of each node being the true source is equal. Hence,
\begin{equation}
\prob(G_T|v_0) = \underbrace{\frac{1}{d_0(d_0-1)^{\delta_H(v_0,v_T)-1}}}_{A_{v_0}}  \, \times \,B_{v_0} \;,
\label{eq:leaf}
\end{equation}
is a constant that does not depend on $v_0$.
This gives $B_{v_0} \propto (d_0-1)^{\delta_H(v_T,v_0)}$.
%
%
From observation ($b$), we have that $B_{\tilde v_0} = B_{v_0}$.
Thus we get that for a $\tilde v_0 \in\tilde G_T\setminus \{\tilde v_T\}$,
\begin{eqnarray*}
	\prob(\tilde G_T |\tilde v_0) &=& A_{\tilde v_0} \, B_{\tilde v_0} \\
		&\propto& \frac{(d_0-1)^{\delta_H(\tilde v_T,\tilde v_0)}}{d_{\tilde v_0} \prod_{\tilde v'\in P(\tilde v_T,\tilde v_0)\setminus \{\tilde v_0,\tilde v_T\}} (d_{\tilde v'}-1)}
\end{eqnarray*}
After scaling appropriately and noting that $|P(\tilde v_T,\tilde v_0)|=\delta_H(\tilde v_T, \tilde v_0)+1$,
this gives the formula in Equation \eqref{eq:mismatch}.
\end{IEEEproof}
%
\floatname{algorithm}{Algorithm}
\begin{algorithm}
\caption{Implementation of ML estimator in \eqref{eq:mismatch}}
\label{alg:ml_msg_passing}
\begin{algorithmic}[1]
\Require infected network $\tilde G_T=(\tilde V_T,\tilde E_T)$, virtual source $\tilde v_T$, time $T$, the spreading model parameter $d_0$
\Ensure $\text{argmax}_{\tilde v \in \tilde V_T}~\prob(\tilde G_T|\tilde v^*=\tilde v)$
\State $P_{\tilde v} \triangleq \prob(\tilde G_T|\tilde v^*=\tilde v)$.
\State $P_{\tilde v_T}\gets 0$
\State $A_{\tilde v} \gets 1$ for $\tilde v\in \tilde V_T \setminus \{\tilde v_T\}$
\State $A_{\tilde v_T}\gets 0$
\State $A\gets$ Degree Message($G_T$, $\tilde v_T$, $\tilde v_T$, $A$)
\State $\prob(G_T|v_{leaf}) \gets \frac{1}{d_0(d_0-1)^{T/2-1}}\prod _{\substack{t<T \\ t\text{ even}}}(1-\alpha_{d_0}(t,\frac{t}{2}))\}$
\ForAll {$\tilde v \in \tilde V_T \setminus \{\tilde v_T\}$}
\State $h\gets \delta_H(\tilde v, \tilde v_T)$
\State $B_{\tilde v}\gets \prob(G_T|v_{leaf})\cdot d_0 \cdot (d_0-1)^{h-1}$
\State $P_{\tilde v}\gets A_{\tilde v}\cdot B_{\tilde v}$
\EndFor
\Return argmax$_{\tilde v\in \tilde V_T}P_{\tilde v}$
\Procedure{Degree Message}{$\tilde G_T$, $\tilde u$, $\tilde v$, $A$}
\ForAll{$\tilde w \in N({\tilde v})\setminus \{\tilde u\}$}
\If{$\tilde v = \tilde u$}
\State $A_{\tilde w} \gets A_{\tilde v} / d_{\tilde w}$
\State Degree Message({$\tilde G_T$, $\tilde v$, $\tilde w$, $A$})
\Else
\If {$\tilde v$ is not a leaf}
\State $A_{\tilde w} \gets A_{\tilde v}\cdot d_{\tilde v} /(d_{\tilde w} \cdot (d_{\tilde v}-1))$
\State Degree Message({$\tilde G_T$, $\tilde v$, $\tilde w$, $A$})
\EndIf
\EndIf
\EndFor
\Return $A$
\EndProcedure
\end{algorithmic}
\end{algorithm}
\floatname{algorithm}{Protocol}

We provide an efficient
message passing algorithm for computing the
ML estimate in Equation \eqref{eq:mismatch}, which is naturally distributed.
We then use this estimator to simulate message spreading for random irregular trees and
show that when $d_0$ exceeds a threshold (determined by the degree distribution), obfuscation is not too sensitive to the choice of $d_0$.

$A_{\tilde v_0}$ can be computed efficiently for irregular graphs with a simple message-passing algorithm.
In this algorithm, each node $\tilde v$ multiplies its degree information by a cumulative likelihood that gets passed from the virtual source to the leaves. Thus if there are $\tilde N_T$ infected nodes in $\tilde G_T$, then $A_{\tilde v_0}$ for every $\tilde v_0 \in \tilde G_T$ can be computed by passing $O(\tilde N_T)$ messages. This message-passing is outlined in procedure `Degree Message' of Algorithm \ref{alg:ml_msg_passing}. For example, consider computing
$A_5$ for the graph in Figure \ref{fig:irregular}. The virtual source $\tilde v_T=3$ starts by setting $A_2=\frac{1}{2}$, $A_4=\frac{1}{2}$, and $A_5=\frac{1}{3}$. This gives $A_5$, but to compute other other values of $A_{\tilde w}$, the message passing continues. Each of the nodes $\tilde v\in N(3)$ in turn sets $A_{\tilde w}$ for \emph{their} children $\tilde w \in N(\tilde v)$; this is done by dividing $A_{\tilde v}$ by $d_{\tilde w}$ and replacing the factor of $\frac{1}{d_{\tilde v}}$ in $A_{\tilde v}$ with $\frac{1}{d_{\tilde v}-1}$. For example, node 5 would set $A_7=\frac{A_5}{2} \cdot \frac{3}{2}$. 
This step is applied recursively until reaching the leaves.

As discussed earlier, $B_{\tilde v_0}$ only depends on $d_0$ and $\delta_H(\tilde v_T,\tilde v_0)$.
If $v_{\text{leaf}} \in G_T$ is a leaf node and $G$ is a regular tree, we get
\begin{equation}
\prob(G_T|v_{\text{leaf}})=\underbrace{\frac{1}{d_0(d_0-1)^{T/2-1}}}_{A_{v_\text{leaf}}}\underbrace{\prod_{\substack{t< T\\t\text{ even}}}(1-\alpha_{d_0}(t,\frac{t}{2})) }_{B_{v_\text{leaf}}}.
\label{eq:leaf}
\end{equation}
If $\tilde v_0$ is $h < T/2$ hops from $\tilde v_T$, then 
for node $v_0$ with $\delta_H(v_0,v_T)=h<T/2$ over a \emph{regular} tree,
\begin{eqnarray}\label{eq:b}
\prob(G_T|v_{0})= \prob(G_T|v_{\text{leaf}}) = \frac{1}{d_0\cdot(d_0-1)^{ h-1}} B_{v_0}. \nonumber
\end{eqnarray}
Finally, $B_{\tilde v_0} = B_{v_0}$. 
So to solve for $B_{5}$ in our example, we compute $\prob (G_T|v_{leaf})$ for a 3-regular graph at time $T=4$. This gives $\prob (G_4|v_{leaf})=A_{v_{leaf}}\cdot B_{v_{leaf}} = \frac{1}{6} \cdot (1-\alpha_3(2,1))=\frac{1}{9}$. Thus $B_5 = \prob (G_4|v_{leaf})\cdot d_0\cdot(d_0-1)^{h-1}=\prob (G_4|v_{leaf})\cdot 3\cdot(2)^{0}=\frac{1}{3}$. This gives
$\prob (\tilde G_4|5)=A_5\cdot B_5 = \frac{1}{9}$. The same can be done for other nodes in the graph to find the maximum likelihood source estimate.

\begin{figure}[t]
	    \centering
  \includegraphics[width=.44\textwidth]{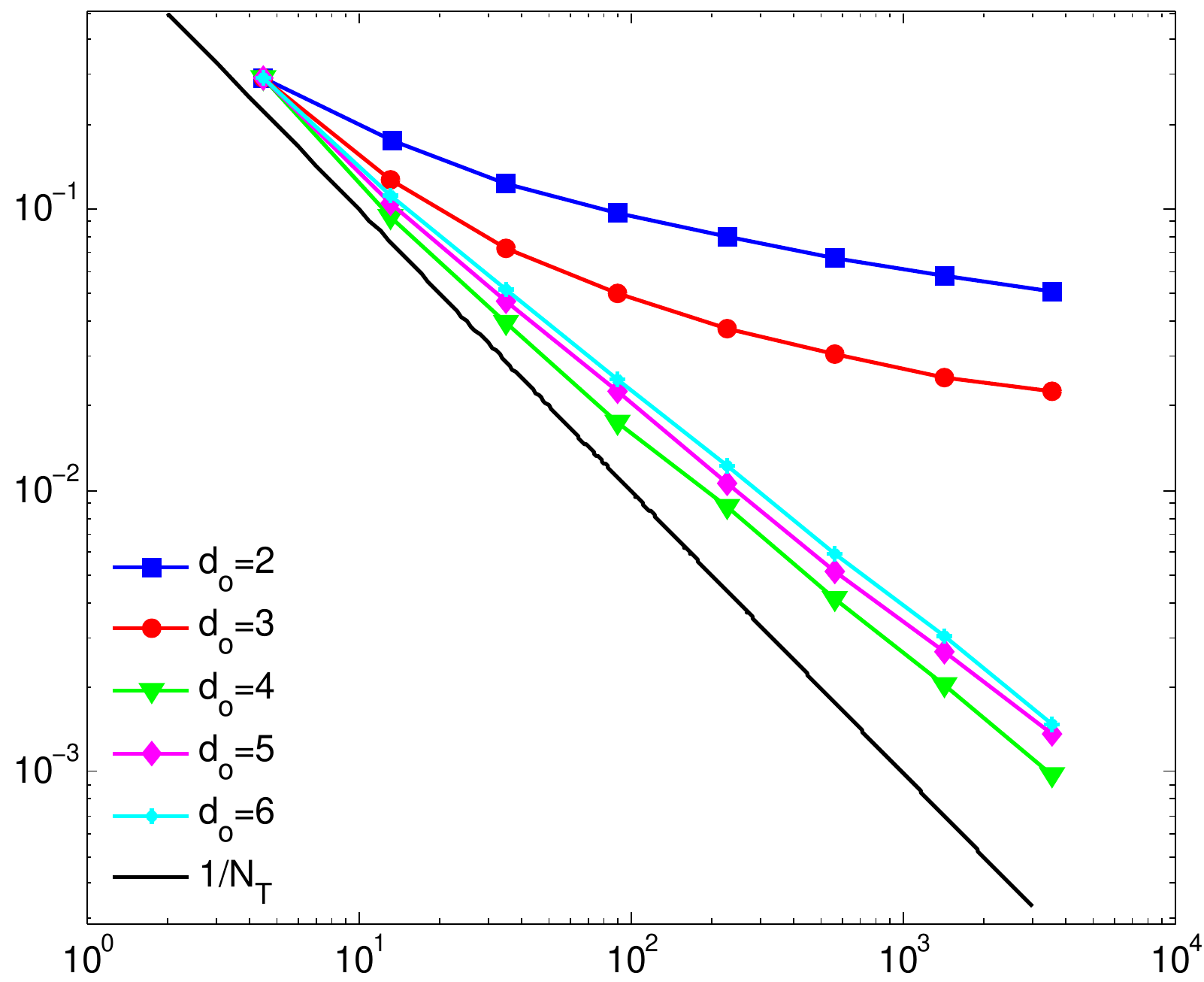}
  \put(-160,-10){Average size of infection, $N_T$}
	\put(-235,65){\rotatebox{90}{\large $\prob(\hv=v^*)$}}
  \caption{The probability of detection by the maximum likelihood estimator depends on the assumed degree $d_0$; the source cannot hide well below a threshold value of $d_0$.}
  \label{fig:irregular_adaptive}
\end{figure}

\vspace{0.1in}
\noindent\textbf{Simulation studies.}
We tested  adaptive diffusion  over random trees in which each node's degree was drawn i.i.d. from a fixed distribution.
Figure \ref{fig:irregular_adaptive} illustrates simulation results for random trees in which each node has degree 3 or 4 with equal probability, averaged over 100,000 trials.
By the law of large numbers, the number of nodes infected scales as $N_T \sim \E[D-1]^{T/2} = 2.5^{T/2}$, where $D$ represents the
degree distribution of the underlying random irregular tree.
The value of $d_0$ corresponds to a regular tree with size scaling as $(d_0-1)^{T/2}$.
Hence, one can expect that for $d_0-1 < 2.5$, the source is likely to be in the center of the infection, and
for $d_0>2.45$ the source is likely to be at the boundary of the infection.
Since the number of nodes in the boundary is exponentially larger than the number of nodes in the center,
the detection probability is lower for $d_0-1>2.5$.
This is illustrated in Figure \ref{fig:irregular_adaptive}, which matches our prediction.
In general, choosing $d_0=1+\lceil \E[D-1] \rceil$ provides the best obfuscation, and it is robust for values above that.
In this plot, data points represent successive even timesteps; their uniform spacing on the (log-scale) horizontal axis implies the message is spreading exponentially quickly. 

\begin{figure}[t]
	    \centering
  \includegraphics[width=.44\textwidth]{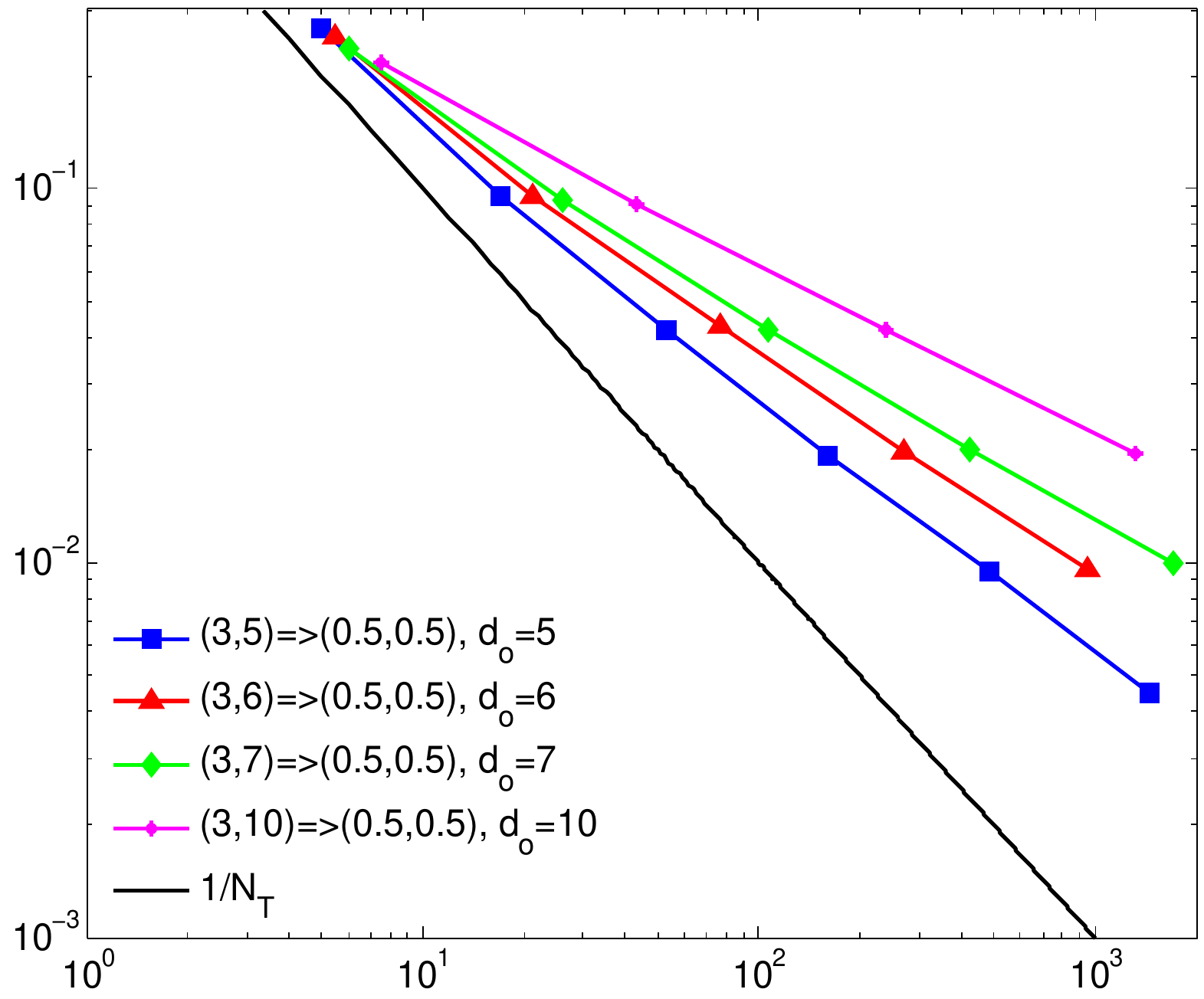}
  \put(-160,-10){Average size of infection, $N_T$}
	\put(-235,65){\rotatebox{90}{\large $\prob(\hv=v^*)$}}
  \caption{Adaptive diffusion no longer provides perfect obfuscation for highly irregular graphs.}
\label{fig:max_degree}
\end{figure}

Figure \ref{fig:max_degree} illustrates the probability of detection as a function of infection size while varying the degree distribution of the underlying tree. The notation $(3,5)=>(0.5,0.5)$ in the legend indicates that each node in the tree has degree 3 or 5, each with probability 0.5. For each distribution tested, we chose $d_0$ to be the maximum degree of each degree distribution.
The average size of infection scales as $N_T \sim \E[D-1]^{T} $ as expected,
whereas the probability of detection scales as $(d_{\rm min}-1)^{-T} = 2^{-T}$, which is independent of the degree distribution.
This suggests that adaptive diffusion fails to provide near-perfect obfuscation when the underlying graph is irregular, and
the gap increases with the irregularity of the graph.
In the next section, we quantify this gap, and gain intuition about how to reduce it.

\vspace{0.1in}
\noindent {\bf Probability of detection.}
In this section, we provide the probability of detection for adaptive diffusion over trees whose node degrees are drawn i.i.d. from some distribution $D$, for $d_0=\infty$. 
However, we cannot exactly use the ML estimator from Equation \ref{eq:mismatch}, which assumes the infinite irregular tree $G$ is given, and the source $v^*$ is chosen randomly from the nodes of $G$.
Equation \ref{eq:mismatch} is the correct ML estimator in any practical scenario,
but analyzing the probability of detection under this model requires a prior on the (infinitely many) nodes of $G$. 
We therefore consider a closely-related random process, in which we fix a source $v^*$ and generate $G$ (and consequently, $G_T$) on-the-fly. 
Specifically, at time $t=0$, $v^*$ draws a degree $d_{v^*}$ from $D$, and generates $d_{v^*}$ child nodes. The source picks one of these neighbors uniformly at random to be the new virtual source.
Each time a node $v$ is infected according to Protocol \ref{alg:adp_diff}, $v$ draws its degree $d_v$ from $D$, then generates $d_v-1$ child nodes.
For example, as soon as $v_2$, neighbor of $v^*$, receives the virtual source token, it draws its degree from $D$ and generates $d_{v_2}-1$ children.
The structure of the underlying, infinite contact network $G$ is independent of $G_T$ conditioned on the uninfected neighbors of the leaves of $G_T$, and need not be considered.
The adversary observes $\cG_T$, which is an unlabeled snapshot including $G_T$ and its uninfected neighbors.
We have that $\prob(\hv_{\rm MAP}=v^*|T)=\sum_{\cG_T}\prob(\cG_T|T)\prob(\hv_{\rm MAP}=v^*|\cG_T)$. We first consider $\prob(\hv_{\rm ML}=v^*|\cG_T)$. 

\subsubsection{Probability of Detection Given a Snapshot} 

The adversary observes this random process at time $T$ (i.e., it observes $\cG_T$, knowing that the interior $G_T$ are the infected nodes),  
and estimates one of the leaf node as an estimate of the true source which started the random process. 
The following theorem  analyzes the probability of detecting the true source for any estimate $\hv$, given a snapshot $\cG_T$. 

\begin{theorem}
	Under the above described random process of adaptive diffusion, 
	 an adversary observes the snapshot $\cG_T$ at an even time $T>0$ and estimates 
	 $\hv \in \partial  G_T$.  
	For any estimator $\hv$, the conditional probability of detection  is 
	\begin{equation}
	\prob(\hv = v^*|\cG_T) \;=\; \frac{1}{d_{v_T}}  \prod\limits_{\substack{w\in \phi(\hv,v_T) \\ \backslash\{v_T,\hv\}}} \frac{1}{ (d_w-1)} \;,
	\label{eq:thm_pd_map}
	\end{equation}
	where $v_T$ is the center of $\cG_T$, $\phi(\hv,v_T)$ is the (unique) path from $\hv$ to $v_T$, 
	$G_T$ is the interior of $\cG_T$ which is the infected sub-tree, 
	and $\partial G_T$ is the set of leaves of $G_T$. 
\label{thm:pd_conditional}
\end{theorem}

A proof is provided in Section \ref{sec:proofs_pd_irregular}. 
Intuitively, Equation \eqref{eq:thm_pd_map} is the probability that 
the virtual source starting from $\hv$ ends up at $v_T$ (up to some constant factor for normalization). 
This gives a simple rule for the adversary to achieve the best detection probability by computing the  MAP estimate: 
\begin{eqnarray} 
	\hv_{\rm MAP}^{(T)} &\in& \arg \max_{\hv} \; \prob(\hv^{(T)}=v^*|\cG_T)\;. 
	\label{eq:defmap}
\end{eqnarray}

\begin{corollary}
	\label{coro:pd_conditional}
	Under the hypotheses of Theorem \ref{thm:pd_conditional}, 
	the MAP estimator in \eqref{eq:defmap} can be computed as  
	\begin{eqnarray}
		\hv^{(T)}_{\rm MAP} &=& \arg \min\limits_{v\in \partial G_T}\prod\limits_{\substack{w\in \phi(v,v_T) \\ \backslash\{v_T,v\}}}(d_w-1)\; ,
		\label{eq:map}
	\end{eqnarray}
	achieving a conditional probability of detection 
	\begin{eqnarray}
		\prob(\hv^{(T)}_{\rm MAP}  =  v^*|\cG_T) &=& 
			\max\limits_{v\in \partial G_T}\frac{1}{d_{v_T} \prod\limits_{\substack{w\in \phi(v,v_T) \\ \backslash\{v_T,v\}}}(d_w-1)}.
	\label{eq:pdmap}
	\end{eqnarray}
\end{corollary}

\begin{figure}[h]
\begin{center}
	\includegraphics[width=.4\textwidth]{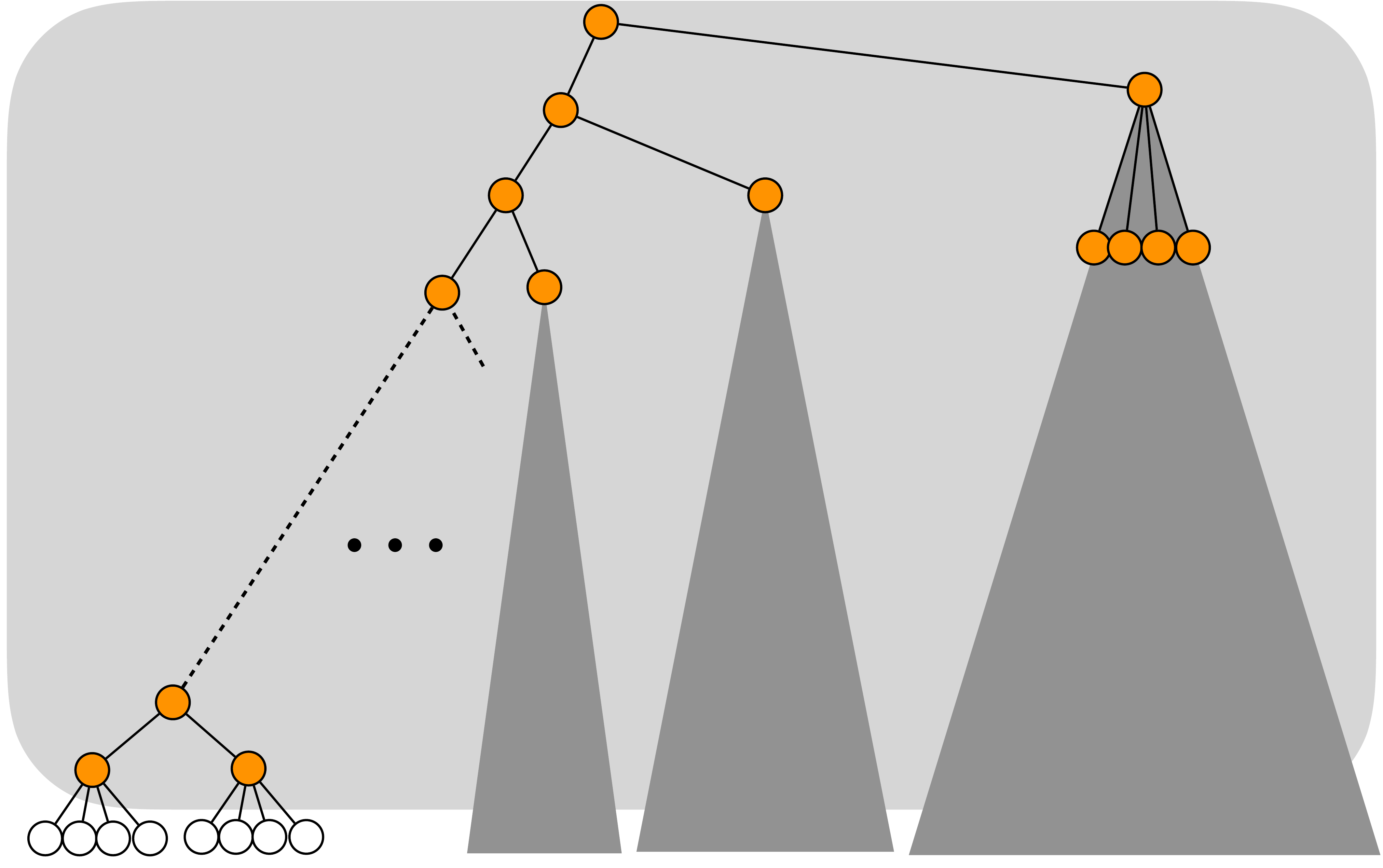}
	\put(-128,120){\small$v_T$}
	\put(-180,110){$G_T$}
	\put(-195,16){$v$}
\end{center}
\caption{An example of a snapshot emphasizing the sub-optimality of adaptive diffusion.}
\label{fig:example}
\end{figure}

When applied to regular trees, this recovers known results of \cite{KFSV14}, which  
confirms that adaptive diffusion provides strong anonymity guarantees under $d$-regular trees. 
But more importantly, Corollary \ref{coro:pd_conditional} characterizes how the anonymity guarantee depends on the general topology of the snapshot. 
We illustrate this in two extreme examples: a regular tree and an extreme example in Figure \ref{fig:example}.

For a $d$-regular tree, where all nodes have the same degree, 
the size of infection at even time $T$ is the number of nodes in a $d$-regular tree of depth $T/2$:  
\begin{eqnarray} 
	N_T\;=\;\frac{d(d-1)^{T/2}}{d-2} + \frac{2}{d-2} \;.
\end{eqnarray}
To achieve a perfect obfuscation, we want the probability of detection to decay as $1/N_T$. 
We can apply 
Corollary \ref{coro:pd_conditional} to this $d$-regular tree and  show the probability of detection is 
$((d-1)/d)(d-1)^{-T/2})$, which recovers one of the known results in \cite[Proposition 2.2]{KFSV14}.  
This confirms that adaptive diffusion achieves near-perfect obfuscation, 
up to a small factor of $(d-1)/(d-2)$. 

On the other hand, when there exists a path to a leaf node consisting of low-degree nodes, 
adaptive diffusion can be sub-optimal, and the gap to optimality can be made  arbitrarily large. 
Figure \ref{fig:example} illustrates such an example. 
This is a tree where all nodes have the same degree $d=5$, except for those nodes along the path from the 
center $v_T$ to a leaf node $v$, including $v_T$ and excluding $v$. 
The center $v_T$ has degree two and the nodes in the path have degree three. 
Hence, the shaded triangles indicate $d$-regular sub-trees of appropriate heights. 
The size of this infection is $N_T=((d-1)^{T/2+1}/(d-2)^2) (1+o(1))$.
Ideally, 
one might hope to achieve a probability of detection that scales as $1/(d-1)^{T/2}$. 
However, Corollary \ref{coro:pd_conditional}  shows that the adaptive diffusion achieves probability of detection $1/2^{T/2}$, 
with the leaf node $v$ achieving this maximum in Equation \eqref{eq:pdmap}. 
Hence, there is a multiplicative gap of $((d-1)/2)^{T/2}$. 
By increasing $d$, the gap can be made arbitrarily large.
On the other hand, such an extreme topology is rare under the i.i.d tree model. 

\subsubsection{Concentration of Probability of Detection}

Depending on the topology, 
adaptive diffusion can be significantly sub-optimal. 
A natural question is 
``what is the typical topology of a graph resulting from the 
random tree model?''  
Under the model introduced previously,  
we give a concrete answer.  
Perhaps surprisingly,  this typical topology can be characterized by solving a simple convex optimization.



We are interested in the following extremal value 
\begin{eqnarray}
	\Lambda_{G_T} &\equiv &
	d_{v_T} \min\limits_{v\in \partial G_T}\prod\limits_{\substack{w\in \phi(v,v_T) \\ \backslash\{v_T,v\}}}(d_w-1) \;,
	\label{eq:deflambda}
\end{eqnarray}
which captures the topology of the snapshot. 
 We want to characterize the typical value of this
function over random tree $G_T$
 resulting from the adaptive diffusion process. 

Observe that the distribution of the balanced tree $G_T$ follows a simple branching process known as Galton-Watson process.   
This is because $G_T$ resulting from adaptive diffusion has the same distribution, independent of the location of the source $v^*$. 
We consider a given degree distribution $D$. 
We use $D$ to denote both a random variable and its distribution---the distinction should be clear from context. 
The random variable $D$ has  support $\boldsymbol f=(f_{1},\ldots,f_{\eta})$ associated with probability $\boldsymbol p=(p_1,\ldots,p_{\eta})$ such that the degree of node $v$ is i.i.d. with 
\begin{eqnarray}	
	d_v &=& \left\{ 
	\begin{array}{rl}
		f_1& \text{with probability }p_1\;,\\
		\vdots& \vdots \\
		f_\eta& \text{with probability }p_\eta\;,
	\end{array}
	\right.
	\label{eq:defD}
\end{eqnarray}
where $2 < f_1 < f_2 < \cdots < f_\eta$ are integers and 
the positive  $p_i$'s sum to one. 
We also assume $D$'s support set has at least two elements, i.e., $\eta\geq 2$.

Note that the adaptive diffusion 
always passes the virtual source token to a uniformly-chosen neighbor. 
It is straightforward to show that adaptive diffusion starting from a leaf node $v^*$ has the same distribution over graphs as the following branching process, denoted $\overline G_T$: 
at time $T=0$ a root node, which we denote as the virtual source $v_T$, 
creates $D$ offspring. At each subsequent even time step, each leaf node in $G_T$ creates 
new offspring independently according to $D-1$ (where we subtract one because each leaf is already connected to its parent).  
This process is repeated until time step $T$, which generates a random tree $G_T$. 
More precisely, the two branching processes are equal in distribution: $\overline G_T {\buildrel D \over =} G_T$.
This can be seen by observing that conditioned on the path of nodes $\phi(v^*,v_T)$, the branching processes are identical. 
Since the node degrees in this path are drawn independently, the path is equally distributed whether it starts from the virtual source $v_T$ or the leaf node $v^*$.

The following theorem provides a concentration inequality on the extremal quantity $\Lambda_{G_T}$, which in turn determines the probability of detection as provided by Corollary \ref{coro:pd_conditional}: 
\begin{eqnarray}
	\prob( \hv^{(T)}_{\rm MAP} = v^*|G_T)  &=&  \frac{1}{\Lambda_{G_T}}\;.
	\label{eq:pd_lambda}
\end{eqnarray}

\begin{theorem}
	For an even $T>0$, suppose a random tree $G_T$ is generated from the root $v_T$ according to the Galton-Watson 
		 process with i.i.d. degree distribution  $D$, where $\boldsymbol f$ and $\boldsymbol p$ are defined as in  \eqref{eq:defD}, 
	then   the following results hold: 
	\begin{itemize}
	\item [$(a)$] If $p_1(f_1-1)>1$, for any positive $\delta>0$, there exists positive constants 
	$C_{D,\delta}$ and $ C'_{D,\delta}$ that depend only on the degree distribution and the choice of $\delta$ such that 	
	\begin{equation}
	\prob \left (\left |   \frac{\log(\Lambda_{G_T})}{T/2} - \log(f_1-1) \right |  > \delta \right )  \leq e^{-C_{D,\delta} T} \;,
	\end{equation}
	for an even time $T \geq C'_{D,\delta}$.
	\item [$(b)$] If $p_1(f_1-1) < 1$, define the mean number of children: 
	\[
	\mu_D \;\equiv\; \sum_{i=1}^{\eta} p_i(f_i-1) \;,
	\]
	and the set 
	\begin{eqnarray}
	{\cal R}_{D} \;=\;  \big\{\,  \boldsymbol r \in S_\eta  ~|~  \log(\mu_D) \geq  D_{\rm KL}(\boldsymbol r\| \boldsymbol \beta ) \,\big\} \, ,
	\label{eq:r_star_set}
	\end{eqnarray}
	where $S_\eta$ denotes the $\eta$-dimensional probability simplex, 
	$D_{\rm KL}(\cdot \| \cdot)$ denotes Kullback-Leibler divergence, and 
	$\boldsymbol \beta$ is a length-$\eta$ probability vector in which $\beta_i=p_{i}(f_{i}-1)/\mu_D$.
	Further, define $\boldsymbol r^*$ as follows:
	\begin{equation}
	\begin{aligned}
	\boldsymbol r^*&=& \underset{\boldsymbol r \in \mathcal R_D}{\arg \min}
	&& &  \big\langle \,\boldsymbol r\,,\,\log (\boldsymbol f-1)\, \big\rangle \;, 
	\end{aligned}
	\label{eq:thm_general_d}
	\end{equation}
	where 
	$\langle \boldsymbol r,\log (\boldsymbol f-1)\rangle = \sum_{i=1}^{\eta} r_{i} \log \left (f_{i}-1\right )$.
	Then for any $\delta>0$, there exists positive constants 
	 $C_{D,\delta}$ and $C'_{D,\delta}$  
	 that only depend on the 	degree distribution $D$ and  the choice of $\delta>0$ such that 
	\begin{eqnarray}
	\prob \left ( \left | \frac{\log(\Lambda_{G_T})}{T/2} - \langle \boldsymbol r^*, \log(\boldsymbol f-1)  \rangle \right |  > \delta \right ) \; \leq \; e^{-C_{D',\delta} T}
	\end{eqnarray}
	for an even time $T\geq C'_{D,\delta}$. 
	\end{itemize}
	\label{thm:product}
\end{theorem}
The results in parts $(a)$ and $(b)$ can be merged, in the sense that 
the solution of \eqref{eq:thm_general_d} is $ \boldsymbol r^* = [1,0,\ldots,0]$ when $p_1(f_1-1)>1$.
A proof of this theorem is provided in Section \ref{sec:proofs_product}.  
Putting it together with \eqref{eq:pd_lambda}, it follows  that the probability of detection concentrates around 
\begin{eqnarray*}
	-\frac{2}{T}\log\big(\, \prob( \hv^{(T)}_{\rm MAP} = v^*)\,\big)  &\simeq& \langle \boldsymbol r^* \,,\, \log(\boldsymbol f-1) \rangle \;,
\end{eqnarray*}
in case $(b)$ and around $\log(f_1-1)$ in case $(a)$. 
Here $\simeq$ indicates concentration for large enough $T$. 
We want to emphasize that $\boldsymbol r^*$ can be computed using off-the-shelf optimization tools, since 
the program in \eqref{eq:thm_general_d} is a convex program of dimension $\eta$. 
This follows from the fact that the objective is linear in $\boldsymbol r$ and 
the feasible region is convex since KL divergence is convex in $\boldsymbol r$. 

\begin{figure}[h]
	    \centering
  \includegraphics[width=.4\textwidth]{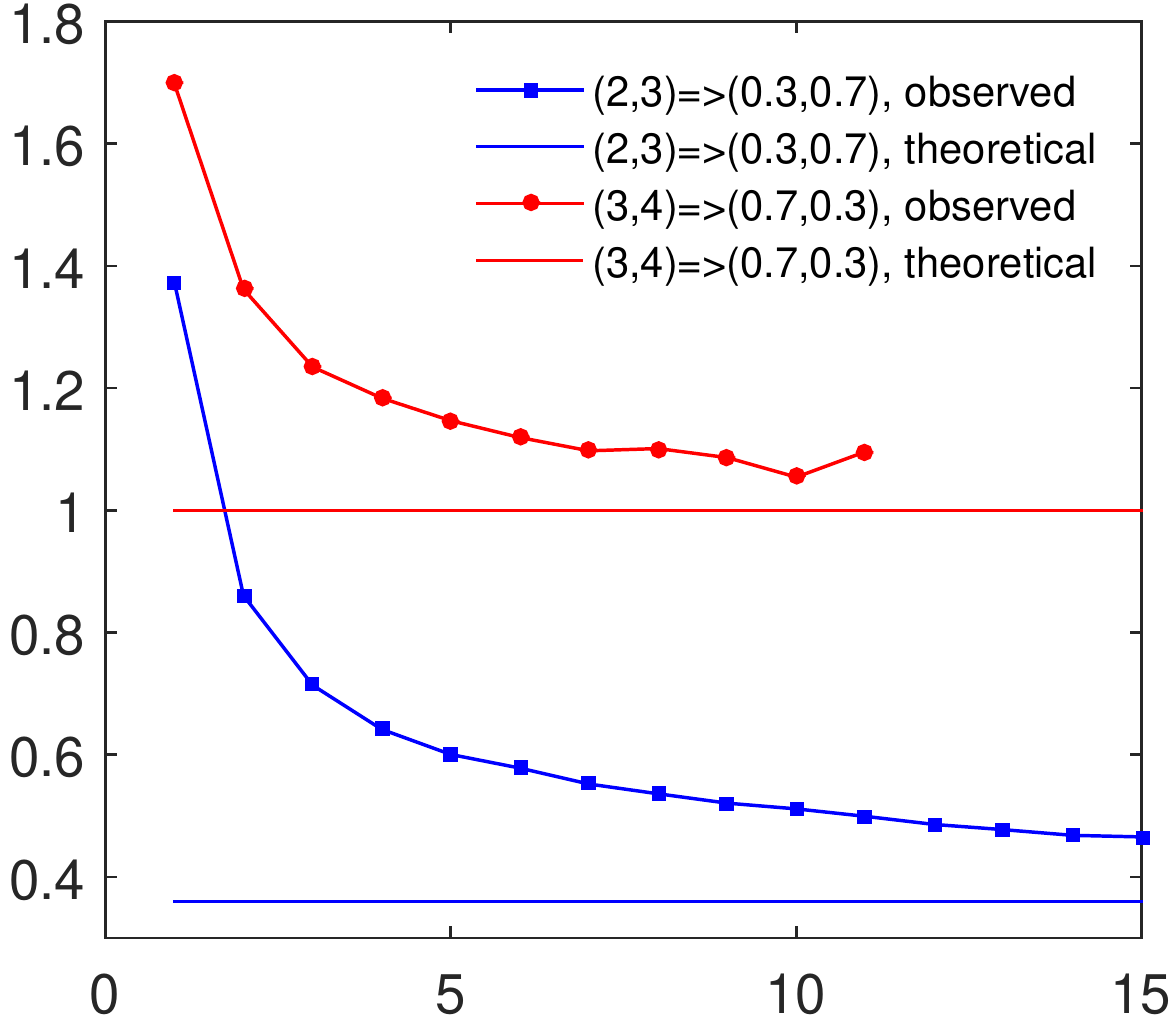}
  \put(-170,-10){Radius of the infected sub-tree, $T/2$}
	\put(-220,38){\rotatebox{90}{ $-(2/T)\log_2(\prob(\hv=v^*))$}}
 	\caption{Empirical verification of Theorems \ref{thm:pd_conditional} and \ref{thm:product}. 
  We observe that the probability of detection converges in time to the predicted values, which depend only on the underlying degree distribution.}
  	\label{fig:irregular_tree_thm}
\end{figure}

For example, if $D$ is $3$ w.p. $0.7$ or $4$ w.p. $0.3$, then this falls under
 case $(a)$. 
The theorem predicts  the probability of detection to decay as $(3-1)^{-T/2}$. On the other hand, if 
 \begin{eqnarray*}
 	D  = \left\{ \begin{array}{rl}
	2& \text{with probability }0.3\\ 
	3& \text{with probability }0.7
	\end{array} 
	\right. \;,
\end{eqnarray*}
 then this falls under case $(b)$ with $\mu_D=1.7$, $\beta_1=0.3/1.7$, and $\beta_2=1.4/1.7$.  
In this case, the exponent is a solution of the following optimization for $\boldsymbol r = [r,1-r]$: 
\begin{eqnarray*}
	\underset{r\in\reals}{\text{minimize}} && r \log 1 + (1-r) \log 2\\
	\text{subject to} &&  r \log \frac{1.7 r}{0.3}  + (1-r)\log \frac{1.7(1-r)}{1.4}  \leq \log (1.7) \\
	&&r \in[0,1]
\end{eqnarray*}
It follows that the optimal solution is $\boldsymbol r^*\simeq [0.64,\; 0.36]$ and the probability of detection decays as 
$2^{-0.36(T/2)}$. 
Figure \ref{fig:irregular_tree_thm} confirms this prediction with simulations for these  examples.

Theorem \ref{thm:product}  provides a simple convex program that 
computes the probability of detection for any degree distribution. 
For random trees, this quantifies the gap between what adaptive diffusion can guarantee
 and the perfect obfuscation one desires. 
We define the rescaled log-multiplicative gap as 
\begin{eqnarray*}
	 \Delta_D &\equiv& \frac{2}{T} \,\log \frac{\prob(v^{(T)}_{\rm MAP}=v^*)}{1/\E[|\partial G_T|]}\;, 
\end{eqnarray*}
where $|\partial G_T|$ is the total number of candidates in a snapshot. 
It is not difficult to show that $\E[|\partial G_T|]=\mu_D^{T/2}$, and it follows that 
$\Delta_D \simeq \log \mu_D - \langle \boldsymbol r^*, \log(\boldsymbol f -1 )\rangle$. 
For example, $\Delta_D=0$ for regular trees, and 
$\Delta_D = \log_2 2.3-\log_2 2 = 0.20$ 
for the first example under case $(a)$
and $\Delta_D = \log_2 1.7 - 0.36 = 0.41$ 
 for the second example under case $(b)$. 





%
%

\vspace{0.1in}
\noindent\textbf{Simulation studies.}
Figure \ref{fig:irregular_tree_thm} empirically checks the predictions in Theorems \ref{thm:pd_conditional} and \ref{thm:product}.
The distribution with support $\boldsymbol f=(3,4)$ with probabilities $\boldsymbol p=(0.5,0.5)$ addresses case 1 from the theorem, where $p_1(f_1-1)>1$. 
The distribution with support $\boldsymbol f=(2,3)$ with probabilities $\boldsymbol p=(0.3,0.7)$ addresses case 2, where $p_1(f_1-1)<1$.
In both examples, we observe that the empirical $\log(\prob(\hat v=v^*))/(T/2)$ converges to the theoretical value predicted in Figure \ref{fig:irregular_tree_thm}. 
However, this convergence may be slow, and the timestep duration of these experiments was limited by computational considerations since the graph size grows exponentially in time.

%

\subsubsection{Preferential Attachment } 
\label{sec:pref_attach}

Our analysis reveals that adaptive diffusion can be significantly sub-optimal, 
when the underlying graph degrees are highly irregular. 
To bridge this gap, 
we introduce a family of protocols we call {\em Preferential Attachment Adaptive Diffusion (PAAD)}. 
We analyze the performance of PAAD and provide numerical simulations showing that PAAD improves over adaptive diffusion when 
 degrees are irregular. 

The reason for this gap is that  
in typical random trees, there are nodes that are 
significantly more likely to be the source, compared to other typical candidate nodes. 
To achieve near-perfect obfuscation, we want all candidate nodes to have 
similar posterior probabilities of being the source. 
To balance the posterior probabilities of leaf nodes, 
we suggest passing the virtual source with higher probability to high-degree nodes. 
We propose a family of protocols based on this idea,  
and make this intuition precise in  Theorem \ref{thm:pd_conditional_pa}. 

PAAD is based on adaptive diffusion, 
but we modify how virtual sources are chosen. 
We parametrize this family of protocols by a non-negative integer  $g$. 
When a new virtual source is to be chosen, 
instead of choosing uniformly among its neighbors (except for the previous virtual source), 
the new virtual source is   selected with probability weighted by the size of its $g$-hop neighborhood. 
Let $\mathcal N_g(v)$ denote the set of $g$-hop neighbors of node $v$, and let $\mathcal N_g(v,w) $ denote the same set, removing any nodes $z$ for which $w \in \phi(z,v)$, where $\phi(z,v)$ denotes the path between $z$ and $v$. 
Then for instance, if $g=1$, then 
each time the virtual source is passed from $v_T$ to $v_{T+2}$, 
it is passed to a neighbor $w\in\mathcal N_1( v_T,v_{T-2})$ with probability proportional to $d_w-1$: 
\begin{eqnarray*}
	\prob (v_{T+2} = w) = \frac{d_w-1}{\sum_{w' \in \mathcal N_1( v_T,  v_{T-2})} (d_{w'}-1) } \;.
\end{eqnarray*}

For general $g$, 
the probability is proportional to the 
size of the candidate $w$'s $g$-hop local neighborhood, excluding those in the direction of the current virtual source $v_T$.  
Each virtual source $v_T$ chooses the next virtual source as follows: 
for any node $w\in \mathcal N_1(v_T,v_{T-2})$,
\[
P(v_{T+2} = w) = \frac{|\mathcal N_g(w,v_T) |}{\sum_{w' \in \mathcal N_1(v_T, v_{T-2})}|\mathcal N_g(w' ,v_T)|}\;. 
\]

PAAD encourages the virtual source to traverse high-degree nodes. 
This balances the posterior probabilities, by strengthening the probability of leaf nodes whose path contain high-degree nodes, 
while weakening those with low-degree nodes. 

This intuition is made precise in the following theorem, 
which  analyzes the probability of detection for a given snapshot. 
Define the probability that the sequence of decisions on choosing the virtual sources results in the path from 
a source $v$ to the current virtual source $v_T$ as 
$
	Q(\mathcal G_T, v) \equiv  \prod_{t=1}^{T/2} \prob(v_{2t} = w_t) \;, 
$
where \\$\phi(v,v_T)=(w_0=v,w_1,w_2,\ldots,w_{T/2-1},w_{T/2}=v_T)$. 
The specific probability depends on the choice of $g$ and the topology of the underlying tree. 
Note that the progression of the virtual source now depends on $g$-hop neighborhood, and we therefore  define $\mathcal G_T$ to include the current infected subgraph $G_T$ and  its
$(g+1)$-hop neighborhood. 

\begin{theorem}
	Suppose a node $v^*$ starts to spread a message at time $t=0$ according to PAAD, where the 
	 underlying irregular tree is generated according to the  random branching process described in Section 	\ref{sec:irregular_trees}. 	
	At a certain even time $T\geq 0$, an adversary observes 
	the snapshot of the infected subtree $\cG_T$ and computes a MAP estimate of the source $v^*$. 
	Then, the following results hold: 
	\begin{itemize}
	\item [$(a)$] The MAP estimator   is 
	\begin{equation}
	\hv_{\rm MAP} \;=\; \arg \max\limits_{v\in \partial G_T} \,d_v\, Q(\cG_T,v)
	\label{eq:MAP_general}
	\end{equation}
	where $\partial G_T$ denotes the leaves of $G_T$.
	\item [$(b)$] The conditional probability of detection achieved by the MAP estimator   is
	\begin{equation}
	\prob(\hv_{\rm MAP}=v^*|\cG_T) \,=\, \frac{\max_{v \in \partial G_T}d_v\,Q(\cG_T,v)}{\sum_{w\in \partial G_T} d_w \,Q(\cG_T,w)}
	\label{eq:thm_pd_MAP_pa}
	\end{equation}
	\end{itemize}
	\label{thm:pd_conditional_pa}
\end{theorem}

The proof relies on the techniques developed for Theorem \ref{thm:pd_conditional}, 
and is omitted due to space limitation. 
The example from Figure \ref{fig:example} illustrates the power of PAAD. 
For this class of snapshots, it is straightforward to show that under adaptive diffusion, 
$P_D^{AD} =2^{-T/2}$, 
whereas under 1-hop PAAD, 
\[
P_D^{PAAD} \leq \frac{2}{\left ( d-1 \right )^{T/2-1}-1}.
\]
Notice from these expressions that $P_D^{PAAD}$ scales as $(d-1)^{-T/2}$, which achieves perfect obfuscation, whereas
regular adaptive diffusion scales as $2^{-T/2}$.


This shows that there exist snapshots where PAAD significantly improves over adaptive diffusion. 
However, such examples are rare under the random tree model, and 
there are also  examples of snapshots where adaptive diffusion can achieve a better obfuscation than PAAD.  
To complete the analysis, we would like to show the analog of Theorem \ref{thm:product} for PAAD. 
However, the observed snapshot is no longer generated by a standard Galton-Watson branching process, 
 due to the preferential attachment. 
 The analysis techniques developed for Theorem \ref{thm:product} do not generalize, 
 and new techniques seem to be needed for a technical analysis. 
 This is outside the scope of this manuscript, but 
we show simulations suggesting that PAAD improves over adaptive diffusion. 

\begin{figure}[t]
	    \centering
  \includegraphics[width=.4\textwidth]{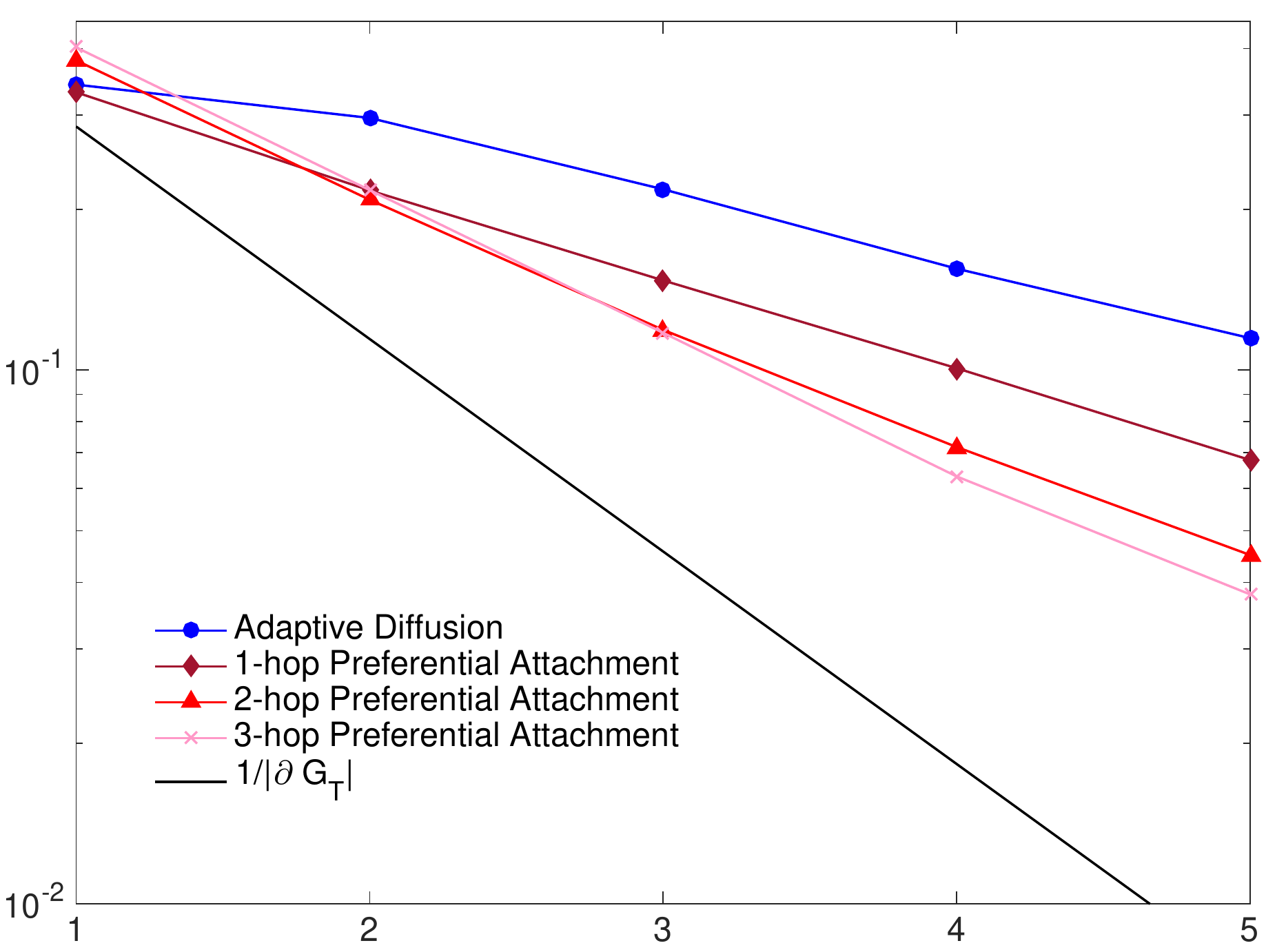}
  \put(-145,-10){Even time step, $T/2$}
	\put(-220,60){\rotatebox{90}{${\large \prob(\hv=v^*)}$}}
  \caption{Probability of detection of regular adaptive diffusion compared to 1-, 2-, and 3-hop preferential attachment adaptive diffusion (PAAD).}
  \label{fig:paad}
\end{figure}

\begin{figure}[t]
	    \centering
  \includegraphics[width=.4\textwidth]{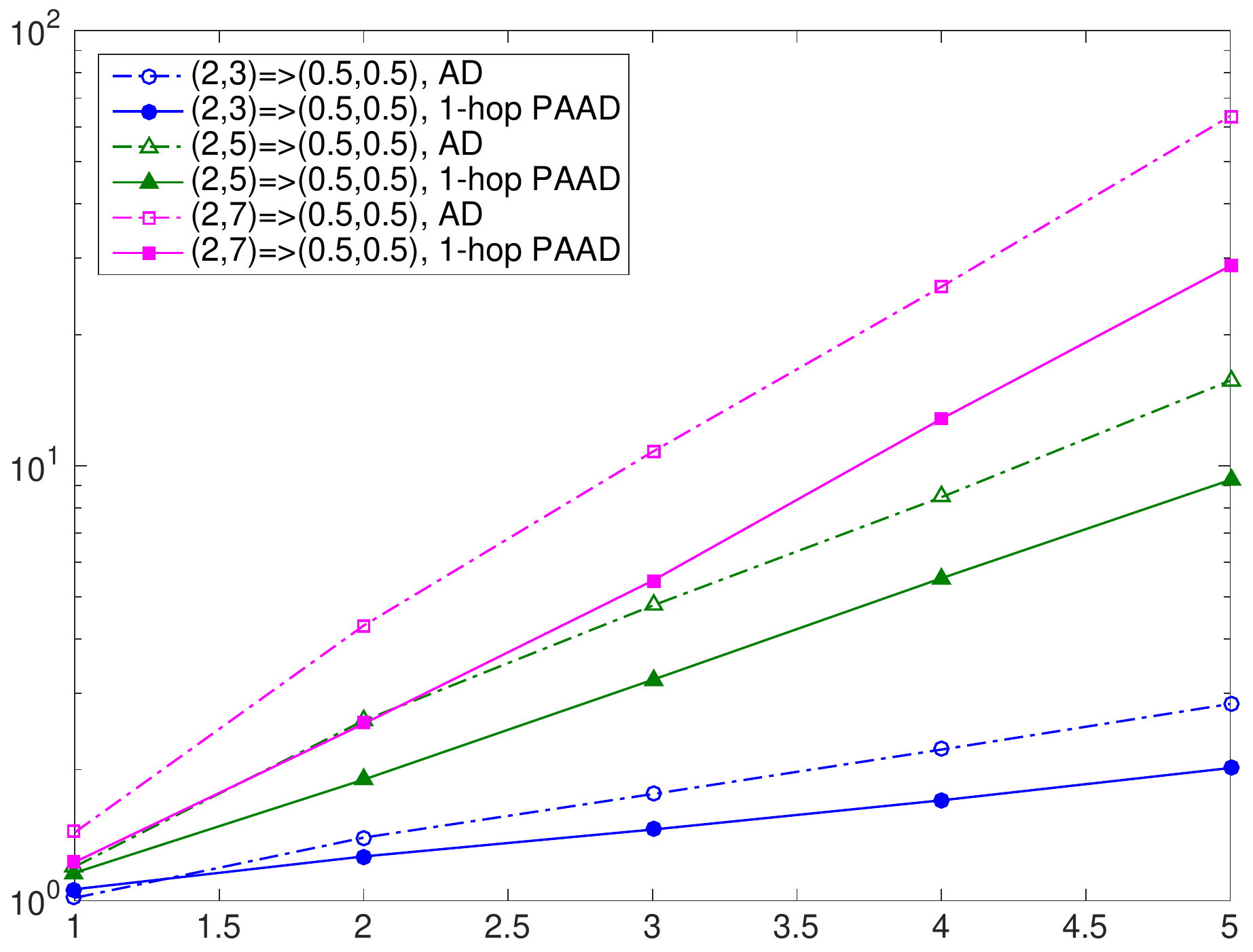}
  \put(-145,-10){Even time step, $T/2$}
	\put(-220,30){\rotatebox{90}{ $\large \prob(\hv=v^*) ~/~ (1 / |\partial G_T|)$}}
  \caption{Ratio of observed probability of detection to lower-bound probability of detection, for a range of degree distributions. PAAD has better anonymity properties than regular adaptive diffusion over random, irregular trees.}
  \label{fig:paad2}
\end{figure}

\vspace{0.1in}
\noindent \textbf{Simulation studies.}
PAAD requires each virtual source to know some information about its local neighborhood on the contact network; in exchange, we observe empirically that it hides the source better than traditional adaptive diffusion. 
Figure \ref{fig:paad} shows the probability of detection over graphs with a degree distribution of support $\boldsymbol f=(2,5)$ with probability $\boldsymbol p=(0.5,0.5)$. 
The results are averaged over 10,000 realizations of the random graph and the spreading sequence.
This plot shows empirically that preferential attachment adaptive diffusion exhibits better hiding properties than regular adaptive diffusion, and that the benefit of preferential attachment increases with the size of the neighborhood considered for preferential attachment (e.g., one-hop vs. two-hop).
Notice that our lower bound on probability of detection is $1/|\partial G_T|$ rather than $1/N_T$, as in \cite{KFSV14}; this is because we constrain the source to always be at one of the leaves of the graph, so $1/|\partial G_T|$ lower bounds the probability of detection.

Figure \ref{fig:paad2} computes the ratio of the observed probability of detection to a lower bound on the probability of detection (i.e., $1/|\partial G_T|$), for both adaptive diffusion (AD) and  one-hop PAAD. Empirically, we observe that the advantage of PAAD is greater when the degree distribution is more imbalanced (i.e., when $f_{\rm max} - f_{\rm min}$ is large).

\subsection{General Graphs}
\label{sec:numerical}
In this section, we demonstrate how adaptive diffusion fares over graphs that involve cycles, irregular degrees, and finite graph size. We provide theoretical guarantees for the special case of two-dimensional grid graphs, and we show simulated results over a social graph dataset.

\subsubsection{Grid graphs}
Here, we derive the optimal parameters $\alpha(t,h)$ for spreading with adaptive diffusion over an infinite \emph{grid graph}, defined as the graph Cartesian product of two infinite line graphs. 
This example highlights challenges associated with spreading over cyclic graphs, while still providing a regular, symmetric structure.
To spread over grids, we make some changes to the adaptive diffusion protocol, outlined in Protocol \ref{alg:grid} (grid adaptive diffusion).

First, standard adaptive diffusion requires the virtual source to know its distance from the true source. Over trees, this information was transmitted by passing a distance counter, $h_t$, that was incremented each time the virtual source changed; since the network was a tree, this distance from the source was non-decreasing 
as long as the virtual source was non-backtracking. 
However, on a cyclic graph (e.g., a grid), the virtual source's non-backtracking random walk could actually cause its distance from the true source to \emph{decrease} with time. 
We wish to avoid this to preserve adaptive diffusion's anonymity guarantees. 

\begin{figure}
	    \centering
  \includegraphics[width=.3\textwidth]{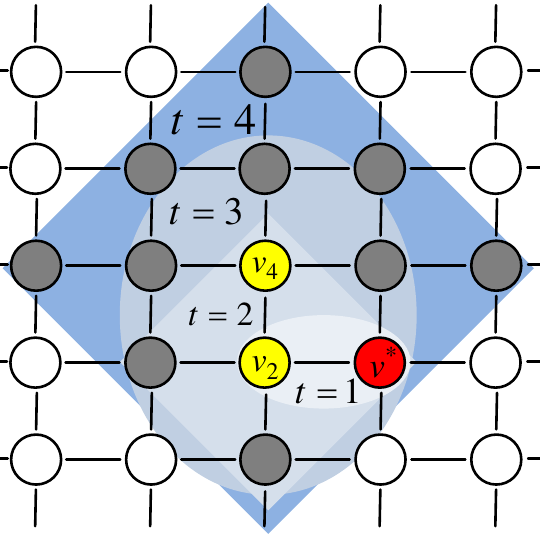}
  \caption{Grid adaptive diffusion spreading pattern. }
\label{fig:grid}
\end{figure}

Therefore, instead of passing the raw hop distance $h_t$ to each new virtual source, grid adaptive diffusion passes \emph{directional} coordinates $(h_t^H, h_t^V)$ detailing the virtual source's horizontal and vertical displacement from the source, respectively.
For example, in Figure \ref{fig:grid}, the virtual source $v_4$ would receive parameters $(h_t^H, h_t^V)=(-1,1)$ because it is one hop west and one hop north of the true source.
This indexing assumes some notion of directionality over the underlying contact network; nodes should know whether they received a message from the north, south, east, or west.
If a virtual source chooses to move, it always passes the token to a node that is further away from the true source, i.e. $|h_{t+1}^H|+|h_{t+1}^V|\geq |h_{t}^H|+|h_{t}^V|$.

To maintain symmetry about the virtual source, we also modify the message-passing algorithm.
Just as in adaptive diffusion over trees, when a \emph{new} virtual source sends out branching messages, it sends them in every direction except that of the old virtual source.
However, unlike adaptive diffusion over trees, each branch message has up to two ``forbidden" directions: the direction of the previous virtual source, and the direction of the node that originated the branching message (these might be the same). Thus, if a branch message is sent west, and the previous virtual source was south of the current virtual source, each node would \emph{only} propagate the message west and/or north. Whenever a node receives a branch message and its neighbors are not all infected, it infects all uninfected neighbors. As in adaptive diffusion over trees, two waves of directional branching messages are sent each time the virtual source moves, in every direction except that of the old virtual source. 
If the virtual source instead chooses to stay fixed, then the same rules hold, except the new virtual source only sends one wave of branch messages, symmetrically in every direction. 

Given the spreading protocol, we can choose $\alpha(t,h)$ to give optimal hiding:
\begin{eqnarray}
\alpha(t,h) = \frac{t-2(h-1)}{t+4}.
\label{eq:alpha_grid}
\end{eqnarray}
Under these conditions, 
the following result shows that we achieve perfect obfuscation, i.e. $\prob(\hv_{\rm ML} = v^*) = 1/N_T + o(1/N_T)$. 

\begin{propo}
\label{propo:grid}
Suppose the contact network is an infinite grid,
and one node $v^*$ in $G$ starts to spread a message according to
Protocol  \ref{alg:grid} (grid adaptive diffusion) at time $t=0$, with $\alpha(t,h)$ chosen according to Equation \eqref{eq:alpha_grid}.
At a certain time $T\geq0$ an adversary estimates the location of the source $v^*$
using the maximum likelihood estimator $\hv_{\rm ML}$.
The following properties hold for Protocol \ref{alg:grid}:

\begin{itemize}
	\item [$(a)$] the number of infected nodes at time $T$ is
		\begin{eqnarray}
		N_{T}\geq \frac{(T+1)^2}{2}
		\label{eq:n_diff_grid}
		\end{eqnarray}
	\item [$(b)$] the probability of source detection for the maximum likelihood estimator at time $T$ is
	\begin{eqnarray}
		\prob\left(\hat{v}_{\rm ML}=v^*\right)  \leq \frac{2}{(T+3)(T-1)}.
		\label{eq:p_diff_grid}
	\end{eqnarray}
\end{itemize}
\end{propo}
(Proof in Section \ref{sec:proofs_snapshot_grid})

The baseline infection rate for deterministic, 
symmetric spreading is $N_T=T^2 + (T+1)^2$. 
Grid adaptive diffusion infection rate is  within a constant factor of this maximum possible rate, 
and it achieves perfect obfuscation over grid graphs. 
The price to pay for this non-tree graph is that $(a)$ 
a significant amount of metadata needs to be transmitted to coordinate the spread---particularly with respect to the directionality of messages; 
and $(b)$ the position of the nodes w.r.t. a global reference needs to be known. 
Hence, the current implementation of the grid adaptive diffusion has a limited scope, and it remains an open question how to avoid such requirements for grids and still
achieve a perfect obfuscation.  

\subsubsection{Real-world social graphs}
In this section, we provide simulation results from running adaptive diffusion over an underlying connectivity network of 10,000 Facebook users, as described by the Facebook WOSN dataset \cite{viswanath-2009-activity}.
We eliminated all nodes with fewer than three friends (this approach is taken by several existing anonymous applications so users cannot guess which of their friends originated the message), which left us with a network of 9,502 users.

Over this underlying network, we selected a node uniformly at random as the rumor source, and spread the message using adaptive diffusion for trees. We did not use grid adaptive diffusion because Protocol \ref{alg:grid} assumes the underlying graph has a symmetric structure with a global notion of directionality, whereas the tree-based adaptive diffusion makes no such assumptions.
We set $d_0 = \infty$, which means that the virtual source is always passed to a new node (i.e., $\alpha_d(t,h)=0$).
This choice is to make the ML source estimation faster; other choices of $d_0$ may outperform this naive choice.
To preserve the symmetry of our constructed trees as much as possible, we constrained each infected node to infect a maximum of three other nodes in each timestep.
We also give the adversary access to the undirected infection {\em subtree}
that explicitly identifies all pairs of nodes for which one node spread the infection to the other.
This subtree is overlaid on the underlying contact network, which is not necessarily tree-structured.
We demonstrate in simulation (Figure~\ref{fig:facebook})  that even with this strong side information, the adversary 
can only identify the true message source with low probability.

Using the naive method of enumerating every possible message trajectory, it is computationally expensive to find the exact ML source estimate since there are $2^T$ possible trajectories, depending on whether the virtual source stayed or moved  at each timestep. If the true source is one of the leaves, we can closely approximate the ML estimate \emph{among all leaf nodes}, using the same procedure as described in \ref{sec:irregular_trees}, with one small modification:
in  graphs with cycles, the term $(d_{v_{j_k}}-1)$ from equation \eqref{eq:likelihood} should be substituted with $(d^u_{v_{j_k}}-1)$, where $d^u_{v_{j_k}}$ denotes the number of uninfected neighbors of $v_{j_k}$ at time $j_k$. Loops in the graph cause this value to be time-varying, and also dependent on the location of $v_0$, the candidate source.
We did not approximate the ML estimate for non-leaves because the simplifications used in Section \ref{sec:irregular_trees} to compute the likelihood no longer hold, leading to an exponential increase in the problem dimension.
\begin{figure}[h]
	    \centering
  \includegraphics[width=.44\textwidth]{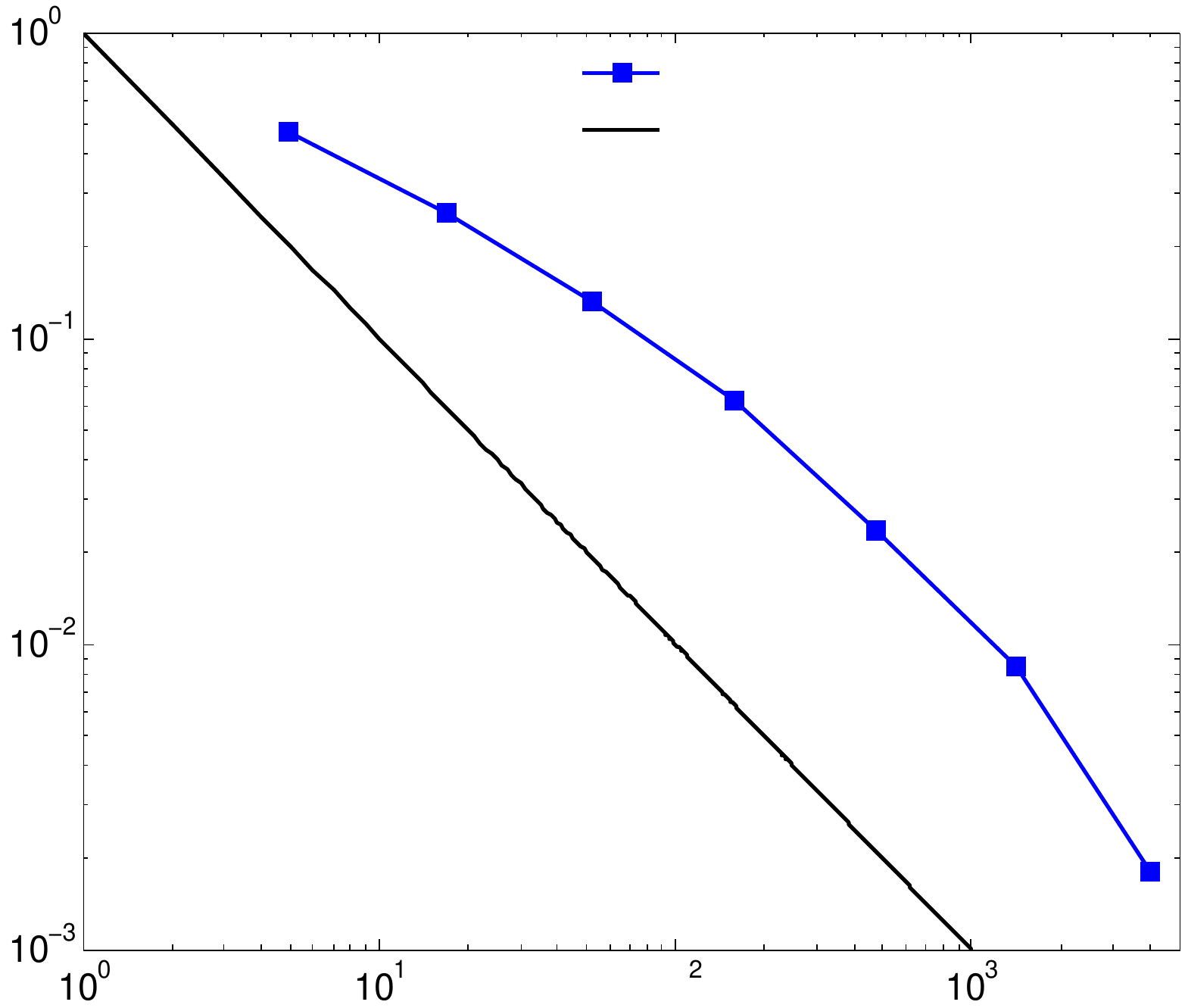}
  \put(-95,177){$\alpha=0$, source = leaf}
  \put(-95,165){Lower bound, $1/N_T$}
  \put(-160,-10){Average size of infection, $N_T$}
	\put(-235,65){\rotatebox{90}{\large $\prob(\hv=v^*)$}}
  \caption{Near-ML probability of detection for the Facebook graph with adaptive diffusion.}
  \label{fig:facebook}
\end{figure}
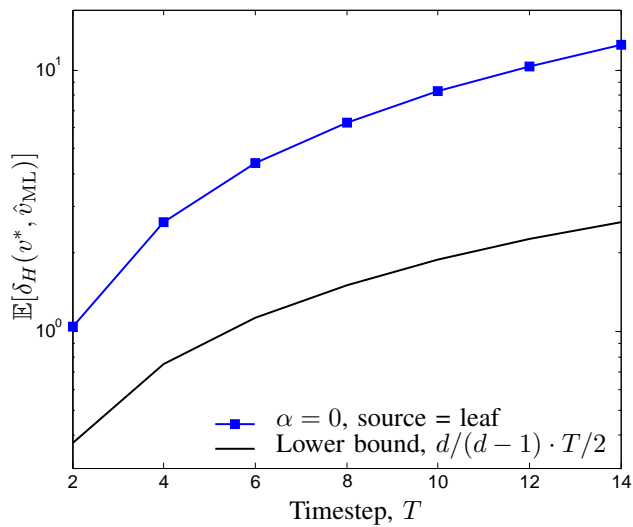
\begin{figure}[h!]
	    \centering
  \includegraphics[width=.44\textwidth]{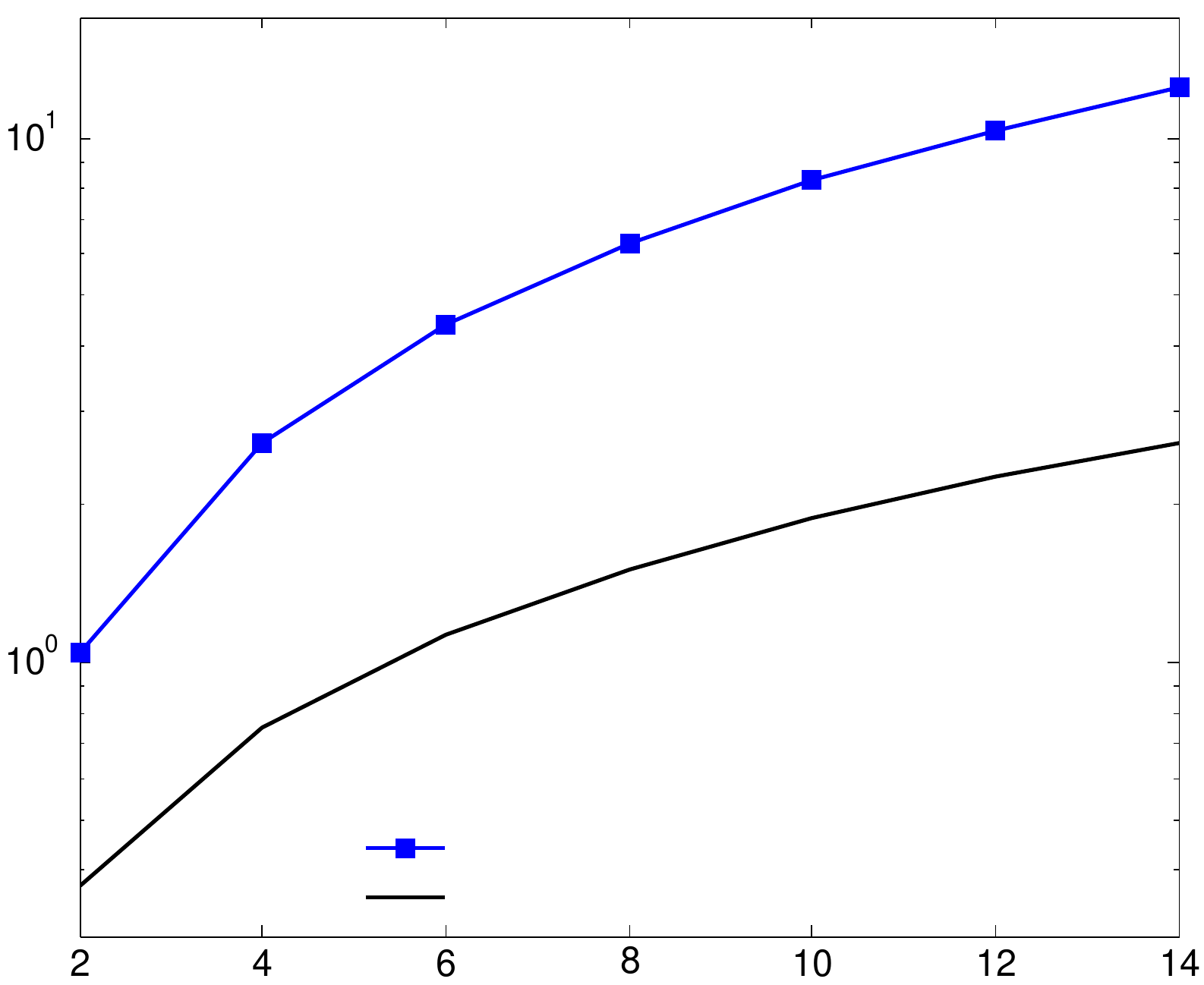}
  \put(-135,25){$\alpha=0$, source = leaf}
  \put(-135,15){Lower bound, $d/(d-1) \cdot T/2$}
  \put(-130,-10){Timestep, $T$}
	\put(-235,65){\rotatebox{90}{$\E[\delta_H(v^*,\hat{v}_{\rm ML})] $ }}
  \caption{ Hop distance between true source and estimated source over infection subtree for adaptive diffusion over the Facebook graph.}
  \label{fig:facebookhopdist}
\end{figure}

This approach is only an approximation of the ML estimate because the virtual source could move in a loop over the social graph (i.e., the same node could be the virtual source more than once, in nonadjacent timesteps).


On average, adaptive diffusion reached 96 percent of the network within 10 timesteps using $d_0=4$.
We also computed the average distance of the true source from the estimated source \emph{over the infected subtree} (Figure \ref{fig:facebookhopdist}). We see that as time progresses, so does the hop distance of the estimated source from the true source.  In social networks, nearly everyone is within a small number of hops (say, 6 hops \cite{ugander2011anatomy}) from everyone else, so this computation is not as informative in this setting.
However, it is relevant in location-based connectivity graphs, which can induce large hop distances between nodes.

\section{ spy-based adversarial model } \label{sec:spies}
The spy-based adversary collects more detailed information than the snapshot adversary, but only for a subset of network nodes.
In this section, we provide some results stating that over $d$-regular trees, choosing $\alpha_d(t,h)=0$ gives asymptotically optimal hiding in $d$. 
While the proofs for these results are not included in this paper (all proofs can be found in \cite{fkorv2016}), 
the results are included for completeness.

For the spy-based adversary, we model each node other than the source as a spy with probability $p$. At some point in time, the source node $v^*$ starts propagating its message over the graph according to some spreading protocol (e.g., diffusion or adaptive diffusion). Each spy node $s_i\in V$ observes: (1) the time $T_{s_i}$ (relative to an absolute reference) at which it receives the message, (2) the parent node $p_{s_i}$ 
that relayed the message, and (3) any other metadata used by the spreading mechanism (such as control signaling in the message header). At some time, spies {\em aggregate} their observations; using the collected metadata and the structure of the underlying graph, the adversary estimates the author of the message, $\hv$.  

To define perfect obfuscation for this adversarial model, we first observe the following:
\begin{propo}[\cite{fkorv2016}]
\label{thm:lb}
Under a spy-based adversary, no spreading protocol can have a probability of detection less than $p$.
\end{propo}

This results from considering the \emph{first-spy estimator}, which returns the parent of the first spy to observe the message.
Regardless of spreading, this estimator returns the true source with probability at least $p$; with probability $p$, the first node (other than the true source) to see the message is a spy.

We therefore say a protocol achieves \emph{perfect obfuscation} against a spy-based adversary if the ML probability of detection conditioned on the spy probability $p$ is bounded by
	\begin{eqnarray}
		\prob\big(\,\hv_{\rm ML}=v^* | p\,\big) &=& p  + o\Big(p\Big)\;.
	\end{eqnarray}


However, when the underlying graph is a $d$-regular tree, the probability of detection increases over time for standard diffusion spreading, since the estimator receives more information. 
Moreover, it is straightforward to show that the probability of detection tends to 1 as degree of the underlying graph $d\rightarrow \infty$:
\begin{propo}[\cite{fkorv2016}]
Suppose the contact network is a regular tree with degree $d$. 
There is a source node $v^*$, and each node other than the source is chosen to be a spy node i.i.d. with probability $p$ 
as described in the spy model.  
In each timestep, each infected node infects each uninfected neighbor independently with probability $q$.
Then the probability of detection $\prob(\hv_{\rm ML}=v^*)\geq 1-(1-qp)^d$. 
\label{pro:degree}
\end{propo}

This bound implies that as degree increases, the probability of detecting the true source of diffusion approaches 1. The proposition also results from analyzing the first-spy estimator. 
These observations suggest that diffusion provides poor anonymity guarantees in real networks; contact networks may be high degree, and the adversary is not time-constrained.



\subsection{Main result (Spy-based adversary)}
\label{sec:main}

In this section, we give results stating that over $d$-regular trees, adaptive diffusion with $\alpha_d(t,h)=0$ achieves asymptotically perfect obfuscation in $d$. 
We also show that adaptive diffusion hides the source better than diffusion over $d$-regular trees, $d>2$. 
However, these results depend on a slightly modified implementation of adaptive diffusion, in which some additional metadata is passed around.
This implementation, which we call the Tree Protocol, facilitates analysis and is also fully distributed, avoiding the explicit notion of a virtual source.

\vspace{0.1in}
\noindent \textbf{Tree Protocol.}
The spreading protocol follows Algorithm 1 (Spreading on a tree) from \cite{fkorv2016}; the goal is to build an infected subtree with the true source at one of the leaves. Whenever a node $v$ passes a message to node $w$, it includes three pieces of metadata: (1) the \emph{parent node} $p_w=v$, (2) a binary \emph{direction} indicator $u_w\in\{\uparrow,~\downarrow\}$, and (3) the node's \emph{level} in the infected subtree $m_w\in\mathbb N$.
The parent $p_w$ is the node that relayed the message to $w$.
The direction bit $u_w$ flags whether node $w$ is a {\em spine} node, responsible for increasing the depth of the infected subtree.
The level $m_w$ describes the hop distance from $w$ to the nearest leaf node in the final infected subtree, as $t\rightarrow \infty$. 

\begin{figure}[h]
	\begin{center}
	\includegraphics[width=.35\textwidth]{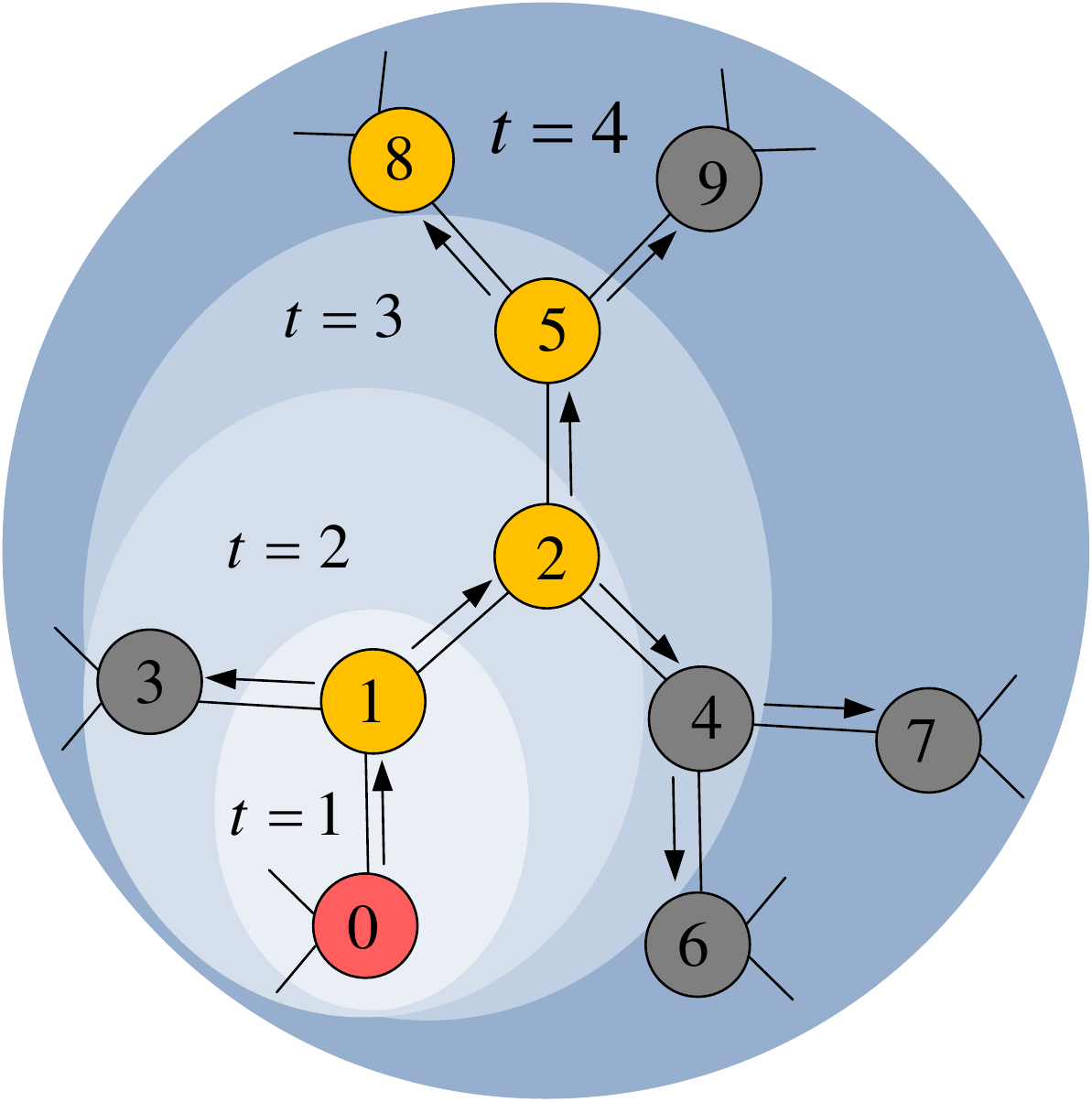}
\end{center}
	\caption{Message spread using the tree protocol from \cite{spyVsSpyArxiv}.}
	\label{fig:treeAlgo}
\end{figure}

At time $t=0$, the source chooses a neighbor uniformly at random (e.g., node 1) and passes the message and metadata ${(p_1=0,~u_1=\uparrow,~m_1 = 1)}$.
Figure \ref{fig:treeAlgo} illustrates an example spread, in which node 0 passes the message to node 1.
Yellow denotes \emph{spine} nodes, which receive the message with $u_w=\uparrow$, and gray denotes those that receive it with $u_w=\downarrow$.
Whenever a node $w$ receives a message, there are two cases.
if $u_w=\uparrow$, node $w$ chooses another neighbor $z$ uniformly at random and forwards the message with `up' metadata: ${(p_z=w,~u_z=\uparrow,~m_z=m_w+1)}$. All of $w$'s remaining neighbors $z'$ receive the message with `down' metadata: ${(p_{z'}=w,~u_{z'}=\downarrow,~m_{z'}=m_w-1)}$.
For instance, in Figure \ref{fig:treeAlgo}, node 1 passes the `up' message to node 2 and the `down' message to node 3.
On the other hand, if $u_w=\downarrow$ \emph{and} $m_w>0$, node $w$ forwards the message to all its remaining neighbors with `down' metadata: ${(p_z=w,~u_z=\downarrow,~m_z=m_w-1)}$.
If a node receives $m_w=0$, it does not forward the message further. Algorithm \ref{alg:tree} describes this process more precisely.

Observe that adaptive diffusion ensures that the infected subgraph is a balanced tree with the true source at one of the leaves. Moreover, unlike regular diffusion, the message does not reach all the nodes in the network under adaptive diffusion (even when $T = \infty$). Even though this may seem like a fundamental drawback for adaptive diffusion, it can be shown that the infected subgraph has a size proportional to $(d-1)^{T/2}$ on regular trees (compared to $(d-1)^T$ under regular diffusion). More critically, real social networks have cycles, so neighbors of nodes with $m_w = 0$ can still get the message from other nodes in the network \cite{fkorv2016}. 

As before, this protocol ensures that the infected subgraph is a symmetric tree with the true source at one of the leaves. The key difference between Protocol \ref{alg:adp_diff} (naive adaptive diffusion) with $\alpha_d(t,h)=0$ and Protocol \ref{alg:tree} (Tree Protocol) is that the latter does not rely on message-passing from the virtual source to control spreading. Instead, it passes enough control information to realize the same spreading pattern in a fully-distributed fashion.

\begin{algorithm}[h]
\caption{Tree Protocol}
\label{alg:tree}
\begin{algorithmic}[1]
\renewcommand{\algorithmicrequire}{\textbf{Input:}}
\renewcommand{\algorithmicensure}{\textbf{Output:}}
\Require contact network $G=(V,E)$, source $v^*$, time $T$
\Ensure infected subgraph $G_T=(V_T,E_T)$

\State $V_0 \gets \{v^*\}$
\State $m_{v^*} \gets 0$ and $u_{v^*} \gets \uparrow$
\State $v^*$ selects one of its neighbors $w$  at random
\State $V_1 \gets V_0 \cup \{w\}$
\State $m_{w} \gets 1$ and $u_{w} \gets \uparrow$
\State $t \gets 2$

\For{$t \leq T$}
\ForAll{$v \in V_{t-1}$ with uninfected neighbors and $m_{v} > 0$}
\If{$u_{v} = \uparrow $}
\State $v$ selects one of its uninfected neighbors $w$ at random
\State $V_t \gets V_{t-1} \cup \{w\}$
\State $m_{w}\gets m_{w}+1$ and $u_{w} \gets \uparrow$
\EndIf
\ForAll{uninfected neighboring nodes $z$ of $v$}
\State $V_t \gets V_{t-1} \cup \{z\}$
\State $u_{z}\gets \downarrow$ and $m_{z} \gets m_{v} - 1$
\EndFor
\EndFor
\State $t \gets t + 1$
\EndFor
\end{algorithmic}
\end{algorithm}

\begin{figure}[h]
	\begin{center}
	\includegraphics[width=.35\textwidth]{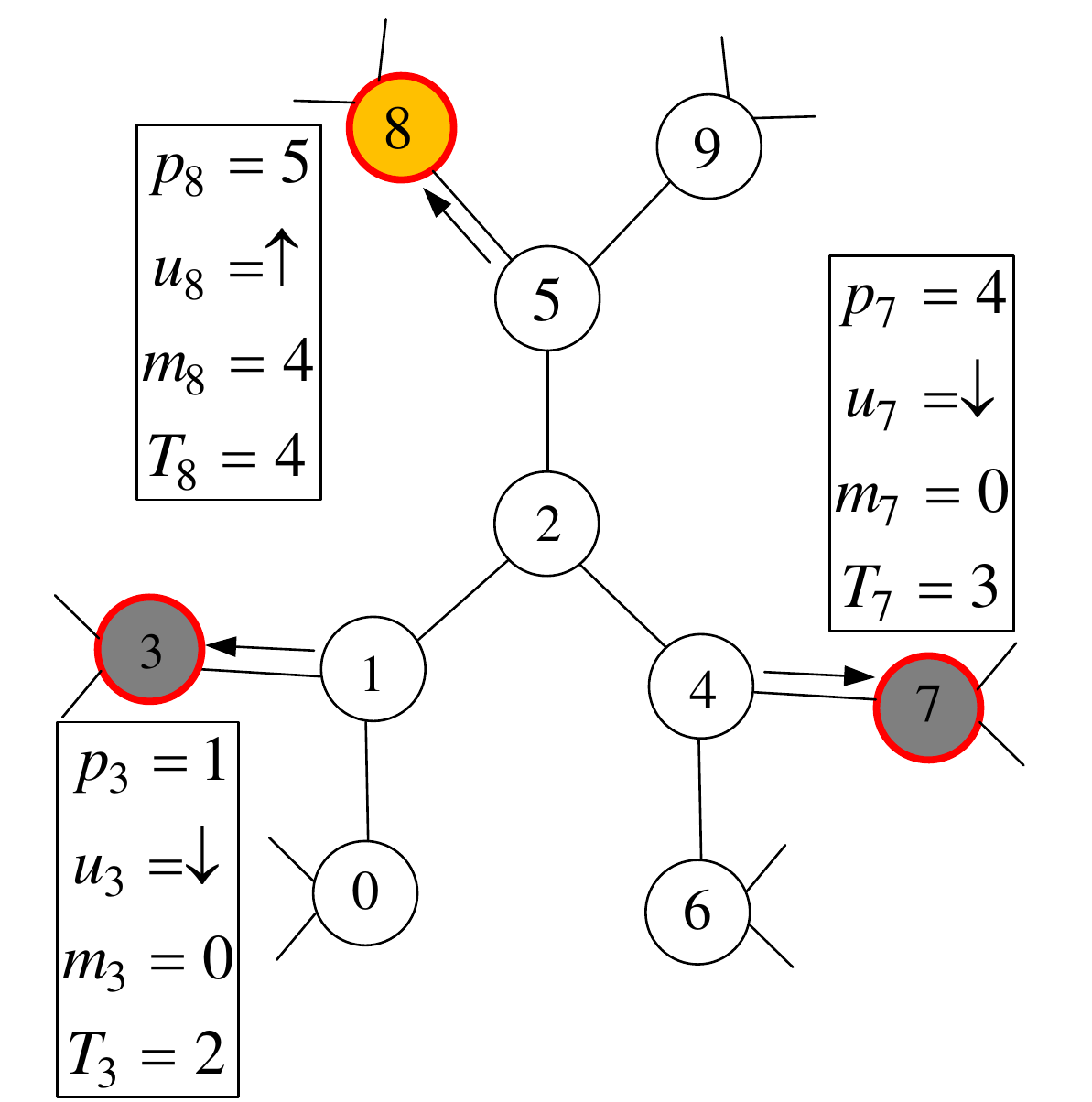}
	\end{center}
	\caption{The information observed by the spy nodes 3, 7, and 8 for the spread in Figure \ref{fig:treeAlgo}. Timestamps in this figure are absolute, but they need not be.}
	\label{fig:spies_info}
\end{figure}

In the spy-based adversarial model, each spy $s_i$ in the network observes any received messages, the associated metadata, and a timestamp $T_{s_i}$.
Figure \ref{fig:spies_info} illustrates the information observed by each spy node, where spies are outlined in red.


\vspace{0.1in}
\noindent \textbf{Source Estimation.}\label{sec:estimation}
The ML source-estimation algorithm for this spreading and adversarial model is described in \cite{fkorv2016}.
The ML estimation algorithm is not necessary to understand this paper's primary contributions. 
We include it in this section for completeness, 
and because the probability of detection for the spy+snapshot adversarial model in Section \ref{sec:spy_snapshot} 
uses terminology that is introduced in this estimator.

To a snapshot adversary, all leaves in the infected subgraph have the same likelihood.
Because adaptive diffusion has deterministic timing, spies only help the estimator discard candidate nodes. 
We assume the message spreads for infinite time. 
There is at least one spy on the spine; consider the first such spy to receive the message, $s_0$. 
This spine spy (along with its parent and level metadata) allows the estimator to specify a \emph{feasible subtree} in which the true source must lie.
In Figure \ref{fig:treeAlgo}, node 8 is on the spine with level $m_8=4$, so the feasible subtree is rooted at node 5 and contains all the pictured nodes except node 8 (9's children and grandchildren also belong, but are not pictured).
Spies outside the feasible subtree do not influence the estimator, because their information is independent of the source conditioned on $s_0$'s metadata. 
Only leaves of the feasible subtree could have been the source---e.g., nodes 0, 3, 6, and 7, as well as 9's grandchildren.

\begin{algorithm}[h]
\caption{ML Source Estimator for Algorithm \ref{alg:tree}}
\label{alg:ml_est}
\begin{algorithmic}[1]
\renewcommand{\algorithmicrequire}{\textbf{Input:}}
\renewcommand{\algorithmicensure}{\textbf{Output:}}
\Require contact network $G=(V,E)$, spy nodes $ S=\{s_0, s_1\ldots\}$ and metadata $s_i:(p_{s_i},m_{s_i},u_{s_i})$
\Ensure ML source estimate $\hat v_{\rm ML}$

\State Let $s_0$ denote the lowest-level spine spy, with metadata $(p_{s_0},m_{s_0},u_{s_0})$.
\State $\tilde V\gets \{v\in V:~\delta_H(v,s_0)\leq m_{s_0}\text{ and }p_{s_0}\in \mathcal P(v,s_0)\}$
\State $\tilde E\gets\{(u,v):(u,v)\in E\text{ and }u,v\in \tilde V\}$
\State Define the feasible subgraph as $F(\tilde V,\tilde E)$
\State $ L\gets \emptyset$ 
\Comment Set of feasible pivots
\State $ K\gets \emptyset$ 
\Comment Set of eliminated pivot neighbors
\ForAll{$s \in  S$ with $s\in \tilde V$}
\State Let $\left [ \begin{array}{c} h_{s,\ell_s}\\ h_{\ell_s,s_0} \end{array}\right ] =\frac{1}{2}\left [ \begin{array}{c c} 1 & -1\\1 &1 \end{array}\right ]\cdot \left [ \begin{array}{c} | P(s,s_0)|\\ T_{s_0}-T_{s} \end{array}\right ]$
\State $\ell_s \gets v\in \mathcal P(s,s_0):~\delta_H(s,\ell_s)=h_{s,\ell_s}$
\State $k_s \gets v\in \mathcal P(s,s_0):~\delta_H(s,k_s)=h_{s,\ell_s}-1$
\State $ L \gets  L \cup \{\ell_s\}$ 
\Comment Add pivot
\State $ K \gets  K \cup \{k_s\}$ 
\Comment Add pivot neighbor
\EndFor
\State Find the lowest-level pivot: $\ell_{min}\gets \text{argmin}_{\ell \in  L}m_{\ell}$
\State $U\gets \emptyset$ 
\Comment Candidate sources
\ForAll{$v \in \tilde V$ where $v$ is a leaf in $F(\tilde V,\tilde E)$}
\If {$\mathcal P(v,\ell_{min}) \cap K =\emptyset$}
\State $U \gets U \cup \{v\}$
\EndIf
\EndFor
\State return $\hat v_{\rm ML}$, drawn uniformly from $U$
\end{algorithmic}
\end{algorithm}

The estimator then uses spies \emph{within} the feasible subtree to prune out candidates.
The goal is to identify nodes in the feasible subtree that are on the spine and close to the source.
For each spy in the feasible subtree, there exists a unique path to the spine spy $s_0$, and at least one node on that path is on the spine; the spies' metadata reveals the identity and level of the spine node on that path with the lowest level---we call this node a \emph{pivot} (details in Algorithm \ref{alg:ml_est}). 
For instance, in Figure \ref{fig:spies_info}, we can use spies 7 and 8 to learn that node 2 is a pivot with level $m_2=2$. 
Estimation hinges on the minimum-level pivot across all spy nodes, $\ell_{min}$.
In the example, $\ell_{min}=1$, since spies 3 and 8 identify node 1 as a pivot with level $m_1=1$.
The true source must lie in a subtree rooted at a neighbor of $\ell_{min}$, with no spies.
In our example, this leaves only node 0, the true source. 

\vspace{0.1in}
\noindent \textbf{Anonymity properties.}
This ML estimation procedure can be analyzed to exactly compute the probability of detection for adaptive diffusion on a $d$-regular tree:

\begin{theorem}[\cite{fkorv2016}]
\label{thm:ad_ub}
Suppose the contact network is a regular tree with degree $d>2$. 
There is a source node $v^*$, and each node other than the source is chosen to be a spy node i.i.d. with probability $p$ 
as described in the spy model.  
Against colluding spies attempting to detect the location of the source, adaptive diffusion achieves the following: 

(a) The probability of detection is 
	\begin{eqnarray}
	\prob(\hv_{\rm ML} = v^*) = p + \frac{1}{d-2} - \sum_{k=1}^\infty \frac{q_k}{(d-1)^k}
	  \;, \label{eq:main1}
	\end{eqnarray}

	where 
	$q_k \equiv  (1 - (1-p)^{{((d-1)^k-1)}/({d-2})} )^{d-1} + \\(1 - p)^{({(d-1)^{k+1}-1})/({d-2})}$.
	
(b) The expected distance between the source and the estimate is bounded by 
	\begin{eqnarray}
	\E[\delta_H(\hv_{\rm ML}, v^*)] \geq 2\sum_{k=1}^\infty  k \cdot r_k
	\label{eq:ed}
	\end{eqnarray}
where  $|T_{d,k}|=\frac{(d-1)^{k}-1}{d-2}$, and \\
$r_k \equiv  \frac{1}{d-1} \Big ( (1-(1-p)^{|T_{d,k}|})^{d-1} + (d-1)(1-p)^{|T_{d,k}|} - \\(d-2)(1-p)^{|T_{d,k}|(d-1)} - 1\Big )
$.
	
\end{theorem}



There are two main observations to note regarding this result:

\textit{(1) Asymptotically optimal probability of detection:} As tree degree $d$ increases, the probability of detection converges to the degree-independent fundamental limit in Proposition \ref{thm:lb}, i.e., ${\prob(V^* = \hv_{\rm ML} )=p}$. 
This is in contrast to diffusion, whose probability of detection tends to 1 asymptotically in $d$.

\textit{(2) Expected hop distance asymptotically increasing:} 
We observe empirically that for regular diffusion, $\E[\delta_H(\hv_{\rm ML},v^*)]$ approaches 0 as $d$ increases.
On the other hand, for adaptive diffusion with a fixed $p>0$, as $d\rightarrow \infty$, $\limsup \E[\delta_H(\hv_{\rm ML},v^*)]=2(1-p)$. 

These observations suggest that adaptive diffusion exhibits provably stronger anonymity properties than standard diffusion on regular trees---a suggestion that is backed up by simulations on irregular trees and the Facebook graph in \cite{fkorv2016}.

\section{spy+snapshot adversarial model} \label{sec:spy_snapshot}

The spy+snapshot adversarial model considers a natural combination of the snapshot and spy-based adversaries. 
At a certain time $T$, the adversary collects a snapshot of the infection pattern, $G_T$. It
also collects metadata from all spies that have seen the message up to (and including) time $T$. Based on these two sets of metadata, the adversary infers the source.  

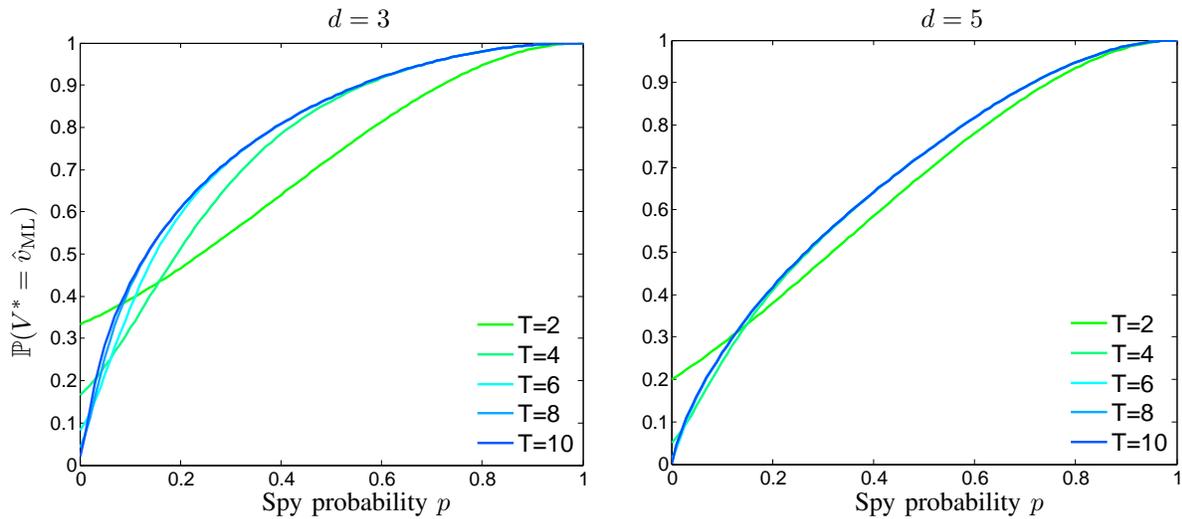
\begin{figure*}
	\begin{center}
	\includegraphics[width=.4\textwidth]{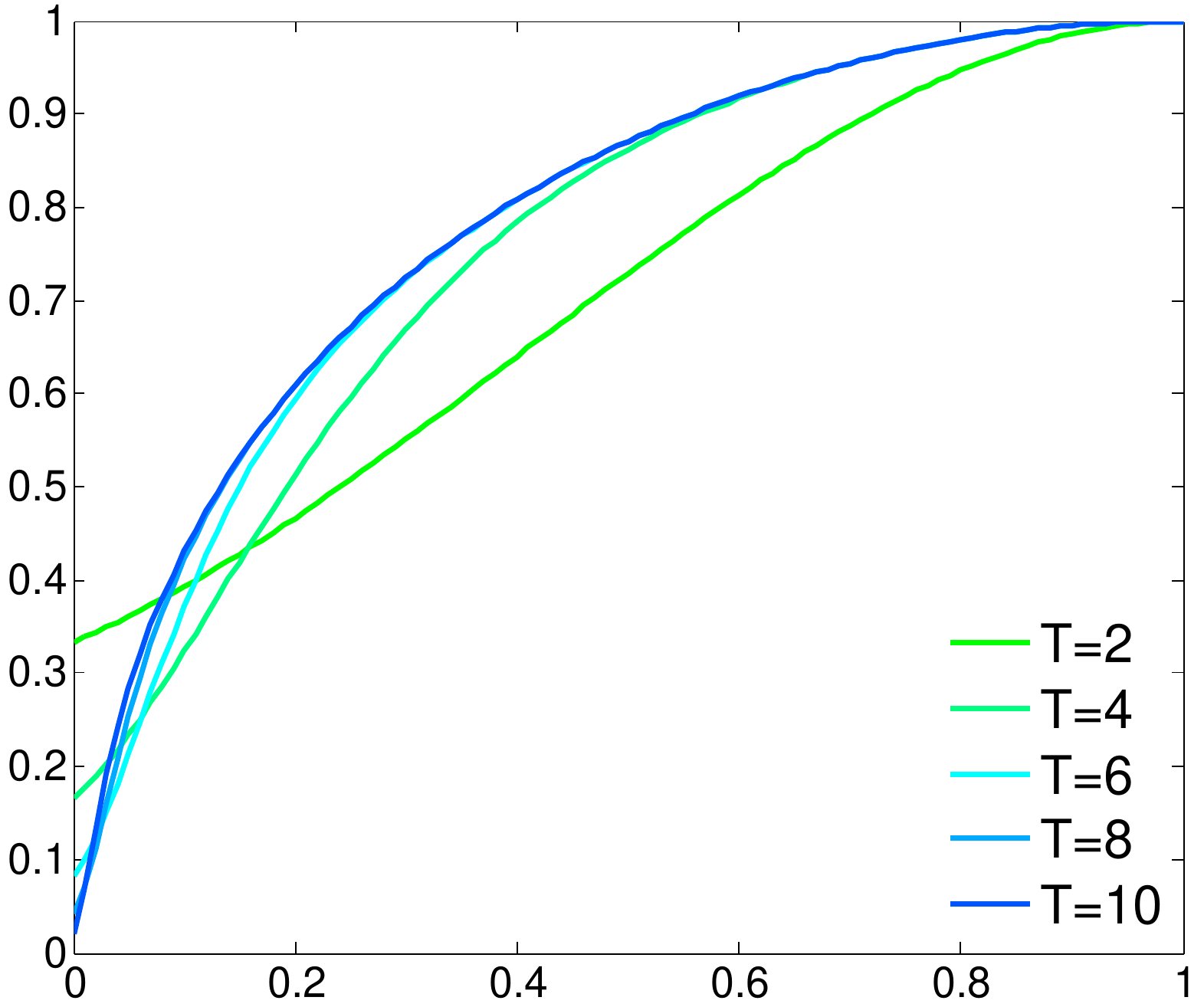}
	\put(-125,-8){Spy probability $p$}
	\put(-100,175){$d=3$}
	\put(-220,50){\rotatebox{90}{$\prob(V^* = \hv_{\rm ML} )$}}
	\hspace{0.2in}
	\includegraphics[width=.4\textwidth]{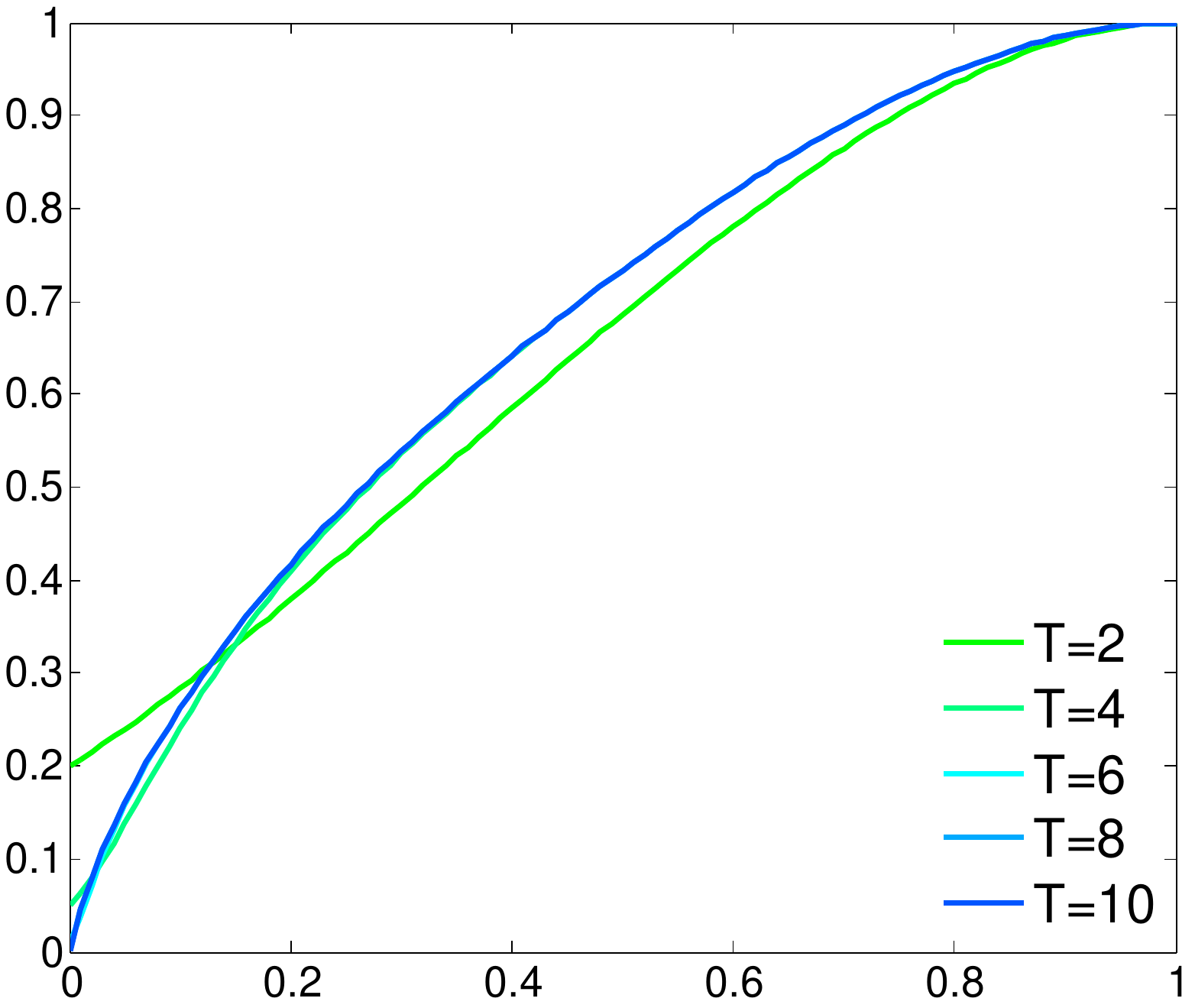}
	\put(-125,-8){Spy probability $p$}
	\put(-100,175){$d=5$}
	\end{center}
	\caption{Probability of detection under the spy+snapshot adversarial model. As estimation time and tree degree increase, the effect of the snapshot on detection probability vanishes.}
	\label{fig:spysnapshot}
\end{figure*}

Notably, this stronger model does not significantly impact the probability of detection as time increases.
The snapshot helps detection when there are few spies by revealing which nodes are true leaves. This effect is most pronounced for small $T$ and/or small $p$. 
The exact probability of detection at time $T$ is given below:
{\small
\begin{align}
	& \prob(\hv_{\rm ML} = v^*) \;=\; \nonumber \\
	&  \underbrace{\frac{(1-p)^{|S_{d,T}|-1}}{|\partial S_{d,T}|}}_{\text{no spy}} \;+\;  
	 \sum_{k=1}^{T/2} \Big\{  \underbrace{\frac{(1-p)^{(|T_{d,k}|-1)}\, p}{|\partial T_{d,k}|}}_{\text{ $\ell_{min}~(k^{th}$ spine node) is a spy}} \;+\; \nonumber \\ 
	& \underbrace{ (1-p)^{|T_{d,k}|} (1-(1-p)^{|S_{d,T}|-|T_{d,k+1}|})\E_X\Big[\frac{\ind(X\neq d-2)}{(X+1)\,|\partial T_{d,k}|}\Big]}_{\text{$\ell_{min}~(k^{th}$ spine node) not a spy}} \;+\; \nonumber\\ 
	& \underbrace{ (1-p)^{|S_{d,T}|-(|T_{d,k+1}|-|T_{d,k}|)} \E_X\Big[ \frac{\ind(X\neq d-2)}{|\partial S_{d,T}| - (d-2-X)|\partial T_{d,k}|} \Big]  }_{\text{all spy descendants of $k$-th spine node}}
	\Big\}  \;, \label{eq:spysnapshot}.
\end{align}
where 
	$X\sim{\rm Binom}(d-2,(1-p)^{|T_{d,k}|})$, 
	$|T_{d,k}|=\frac{(d-1)^{k}-1}{d-2}$ is the number of nodes in each candidate subtree for a pivot at level $k$, and 
	$|\partial T_{d,k}|=(d-1)^{k-1}$ is the number of leaf nodes in each candidate subtree.

}%
This expression can be evaluated numerically, as shown in Figure
\ref{fig:spysnapshot}, which  illustrates the tradeoff between the effect of a snapshot and spy nodes.
The derivation for this expression is included in \cite{fkorv2016}.

\section{Connections to P\'olya's urn processes} \label{sec:Polya}

In this section, 
we make a connection between adaptive diffusion on a line and P\'olya's urn processes. 
In doing so, we highlight a property of P\'olya's urn processes, which inherently provides privacy. 
Further, we apply the Bayesian interpretation of P\'olya's urn processes to design a new implementation 
of adaptive diffusion and analyze the precise cost of 
revealing the control packets to the spy nodes, in terms of leaked anonymity.

To separately characterize the price of timestamp metadata and control packets, we focus on the concrete example of a line graph. 
Consider a line graph in which nodes $0$ and $n+1$ are spies.
One of the $n$ nodes between the spies is chosen uniformly at random as a source,
denoted by $v^* \in \{1,\ldots,n\}$.
We let $t_0$ denote the time the source starts propagating the message according to some global reference clock.
Let $T_{s_1}=T_1+t_0$ and $T_{s_2}=T_2+t_0$  denote the timestamps
when the two spy nodes receive the message, respectively.
Knowing the spreading protocol and the metadata,
the adversary uses the maximum likelihood estimator to optimally estimate the source.

\begin{figure}[h]
\vspace{-0.1in}
	\begin{center}
	\includegraphics[width=.45\textwidth]{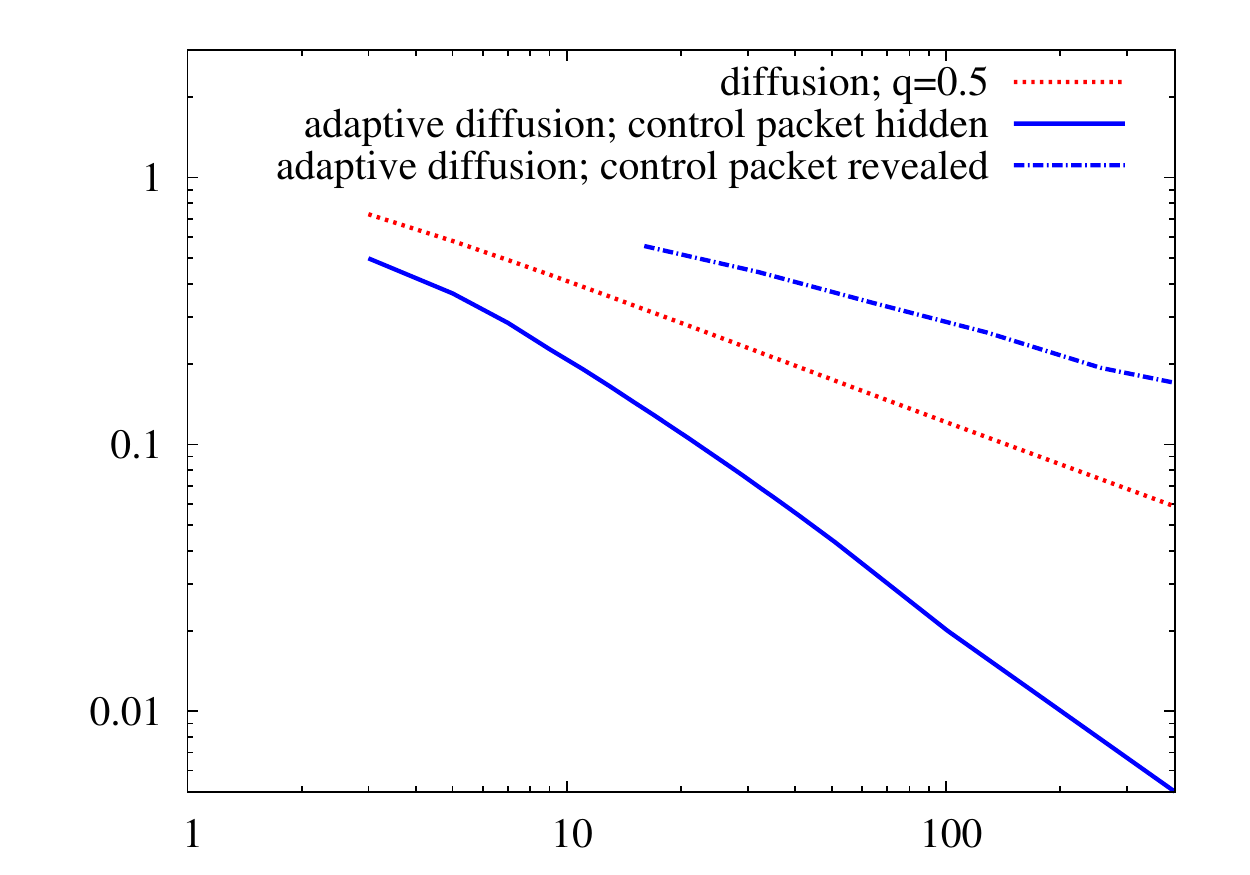}
	\put(-115,-6){number of nodes $n$}
	\put(-230,60){\rotatebox{90}{$\prob(V^* = \hv_{\rm ML} )$}}
	\hspace{.6cm}
	\includegraphics[width=.45\textwidth]{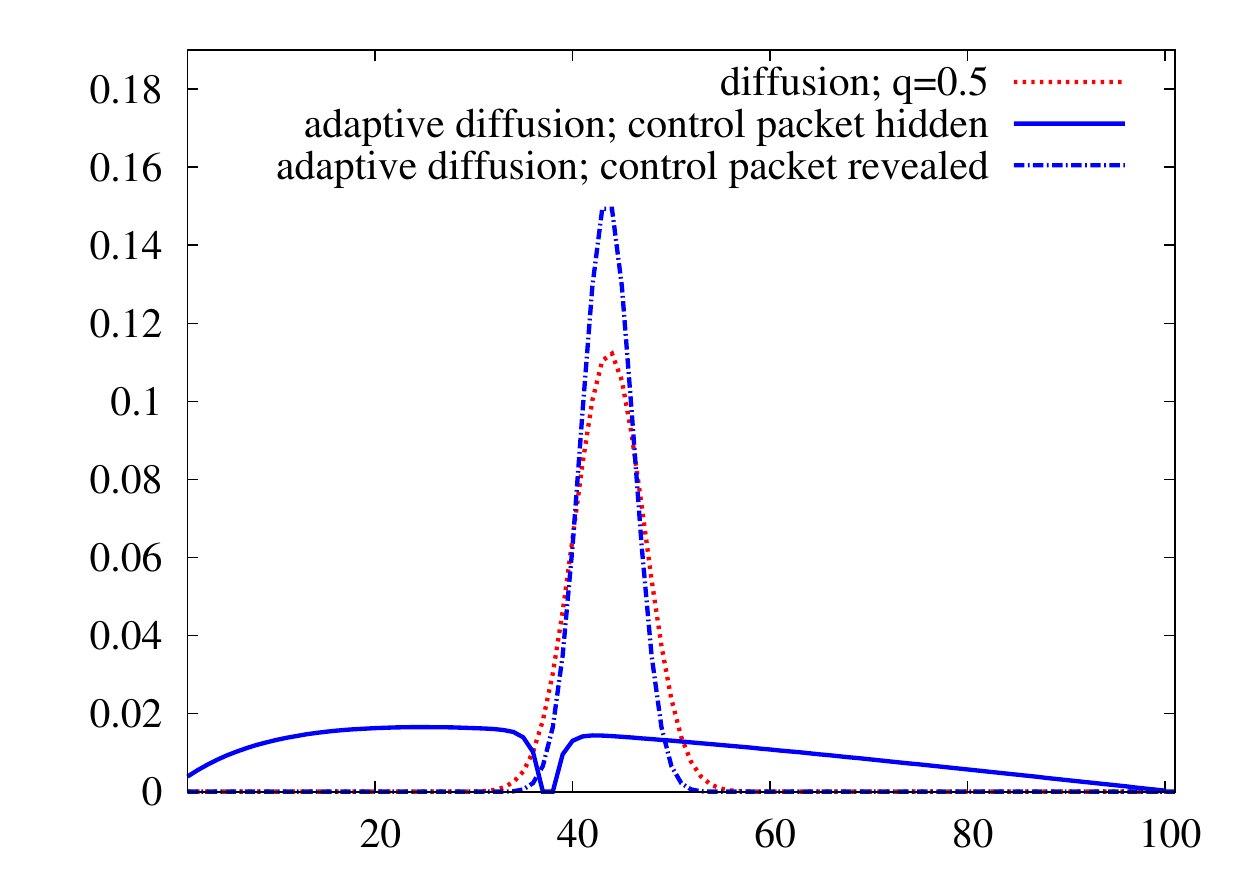}
	\put(-115,-6){candidate node $v$}
	\put(-230,34){\rotatebox{90}{$\prob( V^*=v | T_2-T_1 = 25 )$}}
	\end{center}
	\caption{ Comparisons of probability of detection as a function of  $n$ (top) and the posterior distribution of 
	the source for an example with $n=101$ and $T_2-T_1=25$ (bottom). 
	The line with `control packet revealed' uses the P\'olya's urn implementation. }
	\label{fig:line}
\end{figure}

\vspace{0.1in}
\noindent {\bf Standard diffusion.}
Consider a standard discrete-time random diffusion with a parameter $q\in(0,1)$ where
each uninfected neighbor is infected with probability $q$.
The adversary observes $T_{s_1}$ and $T_{s_2}$. Knowing the value of $q$,
it computes
the ML estimate
$\hv_{\rm ML} = \arg\max _{v\in [n]} \prob_{T_1-T_2|V^*}( T_{s_1}-T_{s_2} | v)$, which is optimal assuming a uniform prior on $v^*$.
Since $t_0$ is not known,
the adversary can only use the difference $T_{s_1}-T_{s_2} = T_1 - T_2$ to estimate the source.
We can exactly compute the corresponding probability of detection;
Figure \ref{fig:line} (bottom panel) illustrates
that the posterior (and the  likelihood) is concentrated around the ML estimate,
and the source can only hide among $O(\sqrt{n})$ nodes.
The detection probability correspondingly scales as $1/\sqrt{n}$ (top panel).

\begin{figure}
    \centering
  \includegraphics[scale=0.6]{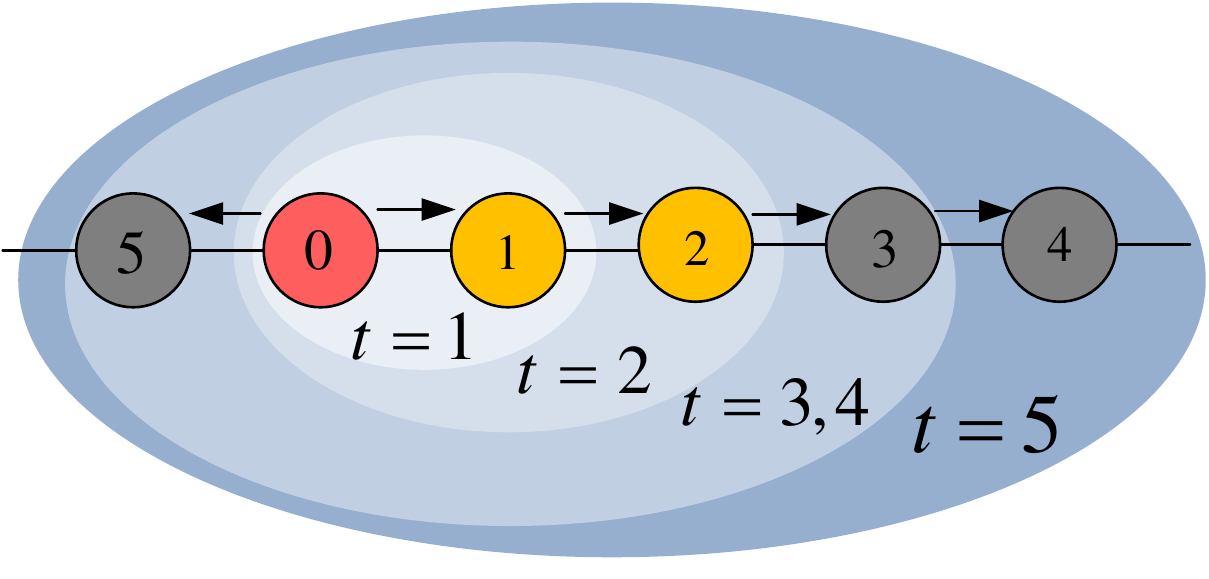}
  \caption{Spreading on a line. The red node is the message source. Yellow nodes denote nodes that have been, are, or will be the center of the infected subtree.
}
  \label{fig:line_example}
\end{figure}

{\bf Adaptive diffusion on a line.}
First, recall the adaptive diffusion (Protocol \ref{alg:adp_diff}) 
with the choice of $\alpha_d(h,t)=\frac{t-2h+2}{t+2}$ (Equation \eqref{eq:alpha})
on a line  illustrated in Figure \ref{fig:line_example}. 
At $t=0$, the message starts at node 0. 
The source passes the virtual source to node 1, so $v_2=1$. The next two timesteps ($t=1,2$) 
are used to restore symmetry about $v_2$. At $t=2$, the virtual source stays with probability 
$\alpha_2(2,1)=1/2$. Since the virtual source remained fixed at $t=2$, at $t=4$ the virtual 
source stays with probability $\alpha_2(4,1)=2/3$. The key property is that if the virtual 
source chooses to remain fixed at the beginning of this random process, it is more likely 
to remain fixed in the future, and vice versa. This is closely related to the well-known 
concept of \emph{P\`olya's urn processes}; we make this connection more precise later in this section.

The protocol keeps the current virtual source with probability $\frac{2 \delta_H(v_t,v^*)}{t+2}$, 
where $\delta_H(v_t,v^*)$ denotes the hop distance between the source and the virtual source, and passes it otherwise.
The control packet therefore contains two pieces of information: $\delta_H(v_t,v^*)$ and $t$.

Suppose spy nodes only observed timestamps and parent nodes but {\em not} control packets.
The adversary could then numerically compute
the ML estimate
$\hv_{\rm ML} = \arg\max _{v\in [n]} \prob_{T_1-T_2|V^*}( T_{s_1}-T_{s_2} | v)$.
We can compute the corresponding detection probability exactly.
Figure \ref{fig:line} shows the posterior is close to uniform (top panel)
and the probability detection would scale as $1/n$ (bottom panel), which is the best one can hope for.
Of course, spies \emph{do} observe control packets,
so they can learn $\delta_H(v^*,v_T)$ and identify the source with probability 1.
We therefore introduce a new adaptive diffusion implementation that is robust to control packet information. 

\vspace{0.1in}
\noindent {\bf Adaptive diffusion via P\'olya's urn.}
The random process governing the virtual source's propagation under adaptive diffusion
is identical to a P\'olya's urn process \cite{polya}.
We propose the following alternative implementation of adaptive diffusion.
At $t=0$ the protocol decides whether to pass the virtual source left $(D=\ell)$ or right $(D=r)$ with probability half. Let $D$ denote this random choice.
Then, a latent variable $q$ is drawn from the uniform distribution over $[0,1]$.
Thereafter, at each even time $t$, the virtual source is passed with probability $q$ or
kept with probability $1-q$.
It follows from the Bayesian interpretation of P\'olya's urn processes that
this process has the same distribution as the adaptive diffusion process.

Further, in practice, the source could simulate the whole process in advance.
The control packet would simply reveal to each node how long it should wait before further propagating the message.
Under this implementation,
spy nodes only observe timestamps $T_{s_1}$ and $T_{s_2}$,
parent nodes, and control packets containing the infection delay for the spy and all its descendants in the infection.
Given this, the adversary can exactly determine
the timing of infection with respect to the start of the infection $T_1$ and $T_2$,
and also the latent variables $D$ and $q$.
A proof of this statement and the following proposition is
provided in Section \ref{sec:line_proof}.
The next proposition 
provides an upper bound on the detection probability for such an adversary.




\begin{propo}
	\label{propo:line_adap}
	When the source is uniformly chosen from $n$ nodes between two spy nodes,
	the message is spread according to adaptive diffusion, and
	the adversary has a full access to the time stamps, parent nodes, and the control packets that is received by
	the spy nodes,
 observations $T_{1}, T_{2},q$ and $D$,
 	the adversary can compute the ML estimate:
\begin{align}
	\hv_{\rm ML} &=& \left\{
	\begin{array}{rl}
	\frac{T_1+2}{2}  + \big\lfloor q\Big(\frac{T_1-2}{2}\Big) \big\rfloor & \text{, if $T_1$ even and $D=\ell$\;,} \\
	 \frac{T_1+3}{2} +\big\lfloor q\Big( \frac{T_1-1}{2}  \Big) \big\rfloor & \text{, if $T_1$ odd and $D=\ell$\;,} \\
	1 + \big\lfloor (1-q) \Big( \frac{T_1-1}{2}  \Big) \big\rfloor& \text{, if $T_1$ odd and $D=r$\;.}
	\end{array}
	\right.\label{eq:line_control_ml}
\end{align}
	where $T_1$ is the time since the start of the spread until $s_1$ receives the message, and $q$ is the hidden parameter of the P\'olya's urn process, and $D$ is the initial choice of direction for the virtual source.
	This estimator
achieves a detection probability upper bounded by \;
	\begin{eqnarray}
		\prob\big(\, V^* = \hv_{\rm ML} \,\big) \leq \frac{\pi \sqrt{8}}{\sqrt{ n  }}  + \frac{2}{n}. 
		\label{eq:line_adap}
	\end{eqnarray}
\end{propo}
Equipped with an estimator, we can also simulate
adaptive diffusion on a line.
 Figure \ref{fig:line} (top) illustrates that even with access to control packets, the adversary achieves probability of detection scaling as $1/\sqrt{n}$ -- similar to standard diffusion.
 For a given value of $T_1$,
 the posterior and the likelihood are concentrated around the ML estimate,
and the source can only hide among $O(\sqrt{n})$ nodes, as shown in the bottom panel for $T_1=58$.
In the realistic adversarial setting where control packets are revealed at spy nodes,
adaptive diffusion can only hide as well as standard diffusion over a line.

\section{Future directions and connections to game theory}

Consider a game-theoretic setting where   
there are two players, the protocol designer and the adversary. 
The designer can choose any strategy to spread the message from a source $v^*$, 
as long as the message is passed one hop at a time. 
The adversary can choose any strategy (computationally expensive or not) to 
compute an estimated source $\hv$ given a some side information on the spread. 
As a result, the source can either be detected or not. 
In terms of the payoff, the protocol designer wants to minimize the probability of detection and 
the adversary wants to maximize it. 

In this static game setting, 
the adaptive diffusion is a (weak) dominant strategy under a certain condition. 
Consider a snapshot-based adversary and a contact network of $d$-regular tree.  
The special condition we impose is that we are only allowed protocols that infect at most, say,  $1+(2(d-1)^{(T+1)/2}-d)/(d-2)$ nodes. In this setting, Theorem~\ref{thm:main} 
implies that adaptive diffusion is dominant up to a vanishing additive factor. 

Following our work \cite{KFSV14}, a game-theoretic formulation of the problem of source obfuscation was recently proposed in   \cite{luo2015rumor}.  
The designer is restricted to use deterministic protocols, 
and the snapshot-based adversary is restricted to use a certain family of estimators based on Jordan centers.  
Under these restrictions, it is shown that 
there is no ``dominant'' protocol in Nash equilibrium sense, 
other than the simple (deterministic) diffusion.  

\begin{figure}[h]
	\includegraphics[width=.5\textwidth]{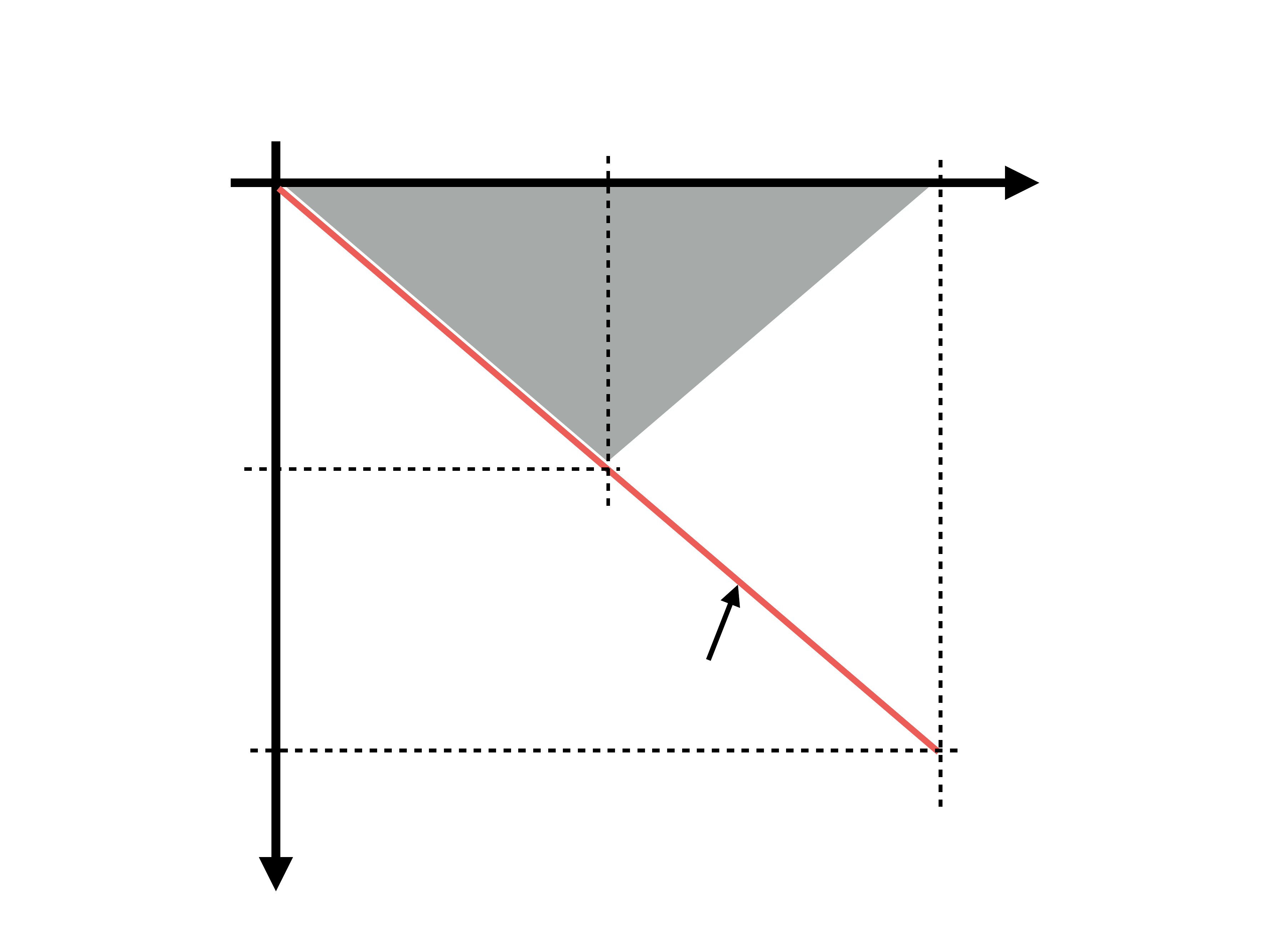}
	\put(-271,39){$-T\log(d-1)$}
	\put(-274,94){$-\frac{1}{2}T\log(d-1)$}
	\put(-80,165){$T\log(d-1)$}
	\put(-150,165){$\frac{1}{2}T\log(d-1)$}
	\put(-120,50){$\frac{1}{N_T}$}
	\put(-170,182){Infection size $(\log N_T)$}
	\put(-280,120){$\log\prob(\text{detection})$}
	\caption{The fundamental limit of $\prob(\text{detection})\geq 1/N_T$ is shown in a solid red line. This is achieved by adaptive diffusion until $ \log(N_T)\leq \frac12 T \log(d-1)$. Infection size at time $T$ is shown on the x-axis in log-scale and the probability of detection on y-axis also in log-scale.  }
	\label{fig:tradeoff}
\end{figure}
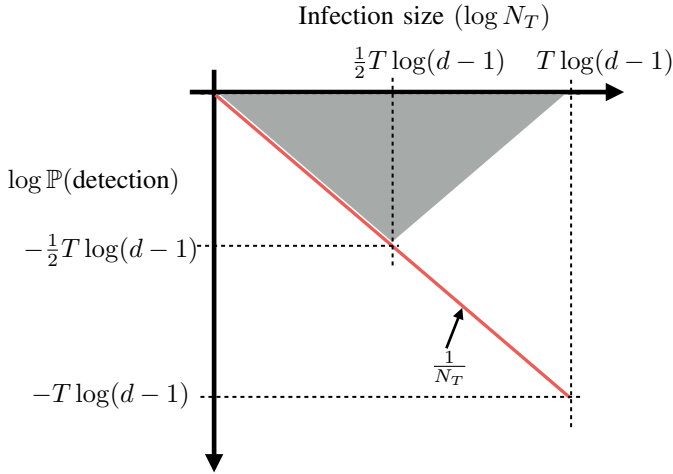

There are several interesting future research directions. 
First, when infecting more nodes is of priority, 
a fundamental question is whether there is a dominant strategy for a given target infection rate. 
Adaptive diffusion achieves the fundamental limit of $\prob(\text{detection})=1/N_T$ until 
$N_T\leq 1+(2(d-1)^{(T+1)/2}-d)/(d-2) \simeq (d-1)^{(T/2)}$ (see Figure \ref{fig:tradeoff}) on $d$-regular trees.  
It is an open question what the fundamental limit is above this threshold, 
and if there is efficient distributed protocol achieving this optimal tradeoff.  
In particular, if we have to spread every time deterministically to achieve the infection speed of 
$N_T \simeq (d-1)^T$, then the source will be trivially detected as the center of infection.
Above the threshold of $\log N_T\simeq \frac12 T \log(d-1)$, A variant of adaptive diffusion can achieve 
the infection rate $\alpha T \log (d-1)$ with probability of detection $(\alpha-1)T\log(d-1)$ for any $\alpha\in[0.5,1]$. Hence, all grey triangular region is achievable by adaptive diffusion in Figure \ref{fig:tradeoff}.

Second, when the same source spreads multiple messages that can be linked, 
this can be posed as a dynamic game. 
If the adversary observes multiple spreads of infection from a single source, 
how much does the probability of detection increase as a function of the multiplicity of the spread? 
One possibility is to spread according to adaptive diffusion the first time, and 
use exactly the same pattern of spread in the consecutive spread of 
the following messages from the same source. Hence, from the meta-data, there is no more information on who the source is. However, this creates a certain permanent bias in the spread, which may be undesirable, depending on the application. 

Next, a set of nodes can collude to spread the exactly same message, but starting from multiple sources simultaneously with possible delays. Unless carefully coordinated, such spread from multiple sources can be easily detected \cite{luo2012identifying} and there is no gain in collusion. 
However, we can consider an alternative strategy of creating a pseudo-source node to make the source hard to find.  
At a certain time (possible $t=0$), the protocol starts another chain of spread starting from a node far away from the infection so far. 
This can improve the detection probability by a factor of the number of such new infections, at 
a price of losing the benefits of social filtering and possibly spamming the users with irrelevant messages. We want to be able to measure such a loss in social filtering and characterize the tradeoff.


\section{Proofs}\label{sec:proofs}

\subsection{Proof of Theorem \ref{thm:main}}\label{sec:proofs_snapshot_main}
\balance
\noindent {\bf Spreading rate.} Under Protocol \ref{alg:adp_diff}, $G_T$ is a complete $(d-1)$-ary tree (with the exception that the root has $d$ children) of depth $T/2$ whenever $T$ is even. Whenever $T$ is odd, with probability $\alpha_d(T,h)$, $G_T$ is again such a $(d-1)$-ary tree of depth $(T+1)/2$. With probability $1-\alpha_d(T,h)$, $G_T$ is made up of two $(d-1)$-ary trees of depth $(T-1)/2$ each with their roots connected by an edge. Therefore, it follows that when $d>2$, $N_T$ is given by
	\begin{equation}
	\label{eq:n_tree2}
	N_T \;=\;
	\left\{
	\begin{array}{rl}
	1, &  T=0,\\
	\frac{2(d-1)^{(T+1)/2}}{d-2}-\frac{2}{d-2},   & T \geq 1, ~T\text{ odd, w.p. }(1-\alpha)\;, \\
	\frac{d(d-1)^{(T+1)/2}}{d-2}-\frac{2}{d-2},   & T \geq 1, ~T\text{ odd, w.p. }\alpha\;, \\
	\frac{d(d-1)^{T/2}}{d-2}-\frac{2}{d-2},   & T \geq 2, ~T\text{ even}\;;
	\end{array}\right.
	\end{equation}
Similarly, when $d=2$, $N_T$ can be expressed as follows:
	\begin{equation}
	\label{eq:n_tree2_d2}
	N_T \;=\;
	\left\{
	\begin{array}{rl}
	1, &  T=0,\\
	T+1,   & T \geq 1, ~T\text{ odd, w.p. }(1-\alpha)\;, \\
	T+2,   & T \geq 1, ~T\text{ odd, w.p. }\alpha\;, \\
	T+2,   & T \geq 2, ~T\text{ even}\;;
	\end{array}\right.
	\end{equation}
 The lower bound on $N_T$ in Equation \eqref{eq:n_diff} follows immediately from the above expressions.

\vspace{0.1in}
\noindent {\bf Probability of detection.} For any given infected graph $G_T$, the virtual source $v_T$ cannot have been the source node, since the true source always passes the token at timestep $t=1$. So $\prob (G_T|v=v_T)=0$. We claim that for any two nodes that are not the virtual source at time $T$, $u, w\in G_T$, $\prob(G_T|u)=\prob(G_T|w)>0$. This is true iff for any non-virtual-source node $v$, there exists a sequence of virtual sources ${v_i}_{i=0}^T$ that evolves according to Protocol \ref{alg:adp_diff} with $v_0=v$ that results in the observed $G_T$, and for all $u,w\in G_T\setminus \{v_T\}$, this sequence has the same likelihood. 
In a tree, a unique path exists between any pair of nodes, so we can always find a valid path of virtual sources from a candidate node $u \in G_T \setminus \{v_T\}$ to $v_T$. We claim that any such path leads to the formation of the observed $G_T$. 
Due to regularity of $G$ and the symmetry in $G_T$, for even $T$, $\prob(G_T| v^{(1)}) =  \prob(G_T| v^{(2)})$  for all $v^{(1)},v^{(2)} \in G_T$ with $\delta_H(v^{(1)},v_T) = \delta_H(v^{(2)},v_T)$. Moreover, recall that the $\alpha_d(t,h)$'s were designed to satisfy the distribution in Equation \eqref{eq:MCp}. Combining these two observations with the fact that we have $(d-1)^h$ infected nodes $h$-hops away from the virtual source, we get that for all $v^{(1)}, v^{(2)} \in G_T\setminus \{ v_T\}$, $\prob(G_T| v^{(1)}) = \prob(G_T| v^{(2)})$. For odd $T$, if the virtual source remains the virtual source, then $G_T$ stays symmetric about $v_T$, in which case the same result holds. If the virtual source passes the token, then $G_T$ is perfectly symmetric about the edge connecting $v_{T-1}$ and $v_T$. Since both nodes are virtual sources (former and present, respectively) and $T>1$, the adversary can infer that neither node was the true source. Since the two connected subtrees are symmetric and each node within a subtree has the same likelihood of being the source by construction (Equation \eqref{eq:MCp}), we get that for all $v^{(1)}, v^{(2)} \in G_T\setminus \{v_T,v_{T-1}\}$, $\prob(G_T| v^{(1)}) = \prob(G_T| v^{(2)})$. Thus at odd timesteps, $\prob(\hat v_{ML}=v^*)\geq 1/(N_T-2)$.

\subsection{Proof of Proposition \ref{pro:tree}}
\label{sec:proofs_tree}
First, under Protocol \ref{alg:adp_diff} (adaptive diffusion) with $\alpha_d(t,h)=0$, $G_T$ is a complete $(d-1)$-ary tree (with the exception that the root has $d$ children) of depth $T/2$ whenever $T$ is even. $G_T$ is made up of two complete $(d-1)$-ary trees of depth $(T-1)/2$ each with their roots connected by an edge whenever $T$ is odd. Therefore, it follows that $N_T$ is a deterministic function of $T$ and is given by
	\begin{equation}
	\label{eq:n_tree2}
	N_T \;=\;
	\left\{
	\begin{array}{rl}
	1, &  T=0,\\
	\frac{2(d-1)^{(T+1)/2}}{d-2}-\frac{2}{d-2},   & T \geq 1, ~T\text{ odd}\;, \\
	\frac{d(d-1)^{T/2}}{d-2}-\frac{2}{d-2},   & T \geq 2, ~T\text{ even}\;;
	\end{array}\right.
	\end{equation}
 The lower bound on $N_T$ in Equation \eqref{eq:n_tree} follows immediately from the above expression.

 For any given infected graph $G_T$, it can be verified that any non-leaf node could not have generated $G_T$ under the Tree Protocol. In other words, $\prob(G_T|v \text{ non-leaf node}) =0$ and $v$ could not have started the rumor. On the other hand, we claim that for any two leaf nodes $v_1,v_2 \in G_T$, we have that $\prob(G_T|v_1)= \prob(G_T|v_2) >0$. This is true because for each leaf node $v \in G_T$, there exists a sequence of state values $\left\{s_{1,u},s_{2,u}\right\}_{u \in G_T}$ that evolves according to the Tree Protocol with $s_{1,v}= 1$ and $s_{2,v}=0$. Further, the regularity of the underlying graph $G$ ensures that all these sequences are equally likely. Therefore, the probability of correct rumor source detection under the maximum likelihood algorithm is given by $\prob_{ML}(T) = 1/N_{l,T}$, where $N_{l,T}$ represents the number of leaf nodes in $G_T$. It can be also shown that $N_{l,T}$ and $N_T$ are related to each other by the following expression
\begin{equation}
N_{l,T} = \frac{(d-2)N_T+2}{d-1}.
\end{equation}
This proves the expression for $\prob\big( \hv_{\rm ML} =v^* \big)$ given in \eqref{eq:pd_tree}.

{\bf Expected distance.}
For any $v^* \in G$ and any $T$, $\E[\delta_H(v^*,\hv_{\rm ML})]$ is given by
\begin{equation}
\label{eq:exp_d}
\E[\delta_H(v^*,\hv_{\rm ML})]= \sum_{v \in G}\sum_{G_T}\prob(G_T|v^*)\prob(\hv_{\rm ML} =v)\delta_H(v^*,v).
\end{equation}
As indicated above, no matter where the rumor starts from, $G_T$ is a $(d-1)$-ary tree (with the exception that the root has $d$ children) of depth $T/2$ whenever $T$ is even. Moreover, $\hv_{\rm ML}=v$ with probability $1/N_{l,T}$ for all $v$ leaf nodes in $G_T$. Therefore, the above equation can be solved exactly to obtain the expression provided in the statement of the proposition.

\subsection{Proof of Proposition \ref{pro:multiple_snap}} \label{sec:proofs_multiple_snap}
We upper bound the probability of detection by assuming that the adversary takes a snapshot at \emph{every} time step after $T$; the adversary can also learn the exact value of $T$ by noting the size of the snapshots in successive time steps.
The structure of all snapshots after $G_T$ depends deterministically on the binary timeseries of choices to either keep the virtual source token, or to pass it, in each time step after $T$---we refer to this timeseries as $\mathcal K_T$.
The timeseries $\mathcal K_T$, in turn, is random, with values that depend probabilistically on only the timestamp (which is known to the adversary), the tree degree (known), and the virtual source's distance from the true source (unknown). 
Because adaptive diffusion does not allow the virtual source to ``backtrack", or move closer to the true source over time, 
the (unique) path from the true source to the virtual source $v_T$ at time $T$
cannot intersect the path comprised of the virtual sources after time $T$---call it $\mathcal P_T$---except possibly at $v_T$ itself.
Therefore, let us consider the first node in $\mathcal P_T$ that is not equal to $v_T$; we call it $v_{T'}$. 
$v_{T'}$ is necessarily a neighbor of $v_T$.
Then let us define the largest possible subtree of $G_T$ that is rooted at $v_{T'}$ and does \emph{not} contain $v_T$; we call this subtree $\mathcal T_T$. 


Now, suppose that by observing the timeseries $\mathcal K_T$, the adversary could learn the distance between $v^*$ and $v_T$ exactly (this is a worst-case assumption).
Let us call that distance $L$. 
Then the source is equally likely to be any node $w$ at a distance of $L$ hops from $v_T$, such that $w\notin \mathcal T_T$.
Therefore, we can upper bound the probability of detection by conditioning on $L$, and counting the number of feasible nodes $w$.

We assume for the sake of simplicity that all snapshots are taken at even time steps (including $G_T$), since snapshots at odd time steps do not contribute any additional information, i.e., if the adversary observed $G_T$ at an odd timestep, it could recover $G_{T-1}$ from the subsequent observed snapshots, which is equivalent to observing the first snapshot at time $T-1$.
Then 
\begin{eqnarray}
\prob(\hat v_{ML}=v^*) = \sum_{\ell=1}^{T/2} \prob(L=\ell) \prob(\hat v_{ML}=v^* | L=\ell)
\end{eqnarray}
From the previous argument, we have 
$$
\prob(\hat v_{ML}=v^* | L=\ell) = \frac{1}{(d-1)^\ell},
$$
instead of $1/d(d-1)^{\ell-1}$,  
since the entire subtree of $G_T$ containing $\mathcal P_T$ is excluded from the set of possible candidate sources.
Additionally, it is straightforward to compute $\prob(L=\ell)$ from the properties of adaptive diffusion:
$$
\prob(L=\ell) = \frac{(d-2)(d-1)^{\ell-1}}{(d-1)^{T/2} - 1},
$$
so the overall probability of detection is
\begin{eqnarray}
\prob(\hat v_{ML}=v^*) &=& \frac{d-2}{d-1}\cdot \frac{1}{(d-1)^{T/2}-1} \cdot \frac{T}{2}.
\end{eqnarray}
Note that
\begin{eqnarray*}
N_T 	&=& \frac{d (d-1)^{T/2}-2}{d-2}  = \frac{(d-1)^{T/2+1}+(d-1)^{T/2}-2}{d-2}.
\end{eqnarray*}
Since $\left((d-1)^{T/2+1}-2\right )\geq (d-1)^{T/2}$ for all $d>2$ and all even $T\geq 2$, it holds that $N_T \geq \frac{2(d-1)^{T/2}}{d-2}$.
From this, we can conclude that $T/2 \leq \log_{d-1}N_T + \log_{d-1}(d/2-1)$.
It also holds that for all $d>2$, $(d-1)^{T/2}-1 \geq \frac{N_T-1}{3}$,
so we have 
\begin{align}
\prob(\hat v_{ML}=v^*) &\leq& \frac{d-2}{d-1}\cdot \frac{3}{N_T-1} \left (\log N_T + \log(d/2-1) \right ), 
\end{align}
which gives the claim.

\subsection{Proof of Theorem \ref{thm:pd_conditional}} \label{sec:proofs_pd_irregular}

We first analyze the probability of detection for any given estimator (see Eq.   \eqref{eq:prob_detect}); we then show that the estimator in \eqref{eq:map} is a MAP estimator, maximizing this probability of detection. Finally, we show that using the MAP estimator in \ref{eq:map} gives the probability of detection in Eq. \eqref{eq:thm_pd_map}.

\begin{figure}[tb]
	\begin{center}
	\includegraphics[width=3.5in]{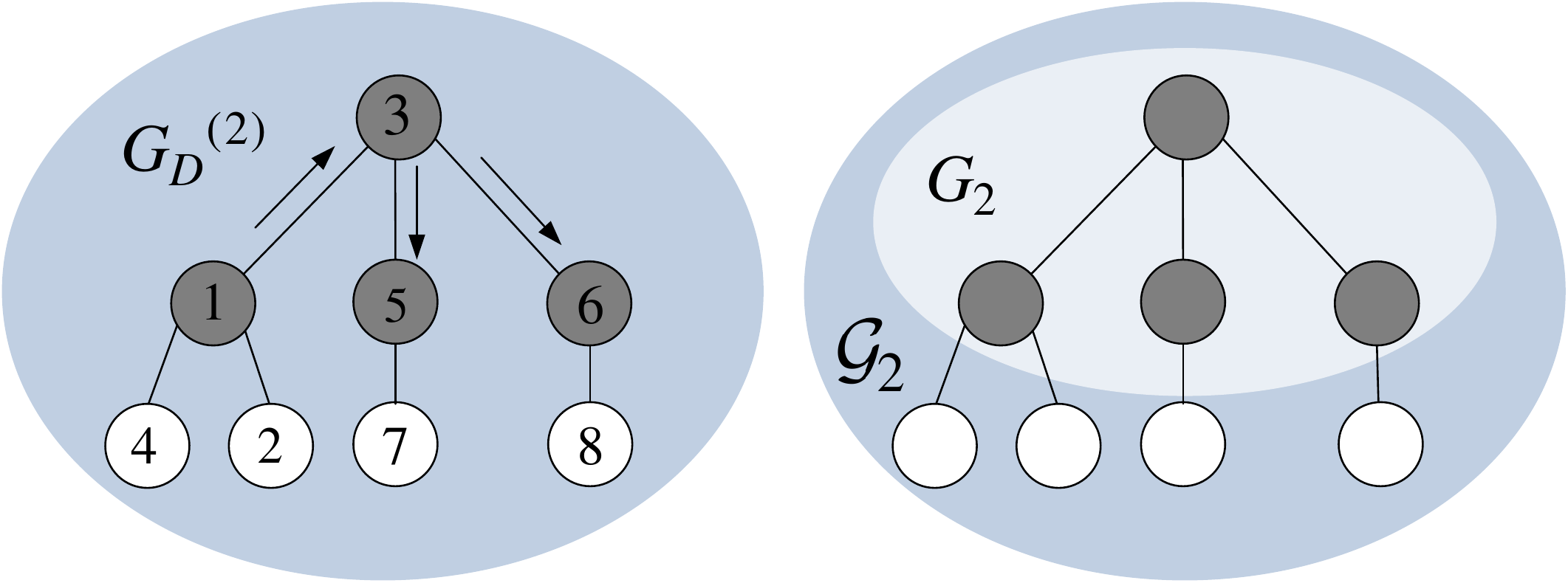}
	\end{center}
	\caption{One realization of the random, irregular-tree branching process. Although each realization of the random process $G_D^{(t)}$ yields a labelled graph, the adversary observes $G_T$ and $\cG_T$, which are \emph{unlabelled}. White nodes are uninfected, grey nodes are infected.}
	\label{fig:rand_proc}
\end{figure}

We begin with some definitions. Consider the following random process, in which we fix a source $v^*$ and generate a (random) labelled tree $G_D^{(t)}$ for each time $t$ and for a given degree distribution $D$. 
At time $t=0$, $G_D^{(t)}$ consists of a single node $v^*$, which is given a label $1$. The source $v^*$ draws a degree $d_1$ from $D$, and generates $d_1$ child nodes, labelled in order of creation (i.e., $2$ through ${d_1+1}$). 
At the next time step, $t=1$, the source picks one of these neighbors uniformly at random to be the new virtual source and infects that neighbor.
According to Protocol \ref{alg:adp_diff}, 
each time a node $v$ is infected, 
$v$ draws its degree $d_v$ from $D$, then generates $d_v-1$ labelled child nodes.
So at the end of time $t=1$, $G_D^{(1)}$ contains the source and its uninfected neighbors, as well as the new virtual source and its uninfected neighbors. An example of $G_D^{(2)}$ is shown in Figure \ref{fig:rand_proc} (left panel) with $d_{1}=3$ and virtual source at node $3$. Grey nodes are infected and white nodes are uninfected neighbors. 
Note that the node labelled $1$ is always exactly one hop from a leaf of $G_D^{(t)}$ for all $t>0$; also, nodes infect their neighbors in ascending order of their labels.
The leaves of $G_D^{(t)}$ represent the uninfected neighbors of infected leaves in standard adaptive diffusion spreading over a given graph.
Define $\Omega_{(t,D)}$ as the set of all labelled trees generated at time $t$ according to this random process. 

\begin{figure}[tb]
	\begin{center}
	\includegraphics[width=3.5in]{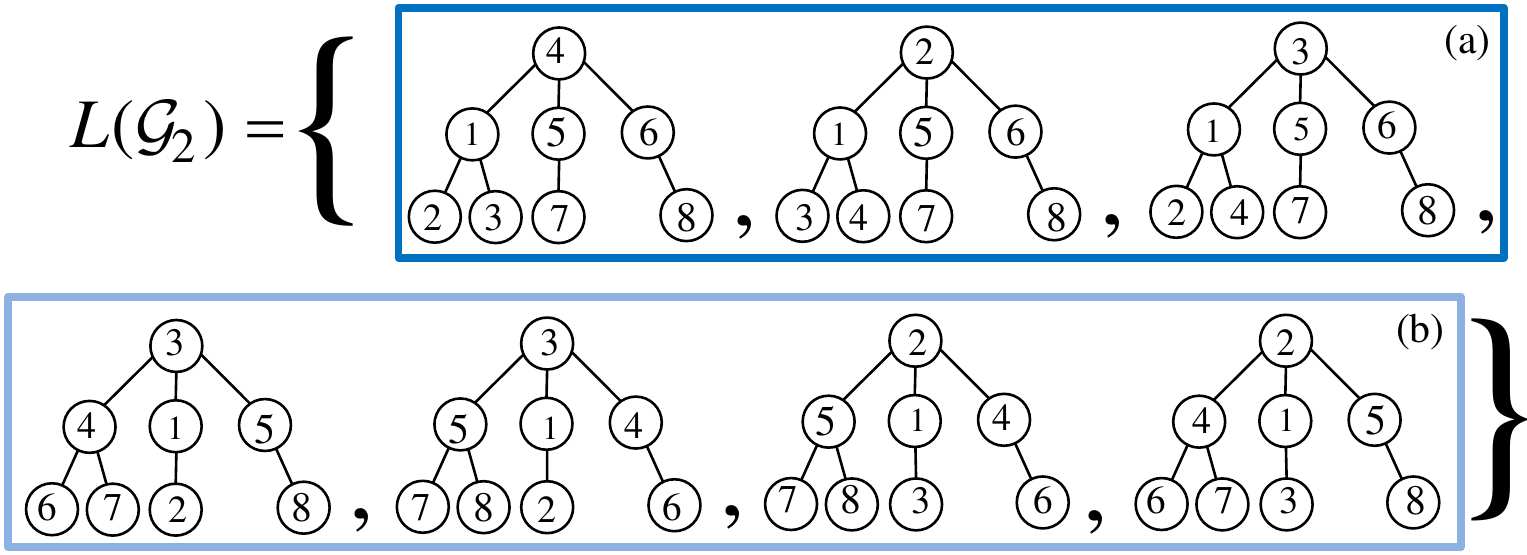}
	\end{center}
	\caption{$L(\cG_2)$ for the snapshot $\cG_2$ illustrated in Figure \protect\ref{fig:rand_proc}. Boxes (a) and (b) illustrate the two families partitioning $L(\cG_2)$.}
	\label{fig:labeling}
\end{figure}

At some time $T$, the 
adversary observes the snapshot of infected subgraph $G_T$. 
Notice that  we do not need to generate the entire contact network, since 
$G_T$ is conditionally independent of the rest of the contact network given its one-hop neighbors. Hence, the we only need to generate (and consider) the one hop neighbors of 
$G_T$ at any given $T$. 
We use $\cG_T$ to denote this random graph that includes $G_T$ and its one hop neighbors as generated according to the previously explained random process.
Notice that the adversary only observes $\cG$, which is 
an \emph{unlabelled} snapshot of the infection and its one hop neighbors (see Figure \ref{fig:rand_proc}, right panel). 
We refer to the  leaves of $G_T$ as `infected leaves', denoted by $\partial G_T$, and the leaves of $\cG_T$ as `uninfected leaves' denoted by $\partial \cG_T$. 
Define
\[
L(\cG_T)\equiv \{\tilde G\in \Omega_{(T,D)}~|~U(\tilde G)=\cG_T\},
\]
i.e., the set of all labelled graphs (generated according to the described random process) whose unlabelled representation $U(\tilde G)$ is equal to the snapshot $\cG_T$.
Figure \ref{fig:labeling} illustrates $L(\cG_T)$ for the graph $\cG_2$ in Figure \ref{fig:rand_proc}.

We define a \emph{family} $C_{\cG_T,v} \subseteq L(\cG_T) $ as 
 the set of all labelled graphs whose labeling could have been generated by breadth-first labeling of $\cG_T$ starting at node $v\in \partial G_T $. 
Here breadth-first labeling is a valid order of traversal for a breadth-first search of $\cG_T$ starting at node $v$. 
We restrict $v$ to be  a valid source for an adaptive diffusion spread---that is, it is an infected leaf in $\partial G_T$.
Note that a BFS labeling starting from two different nodes on the unlabelled tree can yield the same labelled graph. 
In Figure \ref{fig:labeling}, boxes (a) and (b) illustrate the two families contained in $L(\cG_2)$.

Let  $\prob(C_{\cG,v}) \equiv 
\prob( G_D^{(T)} \in C_{\cG,v}  )$ 
denote the probability that the labelled graph $G_D^{(T)}$ whose snapshot is $\cG$ 
is generated from a node $v$.
 From the definition of the random process for generating labelled graphs, we get 
\begin{equation}
\prob(C_{\cG_T,v}) = \underbrace{\left ( \prod_{w\in G_T} \prob_D(d_w) \right )}_{\text{degrees of $G$}} \underbrace{{Q(\cG_T, v)}}_{\text{virtual sources}}\,\underbrace{|C_{\cG_T,v}|}_{\substack{\text{count of}\\\text{isomorphisms}}} 
\label{eq:class_prob}
\end{equation}
where $\prob_D(d)$ is the probability of observing degree $d$ under degree distribution $D$, and
\[
Q(\cG_T, v)=\frac{\mathds 1_{v\in \partial G_T}}{d_{v}\prod_{w\in \Phi_{v,v_T}\setminus \{v,v_T\}}(d_w-1)}
\]
is the probability of passing the virtual source from $v$ to the virtual source 
$v_T$ given the structure of $\cG_T$, where $\Phi_{v,v_T}$ is the unique path from $v$ to $v_T$ in $\cG_T$. 
Eq. \eqref{eq:class_prob} holds because
for all instances in $C_{\cG_T,v}$, the probability of 
the degrees of the nodes and the probability of the path of the virtual source 
remain the same. 


The probability of observing a given snapshot $\cG_T$ is precisely $\prob(G_D^{(T)} \in L(\cG_T))$. 
Notice that $C_{\cG_T,v}$ partitions $L(\cG_T)$ in to family of labelled trees that 
are generated from the same source. This give the following decomposition: 
\begin{equation}
\prob(G_D^{(T)} \in L(G_T)) = \sum_{v\in \mathcal C_{\cG_T} } \prob( C_{\cG_T,v}),
\label{eq:prob_snapshot}
\end{equation}
where we define $\mathcal C_{\cG_T}$ as the set of possible candidates of the source that generate distinct labelled trees, i.e. 
\begin{eqnarray}
	\mathcal C_{\cG_T} \,\equiv\, \{ v \in G_T\,|\, C_{\cG_T,v} \neq C_{\cG_T,v'} ~\forall~ v'\in \mathcal C_{\cG_T},~ v'\neq v \}\;. 
\end{eqnarray} 
Notice that this set is not unique, since there can be multiple nodes that represent the same 
family $C_{\cG_T,v}$. 
We pick one of such node $v$ to represent the class of nodes that can generate the same family of labelled trees. 
We use this $v$ to index these families and  not to denote any particular node in $\partial G_T$. 

Consider an estimate of the source $\hv(\cG_T)$. In general, $\hv(\cG_T)$ is a random variable, potentially selected from a set of candidates. We define detection ($\overline D$) as the event in which  $\hv(\cG_T)=v_1(G_D^{(T)})$; i.e., the estimator outputs the node that started the random process. 
We can partition the set of candidate nodes $\partial G_T$, by grouping together those nodes that are indistinguishable to the estimator into \emph{classes}. Precisely, we define a subset of nodes indexed by $v\in \mathcal C_{\cG_T}$,  
\begin{eqnarray}
\chi_{\cG_T,v} \equiv \{ v' \in \partial G_T \,|\, C_{\cG_T,v} =  C_{\cG_T,v'} \}\;. 
\end{eqnarray}
For a given snapshot, there are as many classes as there are families. 
In Figure \ref{fig:labeling}, the class associated with family (a) has one element---namely, the node labeled `1' in family (a). The class associated with family (b) contains two nodes: the node labeled `1' in family (b), and the node labeled `5' in the rightmost graph of family (b), since both nodes give rise to the same family.

 We consider, without loss of generality, an estimator 
that selects a node in a given class with probability $\prob(\hv(\cG_T) \in \chi_{\cG_T,v})$.  
Notice that $|\chi_{\cG_T,v}|$ denotes the number of (indistinguishable) source candidates in this class.
From Eq. \eqref{eq:prob_snapshot}, the probability of detection given a snapshot is 
\begin{eqnarray}
\prob(\overline D|\cG_T)&=&\frac{\prob \left ( G_D^{(T)} \in L(\cG_T) \land \overline D \right ) }{ \prob ( G_D^{(T)} \in L(\cG_T)  )}. \label{eq:prob_detect}\\
 &=& 
\frac{\sum\limits_{v \in \mathcal C_{\cG_T}}  \prob  ( C_{\cG_T,v})  \prob \left (\overline D\,\big|\, G_D^{(T)} \in C_{\cG_T,v} \right ) }{\sum_{v\in \mathcal C_{\cG_T}}\prob ( C_{\cG_T,v} )}
\label{eq:pd_map}
\end{eqnarray}
where $ \prob (\overline D|G_D^{(T)} \in C_{\cG_T,v}  ) = \prob(\hv(\cG_T) \in \chi_{\cG_T,v}) / |\chi_{\cG_T,v}| $. 
%
%
We use the following observation:
\begin{lemma}
\begin{eqnarray}
\frac{ \prob (C_{\cG_T, v} ) / |\chi_{\cG_T,v}| }
{\sum_{v\in \mathcal C_{\cG_T}}\prob (C_{\cG_T, v})} =
\frac{1}{d_{v_T}\prod\limits_{\substack{w \in \phi(v,v_T) \\ \setminus \{v,v_T\}}}(d_w-1)}.
\label{eq:lemma_prod}
\end{eqnarray}
\label{lem:sum_p_g_v}
\end{lemma}
(Proof in Section \ref{sec:proofs_lem_p_g_v})

Substituting Equation \eqref{eq:lemma_prod} into Equation \eqref{eq:pd_map}, we get that 
\begin{eqnarray*}
\prob(\overline D|\cG_T) &=& \sum_{v \in \mathcal C_{\cG_T}}  \frac{\prob(\hv(\cG_T) \in \chi_{\cG_T,v})}{d_{v_T}\prod\limits_{\substack{w\in \phi(v,v_T)\setminus \\ \{v,v_T\}}}(d_w-1)}.
\end{eqnarray*}
Since each term of this summation is bounded by 
\begin{eqnarray*}
 \frac{\prob(\hv(\cG_T) \in \chi_{\cG_T,v})}{d_{v_T}\prod\limits_{\substack{w\in \phi(v,v_T)\setminus \\ \{v,v_T\}}}(d_w-1)} \leq 
\frac{1}{\min\limits_{v\in \mathcal C_{\cG_T}}  d_{v_T}  \prod\limits_{\substack{w\in \phi(v,v_T) \\ \setminus  \{v,v_T\}}}(d_w-1)},
\end{eqnarray*}
and $\sum_{v\in \mathcal C_{\cG_T}}\prob(\hv(\cG_T) \in \chi_{\cG_T,v})=1$, it must hold that 	
\begin{eqnarray*}
\prob(\overline D|\cG_T) &\leq& \frac{1}{\min\limits_{v\in \mathcal C_{\cG_T}}  d_{v_T}  \prod\limits_{\substack{w\in \phi(v,v_T) \\ \setminus  \{v,v_T\}}}(d_w-1)}.
\end{eqnarray*}
This upper bound on the detection probability is achieved exactly if we choose weight $\prob(\hv(\cG_T) \in \chi_{\cG_T,v})=1$ for the class(es) minimizing the product $\prod_{w\in \phi(v,v_T) \setminus  \{v,v_T\}}(d_w-1)$, i.e.,
\begin{equation*}
\hv(G_T)=\arg \min _{v\in \partial G_T} \prod\limits_{\substack{w\in \phi(v,v_T) \\ \setminus  \{v,v_T\}}}(d_w-1).
\end{equation*}

\subsubsection{Proof of Lemma \protect\ref{lem:sum_p_g_v}}
\label{sec:proofs_lem_p_g_v}

We have that 
\begin{eqnarray*}
\prob(C_{\cG_T,v}) = \underbrace{\left ( \prod_{w\in G_T} \prob_D(d_w) \right )}_{\text{degrees of $G$}} \underbrace{{Q(\cG_T, v)}}_{\text{virtual sources}}\,\underbrace{|C_{\cG_T,v}|}_{\substack{\text{count of}\\\text{isomorphisms}}} 
\end{eqnarray*}
where $v$ is a feasible source for the adaptive diffusion process, i.e., a leaf of the infection $G_T$.

The proof of the lemma proceeds in four steps:
\begin{enumerate}
\item We first  recursively define a function $H(\cG_T,v)$  that is equal to $|C_{\cG_T,v}|$. This function is defined over any balanced, undirected tree and node; the tree need not be generated via the previously-described adaptive diffusion branching process. In addition to $H(\cG_T,v)$, we are interested in $H(\cG_T,v_T)$.  
\item We show that
\begin{eqnarray}
\prob(C_{\cG_T,v})&=&\left ( \prod_{v\in G_T} \prob_D(d_v) \right ) H(\cG_T,v_T) \times \nonumber \\
&& \frac{|\chi_{\cG_T,v}|}{d_{v_T}\prod\limits_{\substack{w\in \phi(v,v_T)\\ \setminus \{v,v_T\}}}(d_w-1)}.
\end{eqnarray}
\item We show that 
\begin{equation}
{\sum_{v \in \mathcal C_{\cG_T,v}}\prob(C_{\cG_T,v})=\left ( \prod_{v\in G_T} \prob_D(d_v) \right ) H (\cG_T,v_T)}.
\label{eq:p_snapshot}
\end{equation}
\item We combine steps (2) and (3) to show the result.
\end{enumerate}

\begin{figure}[tp!]
	\begin{center}
	\includegraphics[width=2.6in]{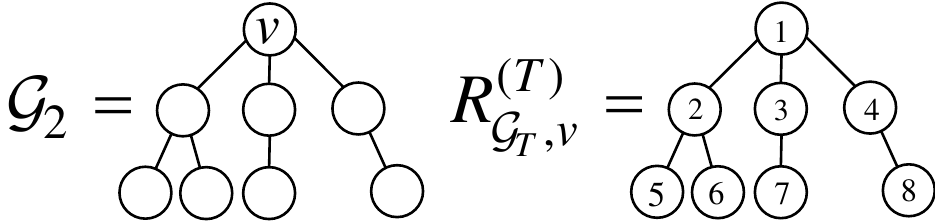}
	\end{center}
	\caption{A realization of the random labeling process given an unlabeled snapshot. }
	\label{fig:labeling_gen}
\end{figure}

\noindent \textbf{Step 1}
We wish to define $H (\cG_T,v)$---a function that counts the number of distinct, isomorphic graphs generated by a breadth-first search of a balanced tree $\cG_T$, rooted at node $v$.   
Consider a random process defined as follows.
Given $\cG_T$ and root node $v$, the process starts at $v$ and labels it $1$. For each neighbor $w$ of node $1$, the process randomly orders $w$'s unlabelled neighbors, and labels them in order of traversal. 
The process proceeds to label nodes in a breadth-first fashion, traversing each node's unlabelled neighbors in a randomly-selected order, until all nodes have been visited. 
Let $R_{\cG_T,v}^{(t)}$ denote a labelled tree generated according to the described random process (see Figure \ref{fig:labeling_gen}).

\begin{figure}[tb]
	\begin{center}
	\includegraphics[width=3.3in]{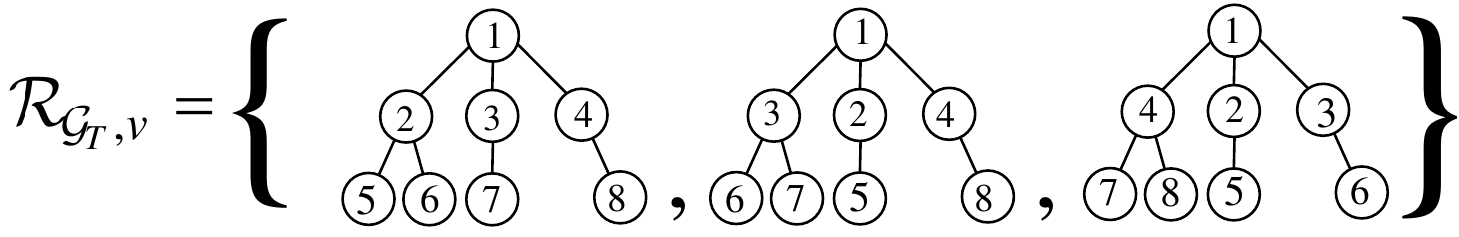}
	\end{center}
	\caption{The set $\mathcal R_{\cG_T,v}^{(T)}$ for the snapshot and node specified in Figure \protect\ref{fig:labeling_gen}.}
	\label{fig:full_set}
\end{figure}

The function $H (\cG_T,v)$ counts the number of distinct graphs that can result from this random process over $\cG_T$ when starting from node $v$.
More precisely, define $\mathcal R_{\cG_T,v}^{(T)}$ as the set of all possible trees $R_{\cG_T,v}^{(T)}$ generated according to this random labeling. 
$H (\cG_T,v)$ is defined as the size of $\mathcal R_{\cG_T,v}^{(T)}$.
Figure \ref{fig:full_set} illustrates $\mathcal R_{\cG_T,v}^{(T)}$ for $\cG_T$ and $v$ shown in Figure \ref{fig:labeling_gen}.
In that example, $H (\cG_T,v)=3$.

Recall that $\cG_T$ is a balanced tree. The Jordan center of this tree is denoted by $v_T$. If $\cG_T$ was generated according to adaptive diffusion, $v_T$ would be the virtual source at time $T$.
Although we say $\cG_T$ is rooted at $v$, we define each node's children with respect to $v_T$. 
That is, node $z$ is among $w$'s children if $z$ is a neighbor of $w$ and $z \notin \phi(w,v_T)$.

Let $\cG_T^{v_i\rightarrow v_j}$ denote the subtree of $\cG_T$ rooted at node $v_j$ with node $v_i$ as parent of $v_j$ (let $\cG_T^{v_1\rightarrow v_1}=\cG_T$). Each node $v_i$ in $\cG_T$  will have some number of child subtrees. 
Some of these subtrees may be identical (i.e., given a realization $R_{\cG_T,v}$ of the labeling random process, they would be isomorphic); let $k_{v}$ denote the number of distinguishable subtrees of node $v$. We use $\Delta_1^{v},\ldots,\Delta_{k_{v}}^{v}$ to denote the \emph{number} of each distinct subtree appearing among the child subtrees of node $v$ (recall children are defined with respect to $v_T$). 
For example, node $v$ in graph $\cG_T$ in Figure \ref{fig:labeling_gen} (left panel) has $\Delta_1^{v}=1$ and $\Delta_2^{v}=2$, since the first of $v$'s child subtrees is equal only to itself, and the second (middle) subtree is isomorphic to the subtree on the right.
If there exists a neighboring, unvisited subtree rooted at a \emph{parent} of $v$, then we say $\Delta_0^v=1$ (by definition, there will only be one such subtree, and it cannot be equal to any child subtrees because $G_T$ is balanced).
Otherwise, we say $\Delta_0^v=0$.
This distinction becomes relevant if $v\neq v_T$.
For example the figure below shows a tree that is rooted at $w\neq v_T$.
In computing $H(\cG_T,w)$, we have $\Delta_0^w=1$ because there is an unvisited branch from $w$ that contains $v_T$, and $\Delta_1^w=2$ because both child subtrees of $w$ are identical.

\begin{figure}[h]
	\begin{center}
	\includegraphics[width=1.2in]{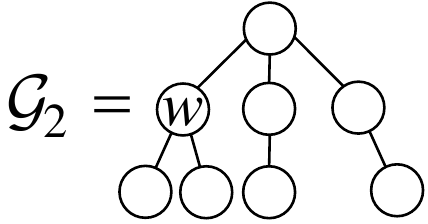}
	\end{center}
	\label{fig:asymmetric}
\end{figure}

Let $\gamma^{v}$ denote the unvisited neighbors of node $v$ in $\cG_T$. 
We give a recursive expression for computing $H (\cG_T,v)$.
\begin{lemma}
\begin{equation}
H (\cG_T,v)= \binom{d_{v}}{\Delta_0^{v}, \Delta_1^{v},\ldots, \Delta_k^{v}} \prod_{w\in \mathcal \gamma^{v}} H(\cG_T^{v\rightarrow w},w).
\label{eq:num_isos}
\end{equation}
\label{lem:isomorphisms}
\end{lemma}

\begin{IEEEproof}
We show this by induction on the depth $\lambda$ of $\cG_T$ (rooted at $v$). For $\lambda=1$, $\cG_T$ has a node $v$ and $d_{v}$ neighbors. Every realization of the random breadth-first labeling of $\cG_T$ will yield an identical graph since the neighbors of $v$ are indistinguishable, so $H (\cG_T,v)=\binom{d_v}{d_v}=1$.

Now suppose Equation \eqref{eq:num_isos} holds for all graph-node pairs $(G_T,v)$ with $\lambda < \lambda_o$; we want to show that it holds for $\lambda=\lambda_o$.
We can represent $\cG_T$ as a root node $v$ and $d_{v}$ subtrees: $\cG_T^{v\rightarrow w}$ for $w\in \gamma^{v}$. 
Since each subtree has depth at most $\lambda_o-1$, we can compute $H(\cG_T^{v\rightarrow w},w)$ for each subtree $\cG_T^{v\rightarrow w}$ using equation \ref{eq:num_isos} (from the inductive hypothesis). 

Suppose we impose (any) valid labeling on $\cG_T$ starting from $v$; we refer to the labeled graph as $R_{\cG,v}$.
Given $R_{\cG,v}$, we order the subtrees of a node in ascending order of their numeric labels.
For any fixed ordering of the $d_v$ subtrees of $v$, we have $\prod_{w\in \gamma^{v}}H(\cG_T^{v\rightarrow w},w)$ nonidentical labelings of $\cG_T$ that respect the ordering of subtrees and are isomorphic to any given realization $R_{\cG_T,v}$. 
At most, there can be $d_v!$ arrangements of the subtrees. However, some of the subtrees are isomorphic, so this value over-counts the number of distinct arrangements. That is, switching the order of two nonidentical, isomorphic subtrees is the same as preserving the order and changing both subtrees to the appropriate nonidentical, isomorphic subtree; this is already accounted for in the product $\prod_{w\in \gamma^{v}}H(G_T^{v\rightarrow w},w)$. 
$\Delta_j^{v}!$ of the $d_v!$ permutations of $v$'s subtrees permute the $j$th unique subtree with isomorphisms of itself. As such, the non-redundant number of different arrangements of the subtrees of node $v$ is $\frac{d_v!}{\Delta_0^{v}!, \Delta_1^{v}!\ldots\Delta_{k_{v}}^{v}!}=\binom{d_v}{\Delta_0^{v}, \Delta_1^{v},\ldots,\Delta_{k_{v}}^{v}}$. This gives the expression in Equation \eqref{eq:num_isos}.
\end{IEEEproof}

\vspace{0.1in}
\noindent \textbf{Step 2.}
We want to show that 
\begin{eqnarray*}
\prob(C_{\cG_T,v})&=& \left ( \prod_{v\in G_T} \prob_D(d_v) \right )  \frac{H(\cG_T,v_T)  |\chi_{\cG_T,v}|}{d_{v_T}\prod\limits_{\substack{w\in \phi(v,v_T)\\ \setminus \{v,v_T\}}}(d_w-1)}.
\end{eqnarray*}
Since $\prob(C_{\cG_T,v}) = \left ( \prod_{v\in G_T} \prob_D(d_v) \right ) Q(\cG_T, v) H (\cG_T,v)$, this is equivalent  to showing that 
\begin{eqnarray*}
\frac{H (\cG_T,v)}{H(\cG_T,v_T)}&=& \frac{|\chi_{\cG_T,v}|}{Q(\cG_T,v) d_{v_T}\prod\limits_{\substack{w\in \phi(v,v_T)\\ \setminus \{v,v_T\}}}(d_w-1)} \\
&=& \frac{d_v}{d_{v_T}}|\chi_{\cG_T,v}|.
\end{eqnarray*}
The expressions for $H(\cG_T,v_T)$ and $H (\cG_T,v)$ differ in that the former starts at the virtual source and counts all subtrees by ``trickling down" the tree (i.e., $\Delta_0^w=0$ for all $w\in G_T$), whereas the latter progresses from an infected leaf $v$ to the virtual source, then recurses over the remaining, unvisited subtrees of $v_T$. Let $P_i$ denote the $i$th node in the path from $v$ to $v_T$, which has length $\ell$. We get
\begin{align*}
H (\cG_T,v)= 
\binom{d_{P_1}}{1,d_{P_1}-1} \times \\
\binom{d_{P_2}-1}{1,\Delta_1^{P_2}-1,\ldots,\Delta_{k_{P_2}}^{P_2} } 
\prod_{w\in \gamma^{P_2}\setminus \{P_1,P_3\}} H(\cG_T^{P_2\rightarrow w},w) \times \\
\ldots \\
\binom{d_{P_{\ell-1}}-1}{1,\Delta_1^{P_{\ell-1}}-1,\ldots,\Delta_{k_{P_{\ell-1}}}^{P_{\ell-1}} } 
\prod\limits_{\substack{w\in \gamma^{P_{\ell-1}} \\ \setminus \{P_{\ell-2},P_\ell\}}} H(\cG_T^{P_{\ell-1}\rightarrow w},w) \times \\
\binom{d_{P_{\ell}}-1}{\Delta_1^{P_{\ell}}-1,\ldots,\Delta_{k_{P_{\ell}}}^{P_{\ell}} } 
\prod_{w\in \gamma^{P_{\ell}}\setminus \{P_{\ell-1}\}} H(\cG_T^{P_{\ell}\rightarrow w},w). 
\end{align*}
where each line corresponds to the terms that result from recursively moving up the path from $v=P_1$ to $v_T=P_\ell$.
Similarly, we have
\begin{align*}
H(\cG_T,v_T)= \binom{d_{P_1}-1}{d_{P_1}-1} \times \\
\binom{d_{P_2}-1}{\Delta_1^{P_2},\ldots,\Delta_{k_{P_2}}^{P_2} } 
\prod_{w\in \gamma^{P_2}\setminus \{P_1,P_3\}} H(\cG_T^{P_2\rightarrow w},w) \times \\
\ldots \\
\binom{d_{P_{\ell-1}}-1}{\Delta_1^{P_{\ell-1}},\ldots,\Delta_{k_{P_{\ell-1}}}^{P_{\ell-1}} } 
\prod\limits_{\substack{w\in \gamma^{P_{\ell-1}} \\ \setminus \{P_{\ell-2},P_\ell\}}} H(\cG_T^{P_{\ell-1}\rightarrow w},w) \times \\
\binom{d_{P_{\ell}}}{\Delta_1^{P_{\ell}},\ldots,\Delta_{k_{P_{\ell}}}^{P_{\ell}} } 
\prod_{w\in \gamma^{P_{\ell}}\setminus \{P_{\ell-1}\}} H(\cG_T^{P_{\ell}\rightarrow w},w). 
\end{align*}
Here we have expanded the expression in terms of the path from $v$ to $v_T$ to make simplification clearer, where $v$ is the node over which we previously computed $H(\cG_T,v)$. 
Computing the ratio of $H (\cG_T,v)$ to $H(\cG_T,v_T)$, all the rightmost products of each line cancel. We are left with the ratio of the combinatorial expressions, which simplify to 
\begin{eqnarray}
\frac{H (\cG_T,v)}{H(\cG_T,v_T)}&=& \frac{d_{P_1}}{d_{P_\ell}}\Delta_1^{P_2}\ldots \Delta_1^{P_{\ell-1}} \Delta_1^{P_\ell} \nonumber \\
&=& \frac{d_{v}}{d_{v_T}}\Delta_1^{v+1}\ldots \Delta_1^{v_T-1} \Delta_1^{v_T}.
\end{eqnarray}
Each $\Delta_1$ denotes the number of child subtrees that are identical to the one containing $v$, for a given root. As such, the product of $\Delta$s above is precisely the number of candidates in the class being considered, or $|\chi_{\cG_T,v}|$. That is, since they are indistinguishable in the unlabelled graph, they generate the same family $C_{\cG_T,v}$.

\vspace{0.1in}
\noindent \textbf{Step 3.}
We have
\begin{align} 
\sum_{v\in \mathcal C_{\cG_T}}\prob(C_{\cG_T,v})&=&\sum_{v\in \mathcal C_{\cG_T}}\left ( \prod_{w\in G_T} \prob_D(d_w) \right ) H(\cG_T,v_T) \nonumber \\
&& \times \frac{| \chi_{\cG_T,v}|}{d_{v_T}\prod_{w\in \phi(v,v_T)\setminus \{v,v_T\}}(d_w-1)} \nonumber  \\
&=&\left ( \prod_{w\in G_T} \prob_D(d_w) \right ) H(\cG_T,v_T) \times   \nonumber \\
&& \sum_{v\in \mathcal C_{\cG_T}} \frac{| \chi_{\cG_T,v}|}{d_{v_T}\prod_{w\in \phi(v,v_T)\setminus \{v,v_T\}}(d_w-1)} \nonumber \\
&=&\left ( \prod_{w\in G_T} \prob_D(d_w) \right ) H(\cG_T,v_T) \times   \nonumber \\
&& \sum_{v\in \partial G_T} \frac{ 1}{d_{v_T}\prod_{w\in \phi(v,v_T)\setminus \{v,v_T\}}(d_w-1)} 
\label{eq:last}
\end{align}
where (\ref{eq:last}) follows because every leaf in the graph is a candidate source in exactly one class.
We wish to show this last summation sums to 1. Consider a random process over $G_T$. The process starts at the virtual source $v_T$, and in each timestep it moves one hop away from $v_T$. It chooses among the (unvisited) children of a node uniformly at random. At time $T$, the process is necessarily at one of the leaves of $G_T$, and the probability of landing at a particular leaf $v$ is precisely $\frac{ 1}{d_{v_T}\prod_{w\in \phi(v,v_T)\setminus \{v,v_T\}}(d_w-1)}$. Therefore, the sum of this quantity over all leaves $v\in \partial G_T$ is 1.

\vspace{0.1in}
\noindent \textbf{Step 4.}
Combining the results from steps 3 and 4, we get that 
\begin{eqnarray*}
\frac{ \prob (C_{\cG_T, v} ) / |\chi_{\cG_T,v}| }
{\sum_{v\in \mathcal C_{\cG_T}}\prob (C_{\cG_T, v})}  = \nonumber \\
\frac{\left ( \prod_{w\in G_T} \prob_D(d_w) \right ) H(\cG_T,v_T)}{\left ( \prod_{w\in G_T} \prob_D(d_w) \right ) H(\cG_T,v_T)} \times \frac{|\chi_{\cG_T,v}|/|\chi_{\cG_T,v}|}{d_{v_T}\prod\limits_{\substack{w\in \phi(v,v_T)\\ \setminus \{v,v_T\}}}(d_w-1)}\\
\frac{1}{d_{v_T}\prod\limits_{w\in \phi(v,v_T) \setminus \{v,v_T\}}(d_w-1)}.
\end{eqnarray*}

\subsection{Proof of Theorem \ref{thm:product}}
\label{sec:proofs_product}
To facilitate the analysis, we consider an alternative random process that generates unlabeled graphs $G_T'$
according to the same distribution as $G_T$ (i.e., the infected, unlabeled subgraph embedded in $U(G_D^{(T)})$ from the proof of Theorem \ref{thm:pd_conditional}). 
For a given degree distribution $D$ and a stopping time $T$, the new process is defined as 
a Galton-Watson process in which the set of offsprings at the first time step is drawn from $D$ and 
the offsprings at subsequent time steps are drawn from $D-1$.  
At time $t=0$, a given root node $v_T$ draws its degree $d_{v_T}$ from $D$, and generates $d_{v_T}$ child nodes. 
The resulting tree now has depth 1. 
In each subsequent time step, the process traverses each leaf $v$ of the tree, draws its degree from $D$, and generates $d_{v}-1$ children.
The random process continues until the tree has depth $T/2$, since under adaptive diffusion, the infected subgraph at even time $T$ has depth $T/2$. 
Because the probability of detection in Equation \eqref{eq:thm_pd_map} does not depend on the degrees of the leaves of $G_T$, the random process stops at depth $T/2$ rather than $T/2+1$.
We call the output of this random process $G_T'$.
The distribution of $G_T'$ is identical to the distribution as the previous random process imposed on $G_T$, which follows from Equation \eqref{eq:p_snapshot} in the proof of Theorem \ref{thm:pd_conditional}.
We therefore use $G_T$ to denote the resulting output in the remainder of this proof.

Distribution $D$ is a multinomial distribution with support $\boldsymbol f=(f_{1},\ldots,f_{\eta})$ and probabilities $\boldsymbol p=(p_1,\ldots,p_{\eta})$. Without loss of generality, we assume $2 \leq f_1<\ldots < f_\eta$.
Let $\mu_D$ denote the mean number of \emph{children} generated by $D$:
\[ \mu_D=\sum_{i=1}^\eta p_i(f_i-1).\]
There are two separate classes of distributions, which we deal with as separate cases. 

\begin{figure}[tp!]
	\begin{center}
	\includegraphics[width=3.4in]{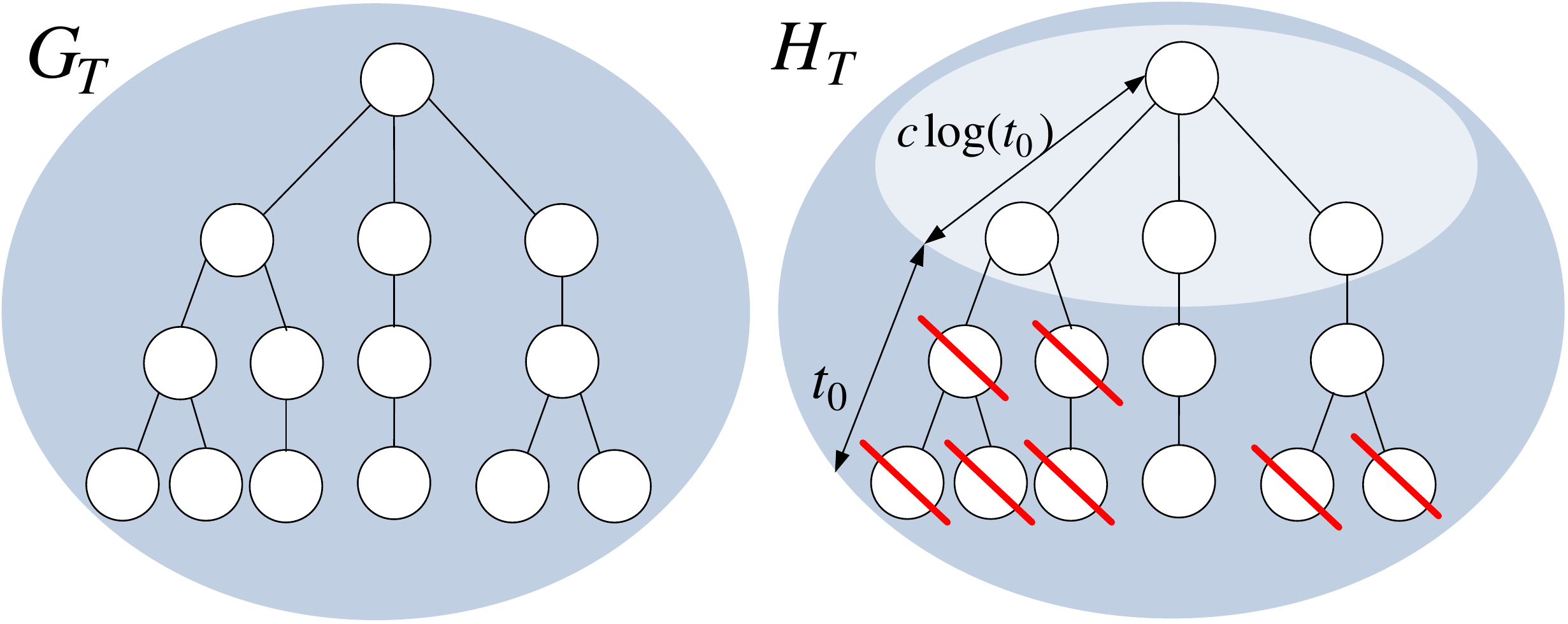}
	\end{center}
	\caption{Pruning of a snapshot. In this example, the distribution $D$ allows nodes to have degree 2 or 3, so we prune all descendants of nodes with degree $3$ that are more than $c\log(t_0)$ hops from the root. In this example, $p_1(f_1-1)<1$ and the pruned random process eventually goes extinct. }
	\label{fig:one_type}
\end{figure}

\bigskip \noindent
{\bf Case 1:} When $p_1(f_1-1)>1$, we claim that 
with high probability, there exists a leaf node $v$ in $\partial G_T$ such that 
on the unique path from the root $v_T$ to this leaf $v$, all nodes in this path have 
the minimum degree $f_1$, 
except for a vanishing fraction. To prove this claim, 
consider a different graph $H_T$ derived from $G_T$ by pruning large degree nodes:
\begin{enumerate}
\item For a fixed, positive $c$, find $t_0$ such that $T/2=t_0 + c\log(t_0)$. 
\item Initialize $H_T$ to be identical to $G_T$.
\item For each node $v\in H_T$, if the hop distance $\delta_H(v,v_T)\leq c\log( t_0)$, do not modify that node.
\item For each node $v\in H_T$, if the hop distance $\delta_H(v,v_T)> c \log(t_0)$ and $d_v > f_1$, prune out all the children of $v$, as well as all their descendants (Figure \ref{fig:one_type}).
\end{enumerate}

We claim that this pruned process survives with high probability. 
The branching process that generates $H_T$ is equivalent to a Galton-Watson process that uses distribution $D-1$ for the first $c \log(t_0)$ generations, and a different degree distribution $D'-1$ for the remaining generations; $D'$ has support $\boldsymbol f'=(f_1,1)$, probability mass $\boldsymbol p'=(p_1, 1-p_1)$, and mean number of children $\mu_{D'}=p_1(f_1-1)$. 

Note that $f_1\geq3$ by the assumption that $p_1(f_1-1)>1$. 
Hence, the inner branching process up to $c\log t_0$ has probability of extinction equal to 0. 
This means that at a hop distance of $t_0$ from $v_T$, there are at least $(f_1-1)^{c\log(t_0)}$ nodes.
Each of these nodes can be thought of as the source of an independent Galton-Watson branching process with degree distribution $D'-1$.
By the properties of Galton-Watson branching processes (\cite{harris2002theory}, Thm. 6.1), since $\mu_{D'}>1$ by assumption, each independent branching process' asymptotic probability of extinction is the unique solution of $g_{D'}(s)=s$, for $s\in[0,1)$, where $g_{D'}(s) =  p_1\,s^{f_1-1} + (1-p_1)$ denotes the probability generating function of the distribution $D'$. 
Call this solution $\theta_{D'}$.
The probability of any individual Galton-Watson process going extinct in the first generation is exactly $1-p_1$. 
It is straightforward to show that $g_{D'}(s)$ is convex, and $g_{D'}(1-p_1)>1-p_1$, which implies that the probability of extinction is nondecreasing over successive generations and  upper bounded by $\theta_{D'}$.
Then for the branching process that generates $H_T$, the overall probability of extinction (for a given time $T$) is at most $\theta_{D'}^{(f_1-1)^{c\log t_0}}$.
Increasing the constant $c$ therefore decreases the probability of extinction.
If there exists at least one leaf at depth $T$ (i.e., extinction did not occur), then there exists at least one path in $H_T$ of length $t_0-c\log t_0$ in which every node (except possibly the final one) has the minimum degree $f_1$. 
This gives
\begin{eqnarray}
 \frac{\log(\Lambda_{H_T})}{T/2} &\leq& \frac{t_0 \log(f_1-1) + c\log(t_0)\log(f_{\eta}-1)}{t_0+c\log(t_0)} \label{eq:gen_fcn1}\\ 
 &\leq& \log (f_1-1) + \frac{c \log t_0}{t_0} \log\frac{f_\eta-1}{f_1-1}\;, \label{eq:gen_fcn2} 
\end{eqnarray}
with probability at least ${1-\theta_{D'}^{(f_1-1)^{c\log t_0}}}
={1-\theta_{D'}^{t_0^{c\log (f_1-1)}}}={1-e^{-C_{D'}t_0}}$, where $C_{D'}=\log(\theta_{D'})$ and
the upper bound in \eqref{eq:gen_fcn1} comes from assuming all the interior nodes have maximum degree $f_\eta$.
Since $H_T$ is a subgraph of a valid snapshot $G_T$, there exists a path in $G_T$ from the virtual source $v_T$ to a leaf of the tree where the hop distance of the path is exactly $T/2$, and at least $t_0$ nodes have the minimum degree $f_1$.
Since the second term in \eqref{eq:gen_fcn2} is $o(t_0)$, the claim follows.
The lower bound $\log(\Lambda_{H_T})/{(T/2)}\geq \log(f_1-1)$ holds by definition.
Therefore, for any $\delta>0$, by setting $T$ (and consequently, $t_0$) large enough, we can make the second term in \eqref{eq:gen_fcn2} arbitrarily small. Thus, for $T \geq C'_{D,\delta}$, where $C'_{D,\delta}$ is a constant that depends only on the degree distribution and $\delta$, the result holds. 

\bigskip\noindent{\bf Case 2: }
Consider the case when $p_1(f_1-1)\leq 1$. 
By the properties of Galton-Watson branching processes (\cite{harris2002theory}, Thm. 6.1), the previous pruned random process that generated graphs $H_T$ goes extinct with probability approaching 1. 
This implies that with high probability there is no path from the root to a leaf that consists
 of only minimum degree nodes.  

Instead, we introduce a Galton-Watson process with multiple types, derived from the original process. 
Our approach is to assign a numeric \emph{type} to each node in $G_T$ according to the number of non-minimum-degree nodes in the unique path between that node and the virtual source. If a node's path to $v_T$ contains too many nodes of high degree, then we prune the node's descendants. 
The challenge is to choose the smallest pruning threshold 
that still ensures the pruned tree will survive with high probability.
Knowing this threshold allows us to precisely characterize $\Lambda_{G_T}$ for most of the instances.

\begin{figure}[tp!]
	\begin{center}
	\includegraphics[width=3.4in]{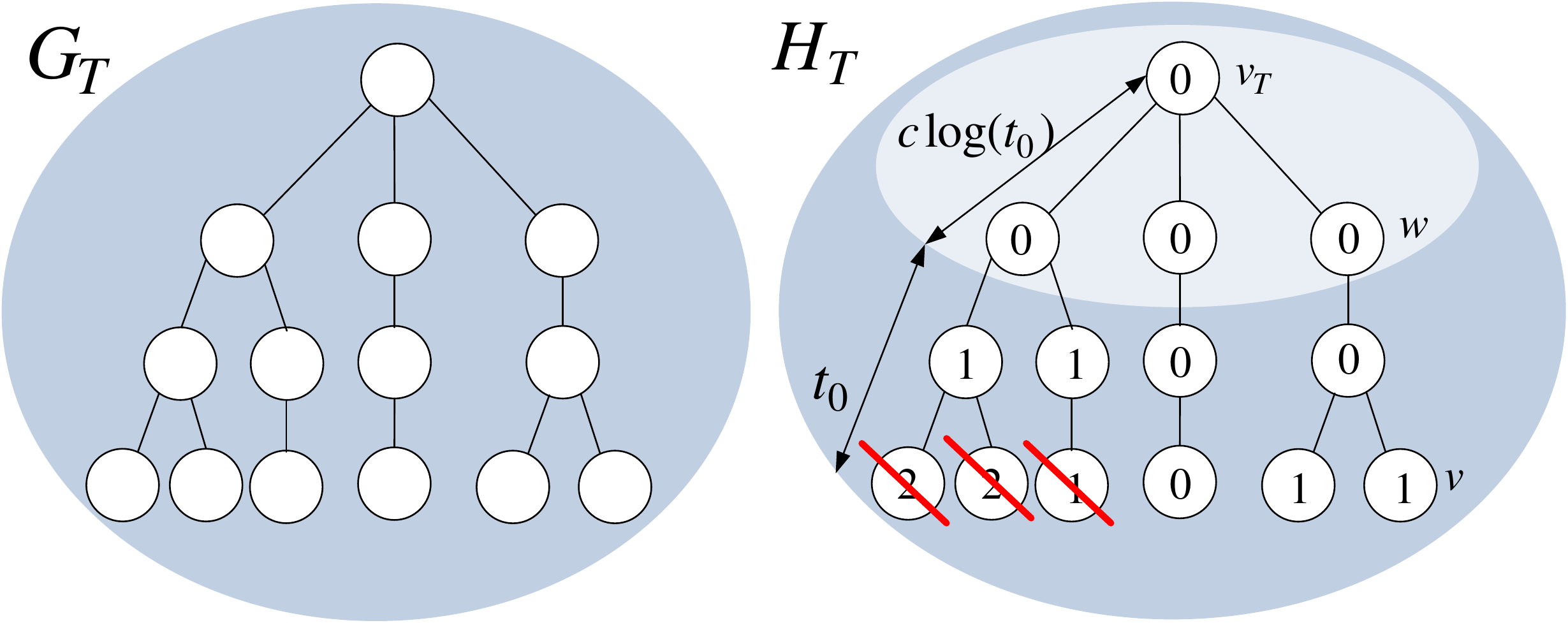}
	\end{center}
	\caption{Pruning of a snapshot using multiple types. In this example, the distribution $D$ allows nodes to have degree 2 or 3. We take $t_0=2$ and $r=0.5$, so all descendants of nodes with type $rt_0=1$ are pruned.}
	\label{fig:multi_type}
\end{figure}

To simplify the discussion, we start by considering a special case in which $D$ allows nodes to take only two values of degrees, i.e., $\eta=2$.
We subsequently extend the results for $\eta=2$ to larger, finite values of $\eta$.
With a slight abuse of a notation, consider a new random process $H_T$ derived from $G_T$ by pruning large degree nodes in the following way:
\begin{enumerate}
\item For a fixed, positive $c$, find $t_0$ such that $T/2=t_0 + c\log(t_0)$. 
\item Initialize $H_T$ to be identical to $G_T$.
\item For each node $v\in H_T$, if the hop distance $\delta_H(v,v_T)\leq c\log( t_0)$, do not modify that node, and assign it type 0.
\item For each node $v\in H_T$, if the hop distance $\delta_H(v,v_T)> c \log(t_0)$, 
assign $v$ a type  $\xi_v$, which is the number of nodes in $\phi(w,v)\setminus\{v\}$ that have the maximum possible degree $f_2$, where $w$ is the closest node in $H_T$ to $v$ such that  $\delta_H(w,v_T)\leq c\log(t_0)$ (Figure \ref{fig:multi_type}).
\item Given a threshold $r\in(0,1)$, if a node $v$ has type $\xi_v \geq rt_0$, prune out all the descendants of $v$. For example, in Figure \ref{fig:multi_type}, if $t_0=2$ and the threshold is $r=0.5$, we would prune out all descendants of nodes with $\xi_v\geq 1$.
\end{enumerate}

We show that for an appropriately-chosen threshold $r$, this pruned tree survives with high probability. 
By choosing the smallest possible $r$, we ensure that $\Lambda_{H_T}$ consists (in all but a vanishing fraction of nodes) of a fraction $r$ nodes with maximum degree, and $(1-r)$ of minimum degree. 
This allows us to derive the bounds on $\log(\Lambda_{H_T})/(T/2)$ stated in the claim, which hold with high probability.

Let $k\equiv rt_0$.
The process that generates $H_T$ is equivalent to a different random branching process that generates nodes in the following manner: set the root's type $\xi_{v_T}=0$.
At time $t=0$, the root $v_T$ draws a number of children according to distribution $D$, and generates $d_{v_T}$ children, all type 0. 
Each leaf generates type 0 children according to child degree distribution $D-1$ until $c\log(t_0)$ generations have passed. 
At that point, each leaf $v$ in this branching process (which necessarily has type 0) reproduces as follows:
if its type $\xi_v > k$, then $v$ does not reproduce. 
Otherwise, it either generates $(f_1-1)$ children with probability $p_1$, each with state $\xi_v$, or it generates $(f_2-1)$ children with probability $p_2$, each with state $\xi_v+1$.
This continues for $t_0$ generations.
Mimicking the notation from Case 1, we use $D'$ to denote the distribution that gives rise to this modified, multi-type random process (in the final $t_0$ generations); this is a slight abuse of notation since the branching dynamics are multi-type, not defined by realizations of i.i.d. degree random variables. 

\begin{lemma}
Consider a Galton-Watson branching process with child degree distribution $D-1$, where each node has at least one child with probability 1, and $\mu_{D-1}>1$. Then the number of leaves in generation $t$, $Z^{(t)}$, satisfies the following:
\[
Z^{(t)}\geq e^{C_\ell t}
\]
with probability at least $1-e^{C_\ell ' t}$, where both $C_\ell$ and $C'_\ell$ are constants that depend on the degree distribution.
\label{lem:num_leaves}
\end{lemma}
(Proof in Section \ref{sec:proofs_num_leaves})

\bigskip
\noindent
The first $c\log(t_0)$ generations ensure that with high probability, we have at least
$e^{C_\ell \log t_0}$ independent multi-type Galton-Watson processes originating from the leaves of the inner subgraph; this follows from Lemma \ref{lem:num_leaves}. 
Here we have encapsulated the constant $c$ from the first $c\log(t_0)$ generations in the constant $C_\ell$.
For example, in Figure \ref{fig:multi_type}, there are 3 independent Galton-Watson processes starting at the leaves of the inner subgraph.
We wish to choose $r$ such that the expected number of new leaves generated by \emph{each} of these processes, at \emph{each} time step, is large enough to ensure that extinction occurs with probability less than one.
For brevity, let $\alpha \equiv p_1(f_1-1)$ and let $\beta \equiv p_2(f_2-1)$.
Let $x^{(t)}$ denote the $(k+1)$-dimensional vector of the expected number of leaves generated with each type from 0 to $k$ in generation $t$. This vector evolves according to the following $(k+1)\times (k+1)$ transition matrix $M$:
\[
x^{(t+1)} =  x^{(t)} \,
\underbrace{
\begin{bmatrix}
	\alpha & \beta & & &\\
	&       & \ddots & & \\
	& & \ddots& \alpha & \beta\\
	& & &  & 0\\
	\end{bmatrix}}_{M}
\;.
\]
The last row of $M$ is 0 because a node with type $k$ does not reproduce.
Since the root of each process always has type 0, we have $x^{(0)}=\boldsymbol e_1$, where $\boldsymbol e_1$ is the indicator vector with a 1 at index 1 and zeros elsewhere. 

Let $Z^{(t)}$ denote the expected number of new leaves created in generation $t$.
This gives
\begin{equation}
\E[Z^{(t)}]=\boldsymbol e_1 M^t \mathds 1_{(k+1)}^\intercal ,
\label{eq:mean_leaves}
\end{equation}
where $^\intercal$ denotes a transpose, and $\mathds 1_{(k+1)}$ is the $(k+1)$ all-ones vector.
When $t<k$, this is a simple binomial expansion of $(\alpha+\beta)^t$. 
For $t\geq k$, this is a truncated expansion up to $k$: 
\begin{equation}
\E[Z^{(t)}]=\sum_{i=0}^k \binom{t}{i} \alpha^{t-i}\beta^i.
\label{eq:truncated}
\end{equation} 

We seek the necessary and sufficient condition on $r$ for non-extinction, such that  $({1}/{t})\log(\E[Z^{(t)}])>0$. 
Consider a binomial random variable $W$ with parameter $\beta/(\alpha+\beta)=\beta/\mu_D$ and $t$ trials. Equation \eqref{eq:truncated} implies that for large $t$, 
\begin{eqnarray}
\E[Z^{(t)}] &=& (\alpha+\beta)^t \, \prob(W\leq k). \label{eq:GaussAppx1} \\
	&=& \mu_D^t \, \exp\Big\{ -t\,D_{\rm KL} \Big( r \,\|\, \frac{\beta}{\mu_D} \Big) + o(t) \Big\} \;, \label{eq:GaussAppx2}
\end{eqnarray}
by Sanov's theorem \cite{cover2012elements}. 
We wish to identify the smallest $r$ for which $(1/t)\log(\E[Z^{(t)}])$ is bounded away from zero.  Such an $r$ is a sufficient (and necessary) condition for the multi-type Galton-Watson process to have a probability of extinction less than 1.  
To achieve this, we define the following set of $r$ such that Eq. \eqref{eq:GaussAppx2} is strictly positive, for some $\epsilon > 0$:
\begin{eqnarray}
{\mathcal R}_{\alpha,\beta}(\epsilon) \;=\;  \big\{\,  r~|~\log(\mu_D) \geq  D_{\rm KL}(r\|\beta/\mu_D) + \epsilon \,\big\} \, ,
\label{eq:lower}
\end{eqnarray}
Suppose we now choose a threshold $r \in {\mathcal R}_{\alpha,\beta}(\epsilon)$.  
This is the regime where the modified Galton-Watson process 
with threshold $r$
has a chance for survival. 
In other words, the probability of extinction $\theta_{D'}$ is strictly less than one. 
Precisely, $\theta_{D'}$ is the unique solution to $s=g_{D'}(s)$, 
where $g_{D'}(s)$ denotes the probability generating function of the described multi-type Galton-Watson process. 
Using the same argument as in Case 1, 
we can construct a process where the probability of extinction is asymptotically zero. 
Precisely, we modify the pruning process such that we do not prune any leaves in the first $c\log(t_0)$ generations. 
This ensures that with high probability, there are at least $e^{C_{\ell}\log(t_0)}$ independent multi-type Galton-Watson processes evolving concurrently after time $c\log(t_0)$, each with probability of extinction $\theta_{D'}$.
Hence with probability at least 
$1-e^{-2C_{D'}t_0}$ (for an appropriate choice of a constant $C_{D'}$ that only depends on the degree distribution $D'$ and the choice of $r$), the overall process does not go extinct. 

Our goal is to find the choice of $r$ with minimum product of degrees ${\log(\Lambda_{G_T})}/({T/2})$ that survives. We define $r_1$ as follows: 
\begin{equation*}
\begin{aligned}
r_1 \equiv & \underset{r \in \mathcal R_{\alpha,\beta}(\epsilon)}{\arg \min}
& & (1-r)\log(1-f_1) + r\log(1-f_2).
\end{aligned}
\end{equation*}
Since $ \mathcal R_{\alpha,\beta}(\epsilon)$ is just an interval and we are minimizing a linear function with a positive slope, the optimal solution is $r_1 = \inf_{r \in \mathcal R_{\alpha,\beta}(\epsilon)} r$.
This is a choice that survives with high probability and has the minimum product of degrees. 
Precisely,  
with probability at least $1-e^{-C_{D'}T}$, where $C_{D'}$ depends on $D'$ and $\epsilon$, 
we have that 
\[
\frac{\log(\Lambda_{G_T})}{T/2} \leq \langle r_1,f \rangle  +  \frac{c \log(t_0)}{t_0} \log \left (f_2-1\right ) 
\]
where with a slight abuse of notation, we define $\langle r_1,\boldsymbol f \rangle  \triangleq (1-r_1)\log(f_1-1) + r_1\log(f_2-1)$. 
It follows that 
\begin{align}
&\frac{\log(\Lambda_{G_T})}{T/2} - \langle r^*,\boldsymbol f \rangle   \leq \nonumber\\
& 
(r_1-r^*)\log \left (\frac{f_2-1}{f_1-1}\right ) + \frac{ c \log(t_0)}{t_0} \log \left (f_2-1\right )
\end{align}
By setting $\epsilon$ small enough and $t_0$ large enough, we can make this as small as we want. 
For any given $\delta>0$,  there exists a positive $\epsilon>0$ such that the first term is bounded by $\delta/2$. 
Further, recall that $T/2 = c\log(t_0)+t_0$. 
For any choice of $\epsilon$, there exists a $t_{D',\epsilon}$ such that 
for all $T\geq t_{D',\epsilon}$ the vanishing term in Eq. \eqref{eq:GaussAppx2} is smaller than $\epsilon$. 
For any given $\delta>0$, 
there exists a positive $t_{D',\delta}$ such that $T\geq t_{D',\delta}$ implies that the second term is upper bounded by $\delta/2$.
Putting everything together (and setting $\epsilon$ small enough for the target $\delta$), we get that 
\begin{align}
	\prob\Big(\frac{\log(\Lambda_{G_T})}{T/2} \geq  \langle r^*,\boldsymbol f  \rangle  + \delta \Big) 
	\; \leq\;  e^{-C_{D',\delta} T}
\end{align}
for all $T\geq C'_{D',\delta}$, where $C_{D',\delta}$ and $C'_{D',\delta}$ are positive constants that only depend on the degree distribution $D'$ and  the choice of $\delta>0$. 


For the lower bound, 
we define the following set of $r$ such that Eq. \eqref{eq:GaussAppx2} is strictly negative:
\begin{eqnarray}
\overline{\mathcal R}_{\alpha,\beta}(\epsilon) \;=\;  \big\{\,  r ~|~\log(\mu_D) \leq  D_{\rm KL}(r\|\beta/\mu_D) -\epsilon \,\big\} \, .
\label{eq:upper}
\end{eqnarray}
Choosing $r \in \overline{\mathcal R}_{\alpha,\beta}(\epsilon)$ 
causes extinction with probability approaching 1. 
Explicitly, $\prob(Z^{(t)}\neq0)$ is the probability of non-extinction at time $t$, and $\prob(Z^{(t)}\neq0)\leq \E[Z^{(t)}]$. By Equation \eqref{eq:GaussAppx2}, we have 
\begin{eqnarray*}
\E[Z^{(t)}] &\leq& e^{t \left ( \log(\mu_D) - D_{\rm KL}(r\|\beta/\mu_D)  + o(t)  \right )}
\end{eqnarray*}
where $\log(\mu_D) - D_{\rm KL}(r\|\beta/\mu_D)\leq -\epsilon$. 
The probability of extinction is therefore at least $1-\E[Z^{(t)}]\geq 1-e^{-t(\epsilon + o(t))}$. 
So defining 
\begin{equation*}
\begin{aligned}
r_2 \equiv & \underset{r \in \overline {\mathcal R}_{\alpha,\beta}(\epsilon)}{\arg \max}
& & (1-r)\log(1-f_1) + r\log(1-f_2),
\end{aligned}
\end{equation*}
we have
\[
\frac{\log(\Lambda_{G_T})}{T/2} \geq \langle r_2,\boldsymbol f \rangle 
+\frac{c\log(t_0)}{t_0}\log(f_1-1)
\]
with probability at least $1-e^{-C_{D',2}T}$ where $C_{D',2}$ is again a constant that depends on $D'$ and $\epsilon$. 
It again follows that 
\begin{align}
&\frac{\log(\Lambda_{G_T})}{T/2} - \langle r^*,\boldsymbol f \rangle   \geq \nonumber\\
& 
(r_2-r^*)\log \left (\frac{f_2-1}{f_1-1}\right ) + \frac{ c \log(t_0)}{t_0} \log \left (f_1-1\right )\;, 
\end{align}
where $r_2-r^*$ is strictly negative.
Again, 
for any given $\delta>0$,  there exists a positive $\epsilon>0$ such that the first term is lower bounded by $-\delta/2$, 
and for any choice of $\epsilon$, there exists a $t_{D',\epsilon}$ such that 
for all $T\geq t_{D',\epsilon}$ the vanishing term in Eq. \eqref{eq:GaussAppx2} is smaller than $\epsilon$. 
Note that this $\epsilon$ might be different from the one used to show the upper bound. 
We ultimately choose the smaller of the two $\epsilon$ values.
For any given $\delta>0$, 
there exists a positive $t_{D',\delta}$ such that $T\geq t_{D',\delta}$ implies that the second term is lower bounded by $-\delta/2$.
Putting everything together (and setting $\epsilon$ small enough for the target $\delta$), we get that 
\begin{align}
	\prob\Big(\frac{\log(\Lambda_{G_T})}{T/2} \leq  \langle r^*,\boldsymbol f  \rangle  - \delta \Big) 
	\; \leq\;  e^{-C_{D',\delta} T}
\end{align}
for all $T\geq C'_{D',\delta}$, where $C_{D',\delta}$ and $C'_{D',\delta}$ are positive constants that only depend on the degree distribution $D'$ and  the choice of $\delta>0$. 
This gives the desired result.

\bigskip
We now address the general case for $D$ with support greater than two. 
We follow the identical structure of the argument. 
The first major difference is that node types are no longer scalar, but tuples. Each node $v$'s type $\xi_v$ is the $(\eta-1)$-tuple listing how many nodes in the path $\phi(w,v)\setminus \{v\}$ had each non-minimum degree from $f_2$ to $f_\eta$, where $w$ is the closest node to $v$ such that $\delta_H(w,v_T) \leq c\log(t_0)$. 
Consequently, the threshold $\boldsymbol r=[r_1,\ldots,r_{\eta-1}]$ is no longer a scalar, but a vector-valued, pointwise threshold on each element of $\xi_v$.
We let $\boldsymbol k=[k_1=r_1t_0,\ldots,k_{\eta-1}=r_{\eta-1} t_0 ]$ denote the time-dependent threshold, and we say $\boldsymbol k<\xi_v$ if $k_i < (\xi_v)_i$ for $1\leq i \leq \eta-1$.
The matrix $M$ is no longer second-order, but a tensor. 
Equation \eqref{eq:mean_leaves} still holds, except $M$ is replaced with its tensor representation. 
For brevity, let $\alpha=p_1(f_1-1)$ and $\beta_i=p_{i+1}(f_{i+1}-1)$. Let $\tilde \beta=\sum_{i=1}^{\eta-1} \beta_i$.
Hence, Equation \eqref{eq:truncated} gets modified as 
\begin{equation}
\E[Z^{(t)}]=\sum_{i_1=0}^{k_1}\ldots \sum_{i_{\eta-1}=0}^{k_{\eta-1}} \binom{t}{i_1,\ldots,i_{\eta-1}} \alpha^{t-\tilde \beta}\beta_1^{i_1}\ldots \beta_{\eta-1}^{i_{\eta-1}}.
\label{eq:truncated2}
\end{equation}

Now we consider a \emph{multinomial} variable $W$ with parameters $\beta_i/\mu_D$ for $1\leq i \leq \eta - 1$ and $t$ trials. Note that $\alpha/\mu_D$ is the `failure' probability (corresponding to a node of degree $f_1$); such events do not contribute to the category count, so the sum of parameters is strictly less than 1. 
As before, equation \eqref{eq:truncated2} can equivalently be written as 
\begin{eqnarray}
\E[Z^{(t)}]\; &=&\;\mu_D^t \, \prob(W\leq \boldsymbol k) \nonumber \\
	&=& \mu_D^t \, \exp\Big\{ -t\,D_{\rm KL} \Big( \boldsymbol r \,\|\, \left(\frac{\boldsymbol \beta}{\mu_D}\right) \Big) + o(t) \Big\} \;, \label{eq:GaussAppx3}
\end{eqnarray}
where $\boldsymbol \beta / \mu_D$ denotes elementwise division.
Once again, we 
wish to obtain bounds on $\prob(W\leq \boldsymbol k)$.
As before, we define the following set of $r$ such that Eq. \eqref{eq:GaussAppx3} is strictly positive, for some $\epsilon > 0$:
\begin{eqnarray}
{\mathcal R}_{\alpha,\boldsymbol \beta}(\epsilon) \;=\;  \big\{\,  \boldsymbol r~|~\log(\mu_D) \geq  D_{\rm KL}(\boldsymbol r\|\left(\frac{\boldsymbol \beta}{\mu_D}\right) ) + \epsilon \,\big\} \, ,
\label{eq:lower}
\end{eqnarray}

We now choose a threshold $\boldsymbol r \in {\mathcal R}_{\alpha,\beta}(\epsilon)$.  
Using the same argument as before, 
we can construct a process where the probability of extinction is asymptotically zero. 
We again do not prune any leaves in the first $c\log(t_0)$ generations. 
This ensures that with high probability, there are at least $e^{C_{\ell}\log(t_0)}$ independent multi-type Galton-Watson processes evolving concurrently after time $c\log(t_0)$, each with probability of extinction $\theta_{D'}$.
Hence with probability at least 
$1-e^{-2C_{D'}t_0}$ (for an appropriate choice of a constant $C_{D'}$ that only depends on the degree distribution $D'$ and the choice of $\boldsymbol r$), the overall process does not go extinct. 

We define $\boldsymbol r_1$ analogously to the $\eta=2$ case: 
\begin{equation*}
\begin{aligned}
\boldsymbol r_1 \equiv & \underset{\boldsymbol r \in \mathcal R_{\alpha,\boldsymbol \beta}(\epsilon)}{\arg \min}
& & \langle \boldsymbol r, \boldsymbol f\rangle \;, 
\end{aligned}
\end{equation*}
where we now define $\langle \boldsymbol r, \boldsymbol f\rangle\equiv (1-\sum_i r_i)\log(f_1-1) + \sum_{j=1}^{\eta-1}r_j \log(f_{j+1}-1)$.
Therefore  
with probability at least $1-e^{-C_{D'}T}$, where $C_{D'}$ depends on $D'$ and $\epsilon$, 
we have that 
\[
\frac{\log(\Lambda_{G_T})}{T/2} \leq \langle \boldsymbol r_1,\boldsymbol f \rangle  +  \frac{c \log(t_0)}{t_0} \log \left (f_\eta-1\right ).
\]
It follows that 
\begin{align}
&\frac{\log(\Lambda_{G_T})}{T/2} - \langle \boldsymbol r^*,\boldsymbol f \rangle   \leq \nonumber\\
& 
\sum_{j=1}^{\eta-1} ((r_1)_j-r^*_j)\log \left (\frac{f_{j+1}-1}{f_1-1}\right ) + \frac{ c \log(t_0)}{t_0} \log \left (f_\eta-1\right )\;. 
\label{eq:highdim_upper}
\end{align}
By setting $\epsilon$ small enough and $t_0$ large enough, we can make this as small as we want. 
For any given $\delta>0$,  there exists a positive $\epsilon>0$ such that each term in the summation in \eqref{eq:highdim_upper} is bounded by $\delta/\eta$. 
Further, recall that $T/2 = c\log(t_0)+t_0$. 
For any choice of $\epsilon$, there exists a $t_{D',\epsilon}$ such that 
for all $T\geq t_{D',\epsilon}$ the vanishing term in Eq. \eqref{eq:GaussAppx2} is smaller than $\epsilon$. 
For any given $\delta>0$, 
there exists a positive $t_{D',\delta}$ such that $T\geq t_{D',\delta}$ implies that the second term of \eqref{eq:highdim_upper} is upper bounded by $\delta/\eta$.
Putting everything together (and setting $\epsilon$ small enough for the target $\delta$), we get that 
\begin{align}
	\prob\Big(\frac{\log(\Lambda_{G_T})}{T/2} \geq  \langle \boldsymbol r^*,\boldsymbol f  \rangle  + \delta \Big) 
	\; \leq\;  e^{-C_{D',\delta} T}
\end{align}
for all $T\geq C'_{D',\delta}$, where $C_{D',\delta}$ and $C'_{D',\delta}$ are positive constants that only depend on the degree distribution $D'$ and  the choice of $\delta>0$. 

For the lower bound, 
we again define a set of $r$ such that Eq. \eqref{eq:GaussAppx2} is strictly negative:
\begin{eqnarray}
\overline{\mathcal R}_{\alpha,\boldsymbol \beta}(\epsilon) \;=\;  \big\{\,  \boldsymbol r ~|~\log(\mu_D) \leq  D_{\rm KL}(\boldsymbol r\|\left ( \frac{\boldsymbol \beta}{\mu_D} \right )) -\epsilon \,\big\} \, .
\label{eq:upper}
\end{eqnarray}
Choosing $\boldsymbol r \in \overline{\mathcal R}_{\alpha,\boldsymbol \beta}(\epsilon)$ 
causes extinction with probability approaching 1. 
Explicitly, $\prob(Z^{(t)}\neq0)$ is the probability of non-extinction at time $t$, and $\prob(Z^{(t)}\neq0)\leq \E[Z^{(t)}]$. By Equation \eqref{eq:GaussAppx2}, we have 
\begin{eqnarray*}
\E[Z^{(t)}] &\leq& e^{t \left ( \log(\mu_D) - D_{\rm KL}(\boldsymbol r\| \boldsymbol \beta/ \mu_D)  + o(t)  \right )}
\end{eqnarray*}
where $\log(\mu_D) - D_{\rm KL}(\boldsymbol r\| \boldsymbol \beta/\mu_D)\leq -\epsilon$. 
The probability of extinction is therefore at least $1-\E[Z^{(t)}]\geq 1-e^{-t(\epsilon + o(t))}$. 
So defining 
\begin{equation*}
\begin{aligned}
\boldsymbol r_2 \equiv & \underset{\boldsymbol r \in \overline {\mathcal R}_{\alpha,\beta}(\epsilon)}{\arg \max}
& & \langle \boldsymbol r,\boldsymbol  f \rangle \;,
\end{aligned}
\end{equation*}
we have
\[
\frac{\log(\Lambda_{G_T})}{T/2} \geq \langle \boldsymbol r_2,\boldsymbol f \rangle 
+\frac{c\log(t_0)}{t_0}\log(f_1-1)
\]
with probability at least $1-e^{-C_{D',2}T}$ where $C_{D',2}$ is again a constant that depends on $D'$ and $\epsilon$. 
It follows that 
\begin{align}
&\frac{\log(\Lambda_{G_T})}{T/2} - \langle \boldsymbol r^*,\boldsymbol f \rangle   \geq \nonumber\\
& 
\sum_{j=1}^{\eta-1} ((r_2)_j-r^*_j)\log \left (\frac{f_{j+1}-1}{f_1-1}\right ) + \frac{ c \log(t_0)}{t_0} \log \left (f_\eta-1\right )\;. 
\label{eq:highdim_lower}
\end{align}
where $(r_2)_j-r^*_j$ is strictly negative.
Again, 
for any given $\delta>0$,  there exists a positive $\epsilon>0$ such that each term in the summation in \eqref{eq:highdim_lower} is lower bounded by $-\delta/\eta$, 
and for any choice of $\epsilon$, there exists a $t_{D',\epsilon}$ such that 
for all $T\geq t_{D',\epsilon}$ the vanishing term in Eq. \eqref{eq:GaussAppx2} is smaller than $\epsilon$. 
We again choose the smaller of the two $\epsilon$ values from the upper and lower bound.
For any given $\delta>0$, 
there exists a positive $t_{D',\delta}$ such that $T\geq t_{D',\delta}$ implies that the second term is lower bounded by $-\delta/\eta$.
Putting everything together (and setting $\epsilon$ small enough for the target $\delta$), we get that 
\begin{align}
	\prob\Big(\frac{\log(\Lambda_{G_T})}{T/2} \leq  \langle \boldsymbol r^*,\boldsymbol f  \rangle  - \delta \Big) 
	\; \leq\;  e^{-C_{D',\delta} T}
\end{align}
for all $T\geq C'_{D',\delta}$, where $C_{D',\delta}$ and $C'_{D',\delta}$ are positive constants that only depend on the degree distribution $D'$ and  the choice of $\delta>0$. 
This gives the desired result.

\subsubsection{Proof of Lemma \ref{lem:num_leaves}}
\label{sec:proofs_num_leaves}
If $f_1>2$, then the claim follows directly, because each leaf generates at least 2 children in each generation.

If $f_1=2$, then
for parameters $\rho>0$ and $\lambda>0$, we use the Markov inequality to get 
\begin{eqnarray*}
\prob(Z^{(t)} \leq \rho) 
&\leq& \E[e^{-\lambda Z^{(t)}}]e^{\rho \lambda}\\
&=& g_{D-1}^{(t)}(e^{-\lambda})e^{\rho \lambda}\;, 
\end{eqnarray*}
where $g_{D-1}(s)= \E[e^{s(D-1)}]$ is the probability generating function of $D-1$, and $g_{D-1}^{(t)}(s)$ is the $t$-fold composition of this function. The goal is to choose parameters $\rho$ and $\lambda$ such that this quantity approaches zero exponentially fast. 
The challenge is understanding how $g_{D-1}^{(t)}(e^{-\lambda})$ behaves for a given choice of $\lambda$.

\begin{figure}[h]
	\begin{center}
	\includegraphics[width=3.1in]{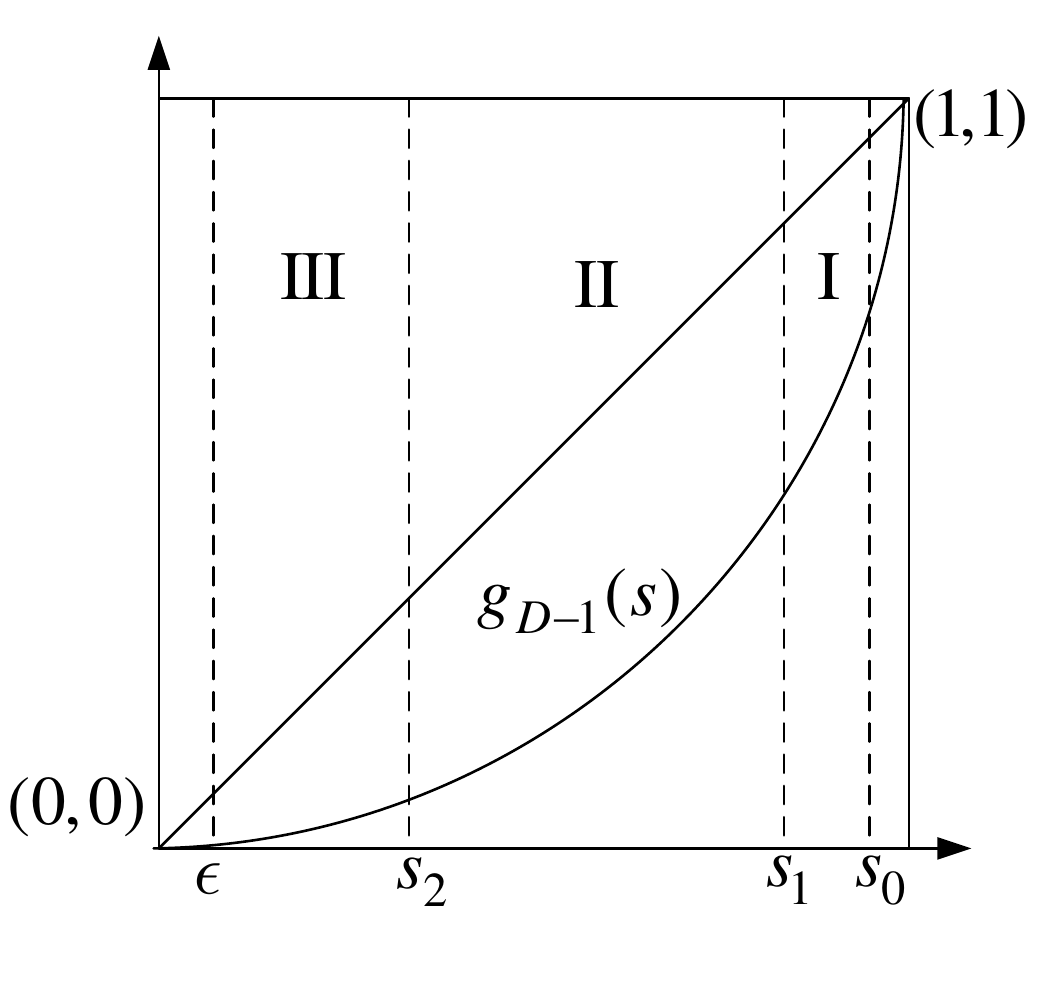}
	\end{center}
	\caption{Regions of the probability generating function, in which we bound the rate of convergence.}
	\label{fig:regions}
\end{figure}

Figure \ref{fig:regions} illustrates $g_{D-1}(s)$. 
Because each node always has at least one child, the probability of extinction for this branching process is 0.
As such, the probability generating function is convex, with $g_{D-1}(0)=0$ and $g_{D-1}(1)=1$. 
This implies that for any starting point $e^{-\lambda}$, the fixed-point iteration method approaches 0. 
We characterize the rate at which $g_{D-1}^{(t)}(s_0)$ approaches 0 by separately bounding the rate of convergence in three different regions of $s$ (Figure \ref{fig:regions}). First, we choose a starting point $s_0=e^{-\lambda}$. We pick any value $s_1<1$, such that the slope is strictly larger than one, i.e. $g_{D-1}'(s_1)>1$. There may be multiple points that satisfy this property; we can choose any one of them, since it only changes the constant factor in the exponent. 
Without loss of generality, we assume that $s_0>s_1$, since otherwise we can start the analysis from the region III.  
Then region I consists of all $s\in [s_1,s_0]$. 
To define $s_2$, we draw a line segment parallel to the diagonal from $s_1$. 
The intersection is defined as $(s_2,g_{D-1}(s_2))$. Region II consists of all $s\in [s_2,s_1)$. Finally, we choose a threshold $\epsilon$, below which we say the process has converged. Then region III consists of all $s\in [\epsilon,s_2)$. We wish to identify a time $t$ that guarantees, for a given $\epsilon$ and $\lambda$, that $g_{D-1}^{(t)}(e^{-\lambda})\leq \epsilon$.

To begin, we split the time spent in each region into $t_1$, $t_2$, and $t_3$, with $t_1+t_2+t_3=t$. 
We first characterize $t_1$.
Note that $g_{D-1}(s_0)\leq 1-g_{D-1}'(s_1)(1-s_0)$ for $s_0$ in region I. 
This holds because $s_1$ has the lowest slope of all points in region I.
Applying this recursively, we get that 
\begin{eqnarray*}
g_{D-1}^{(t_1)}(s) &\leq& \max \left \{ 1- g_{D-1}'(s_1)^{t_1}(1-s), g_{D-1}(s_1) \right \} \end{eqnarray*}
for all $s$ in region I.
In region II, we instead upper bound $g_{D-1}(s)$ by the line segment joining $g_{D-1}(s_1)$ and $g_{D-1}(s_2)$.
This line has slope 1, giving 
\begin{eqnarray*}
g_{D-1}^{(t_2)}(s) &\leq& \max \left \{ g_{D-1}(s_1) - (s_1 - g_{D-1}(s_1))t_2, g_{D-1}(s_2)\right \}. 
\end{eqnarray*}
In region III, we upper bound $g_{D-1}(s)$ by the line $y(s)=g_{D-1}'(s_2)s$. We have that $g_{D-1}(s) < g_{D-1}'(s_2)\cdot s$ for $s$ in region III. Recursing this relation gives 
\begin{eqnarray*}
g_{D-1}^{(t_3)}(s) &\leq& \max \left \{ g_{D-1}'(s_2)^{t_3}\cdot s, \epsilon \right \}. 
\end{eqnarray*}

Thus, if $t\geq 3\max \{t_1,t_2,t_3\}$, then $g_{D-1}^{(t)}(e^{-\lambda})\leq \epsilon$. In particular, we choose 
\begin{eqnarray}
t &\geq& 3\max \big \{ \frac{\log((1-g_{D-1}(s_1))/(1-e^{-\lambda}))}{\log(g_{D-1}'(s_1))}, \nonumber \\ 
&& \frac{g_{D-1}(s_1)-g_{D-1}(s_2)}{s_1-g_{D-1}(s_1)}, \frac{\log(\epsilon)}{s_2 \log(g_{D-1}'(s_2))}\big \}.
\label{eq:t_bounds}
\end{eqnarray}
So for sufficiently large $t$, we have $\prob(Z^{(t)} \leq \rho) \leq \epsilon \cdot e^{\rho \lambda}$. 
By choosing 
\[
\epsilon = g_{D-1}'(s_2)^{s_2 t/3},
\]
we ensure that the third bound on $t$ is always true, and the other two are constant.
Similarly, we choose
\[
e^{-\lambda} = 1- \frac{1-s_2}{g_{D-1}'(s_1)^{t/3}} ,
\]
giving
\begin{eqnarray*}
\prob(Z^{(t)} \leq \rho) &\leq& s_2 \cdot g_{D-1}'(s_2)^{t/3} \left ( 1- \underbrace{\frac{1-s_2}{g_{D-1}'(s_1)^{t/3}}}_{B} \right )^{-\rho} \\
&=& s_2 \cdot g_{D-1}'(s_2)^{t/3} \left ( 1- B \right )^{- \frac{1}{B}B\rho} \\
&\leq& s_2 \cdot g_{D-1}'(s_2)^{t/3} e^{\rho (1-s_2) g_{D-1}'(s_1)^{-t/3}}.
\end{eqnarray*}
Choosing $\rho = g_{D-1}'(s_1)^{t/3}/(1-s_2)$, we observe that for $t$ larger than the bound in \eqref{eq:t_bounds}, the number of leaves is lower bounded by an exponentially growing quantity ($\rho$) with probability approaching 1 exponentially fast in $t$.

\subsection{Proof of Proposition \ref{propo:grid}}
\label{sec:proofs_snapshot_grid}

\begin{algorithm}[ht!]
\caption{Grid adaptive diffusion}
\label{alg:grid}
\begin{algorithmic}[1]
\Require grid contact network $G=(V,E)$, source $v^*$, time $T$
\Ensure set of infected nodes $V_T$
\State $V_0 \gets \{v^*\}$, $h \gets 0$, $v_0 \gets v^*$
\State $\mathcal K \gets \{N,S,E, W\}$
\Comment Cardinality directions
\State let $k_v(u)$ denote $u$'s direction with respect to $v$
\State $v^*$ selects one of its neighbors $u$ at random
\State $V_1 \gets V_0 \cup \{u\}$, $v_1 \gets u$
\State $h^H=\mathds 1_{\{k_v(u)=E\}}-\mathds 1_{\{k_v(u)=W\}}$
\State $h^V = \mathds 1_{\{k_v(u)=N\}}-\mathds 1_{\{k_v(u)=S\}}$
\State let $N^{K}(u)$ represent $u$'s neighbors in directions $K \subseteq \mathcal K$
\State $V_2 \gets V_1 \cup N^{\mathcal K}(u)\setminus \{v^*\}$, $v_2 \gets v_1$
\State $t \gets 3$
\For{$t \leq T$}
\State $v_{t-1}$ selects a random variable $X \sim U(0,1)$
\If{$X \leq \alpha(t-1, |h^V|+|h^H|)$}
\ForAll{$v \in N({v_{t-1}})$}
\State Infection Message({$G$,$v_{t-1}$,$v$,$\{k_v(v_{t-1})\}$, $G_t$})
\EndFor
\Else
\State $K\gets \emptyset$
\If{$h^H<0$}
\State $K\gets K \cup \{E\}$
\ElsIf{$h^H>0$}
\State $K\gets K \cup \{W\}$
\EndIf
\If{$h^V<0$}
\State $K\gets K \cup \{N\}$
\ElsIf {$h^V>0$}
\State $K\gets K \cup \{S\}$
\EndIf
\State $v_{t-1}$ randomly selects $u \in N^{\mathcal K \setminus K}({v_{t-1}})$
\State $h^H=h^H+\mathds 1_{\{k_v(u)=E\}}-\mathds 1_{\{k_v(u)=W\}}$
\State $h^V = h^V+\mathds 1_{\{k_v(u)=N\}}-\mathds 1_{\{k_v(u)=S\}}$
\State $v_{t} \gets u$
\ForAll{$v \in N^{\mathcal K \setminus \{k_{v_{t-1}}(v)}({v_{t}})\}$}
\State Infection Message({$G$,$v_{t}$,$v$,$\{k_{v_t}(v_{t-1}),k_v(v_t)\}$,$V_t$})
\If{$t+1>T$}
\State \text{break}
\EndIf
\State Infection Message({$G$,$v_{t}$,$v$,$\{k_{v_t}(v_{t-1}),k_v(v_t)\}$,$V_t$})
\EndFor
\EndIf
\State $t \gets t + 2$
\EndFor
\Procedure{Infection Message}{$G$,$u$,$v$,$K$,$V_t$}
\If{$v \in V_t$}
\ForAll{$w \in N^{\mathcal K \setminus K}({v})$}
\State Infection Message({$G$,$v$,$w$,$K$,$G_t$})
\EndFor
\Else
\State $V_t \gets V_{t-2} \cup \{v\}$
\EndIf
\EndProcedure
\end{algorithmic}
\end{algorithm}

\noindent \textbf{Number of nodes.} $T$ is either even or odd. At each even $T$, $G_T$ is a ball (defined over a grid graph) centered at the virtual source with radius $T/2$; that is, $G_T$ consists of all nodes whose distance from the virtual source is at most $T/2$ hops. Thus at each successive even $T$, $G_T$ increases in radius by one. The perimeter of such a ball (over a two-dimensional grid) is $4\frac{T}{2}$. The total number of nodes is therefore $1+\sum_{i=1}^{T/2}4i=\frac{1}{2}(T^2+2T+2)$.

When $T$ is odd, there are two cases. Either the virtual source did not move, in which case $N_T=N_{T+1}$ (because all the spreading occurs in one time step), or the virtual source did move, so spreading occurs over two timesteps. In the latter case, the odd timestep adds a number of nodes that is at least half plus one of the previous timestep's perimeter nodes: $N_T \geq N_{T-1}+2\frac{T-1}{2}+1=\frac{1}{2}(T^2+2T+1)$. This is the smaller of the two expressions, so we have $N_T\geq (T+1)^2/2$.

\noindent \textbf{Probability of detection.}
At each even $T$, $G_T$ is symmetric about the virtual source. 
We reiterate that the snapshot adversary can only see which nodes are infected---it has no information about who infected whom.  

In order to ensure that each node is equally likely to be the source, we want the distribution of the (strictly positive) distance from the virtual source to the true source to match exactly the distribution of nodes at each viable distance from the virtual source:
\begin{eqnarray}
	\label{eq:MCp_grid}
	p^{(t)} = \frac{4}{t(\frac{t}{2}+1)}\begin{bmatrix}
		1\\
		2\\
		\vdots\\
		t/2
	\end{bmatrix} \;\in\reals^{t/2} \;.
\end{eqnarray}
There are $4h$ nodes at distance $h$ from the virtual source,
and by symmetry all of them are equally likely to have been the source, giving:
\begin{eqnarray}
	\prob(G_T| v^*, \delta_H(v^*,v_t)=h) &=& \frac{1}{4h} p^{(t)}_{h} \nonumber\\
		&=& \frac{1}{t(\frac{t}{2}-1)} \;,\nonumber
\end{eqnarray}
which is independent of $h$. Thus all nodes in the graph are equally likely to have been the source. The claim is that by choosing $\alpha(t,h)$ according to Equation \eqref{eq:alpha_grid}, we satisfy the distribution in \ref{eq:MCp_grid}.

The state transition can be represented as  the usual $((t/2)+1)\times (t/2)$ dimensional column stochastic matrix:
\begin{eqnarray*}
	\label{eq:MC}
	\mathclap{p^{(t+2)} = \begin{bmatrix}
	\alpha(t,1) & & & \\
	1-\alpha(t,1) &\alpha(t,2) & & \\
	& 1-\alpha(t,2)& \ddots& \\
	& & \ddots& \alpha(t,t/2) \\
	& & &  1-\alpha(t,t/2) \\
	\end{bmatrix}p^{(t)}}\;.
\end{eqnarray*}
This relation holds because we have imposed the condition that the virtual source never moves closer to the true source. We can solve directly for $\alpha(t,1)=t/(t+4)$, and obtain a recursive expression for $\alpha(t,h)$ when $h>1$:
\begin{eqnarray}
\alpha(t,h)=\frac{t}{t+4}-\frac{h-1}{h}(1-\alpha(t,h-1))\;.
\end{eqnarray}
We show by induction that this expression evaluates to Equation \eqref{eq:alpha_grid}. For $h=2$, we have $\alpha(t,2)=\frac{t}{t+4}-\frac{1}{2}\frac{4}{t+4}=\frac{t-2}{t+4}$. Now suppose that Equation \eqref{eq:alpha_grid} holds for all $h<h_0$. 
We then have 
\begin{align*}
\alpha(t,h_0)&=&\frac{t}{t+4}-\frac{h_0-1}{h_0}(1-\frac{t-2(h_0-1)}{t+4}) \\
&=& \frac{t-2(h_0-1)}{t+4}\;,
\end{align*}
which is the claim.

By construction the ML estimator for even $T$ is to choose any node except the virtual source uniformly at random. For odd $T$, there are two options: either the virtual source stayed fixed or it moved. If the former is true, then spreading occurs in one timestep, so the ML estimator once again chooses a node other than the virtual source uniformly at random. If the virtual source moved, then $G_T$ is symmetric about the edge connecting the old virtual source to the new one. Since the adversary only knows that virtual sources cannot be the true source, the ML estimator chooses one of the remaining $N_T-2$ nodes uniformly at random. This gives a probability of detection of $1/(N_T-2)$. The claim follows from observing that $N_T\geq \frac{1}{2}(T+1)^2-2=\frac{(T+3)(T-1)}{2}$.

\subsection{Proof of Proposition \ref{propo:line_adap}}
\label{sec:line_proof}

The control packet at spy node $s_1$ includes the amount of delay at $s_1=0$ and 
all descendants of $s_1$, which is the set of nodes $\{-1,-2,\ldots\}$. 
The  control packet at spy node $s_2$ includes the amount of delay at $s_2=n+1$ and all 
descendants of $s_2$, which is the set of nodes $\{n+2,n+3,\ldots\}$. 
Given this, it is easy to figure out the whole trajectory of the virtual source for time $t\geq T_1$. 
Since the virtual source follow i.i.d. Bernoulli trials with probability $q$, one can exactly figure out $q$ from the infinite Bernoulli trials. 
Also the direction $D$ is trivially revealed. 

To lighten the notations, let us suppose that $T_{1}\leq T_{2}$ (or equivalently $T_{s_1} \leq T_{s_2} $). 
Now using the difference of the observed time stamps $T_{s_2}-T_{s_1}$ 
and the trajectory of the virtual source between $T_{s_1}$ and $T_{s_2}$, the adversary can also figure out 
the time stamp $T_1$ with respect to the start of the infection. 
Further, once the adversary figures out $T_1$ and the location of the virtual source $v_{T_1}$,  the timestamp $T_2$ does not provide any more information. 
Hence, the adversary performs ML estimate using $T_1,D$ and $q$. 
Let $B(k,n,q) = { n \choose k} q^k (1-q)^{n-k}$ denote the pmf of the binomial distribution. 
Then, the likelihood can be  computed for $T_1$ as \\
$	 \prob^{\rm (adaptive)}_{ T_1 |V^*,Q,D} \big(\,t_1  \big| v^*, \q ,\ell \big) =$
\begin{align}
~~ \left\{ 
		\begin{array}{rl}
			 \,q\, B( v^* - \frac{t_1}{2}-2 , \frac{t_1}{2}-2, q) 	
 			\, \ind_{(v^* \in [2+\frac{t_1}{2},t_1])}  & \text{, if $t_1$ even}\;, \\
			 \,B( v^* - \frac{t_1+3}{2},\frac{t_1-3}{2},  q)\, \ind_{(v^*\in[\frac{t_1+3}{2},t_1])}
			 & \text{, if $t_1$ odd} \;,
		\end{array}
	\right. \label{eq:line_control_likelihood1}
\end{align}
$	\prob^{\rm (adaptive)}_{ T_1 |V^*,Q,D} \big(\,t_1 \big|\, v^*, \q ,r \big) =$
\begin{align}
~~ \left\{ 
		\begin{array}{rl}
			0  & \text{, if $t_1$ even} \;,\\
			 (1-q)\, B(\frac{t_1-1}{2}-v^*, \frac{t_1-3}{2},q)   \,
			\ind_{(v^*\in[1,\frac{t_1-1}{2}] )}
			 & \text{, if $t_1$ odd}\;. 
		\end{array}
	\right.\label{eq:line_control_likelihood2}
\end{align}
This follows from the construction of the adaptive diffusion.  
The protocol follows a binomial distribution with parameter $q$ until $(T_1-1)$. 
At time $T_1$, 
one of the following can happen: 
the virtual source can only be passed 
(the first equation in \eqref{eq:line_control_likelihood1}), 
it can only  stay (the second equation in \eqref{eq:line_control_likelihood2}), or 
both cases are possible (the second equation in  \eqref{eq:line_control_likelihood1}). 

Given $T_1$, $Q$ and $D$, which are revealed under the adversarial model we consider, 
the above formula implies that the posterior distribution of the source also follows a binomial distribution. 
Hence,  the ML estimate 
is the mode of a binomial distribution with a shift, for example 
 when $t_1$ is even,  ML estimate is the mode of  $2+(t_1/2) + Z$ where 
$Z\sim \text{Binom}((t_1/2)-2,q)$.  
	The adversary can compute the ML estimate: 
\begin{eqnarray}
	\hv_{\rm ML} &=& \left\{ 
	\begin{array}{rl}
	\frac{T_1+2}{2}  + \big\lfloor q\Big(\frac{T_1-2}{2}\Big) \big\rfloor & \text{if $T_1$ even, $D=\ell$\;,} \\
	 \frac{T_1+3}{2} +\big\lfloor q\Big( \frac{T_1-1}{2}  \Big) \big\rfloor & \text{if $T_1$ odd, $D=\ell$\;,} \\
	1 + \big\lfloor (1-q) \Big( \frac{T_1-1}{2}  \Big) \big\rfloor& \text{if $T_1$ odd, $D=r$\;.} 
	\end{array}
	\right.\label{eq:line_control_ml}
\end{eqnarray}

Together with the likelihoods in Eqs. \eqref{eq:line_control_likelihood1} and \eqref{eq:line_control_likelihood2}, this gives \\ 
$	\prob^{\rm (adaptive)}_{ T_1,D |V^*,Q} \big(\,t_1, r, \hv_{\rm ML} = v^*  \big| v^*, \q  \big) =$
\begin{align} 
	\frac12 (1-q)\, B\Big(\frac{t_1-1}{2}-v^*, \frac{t_1-3}{2},q\Big)   \,
			\ind_{(\hv_{\rm ML} = v^*)} \, \ind_{(\text{$t_1$ is odd})}
\end{align}
$	\prob^{\rm (adaptive)}_{ T_1,D |Q} \big(\,t_1, r, V^*=\hv_{\rm ML}   \big|  \q  \big) =$
\begin{align}
	&=&\frac{1}{2 \,n} (1-q)\, B\Big(\frac{t_1-1}{2}-\hv_{\rm ML}, \frac{t_1-3}{2},q\Big)   \,
			  \, \ind_{(\text{$t_1$ is odd})} \\
	&\leq& \frac{(1-q)}{2\, n} \Big( \frac{\sqrt{2}\,\ind_{\text{($t_1$ is odd and $t_1>3$)}}}{\sqrt{\frac{t_1-3}{2}q(1-q)}}
			 	\,+\,\ind_{(t_1=3)} \Big)
\end{align}
where $\hv_{\rm ML}= \hv_{\rm ML}(t_1,q,r)$ is provided in \eqref{eq:line_control_ml}, 
and the bound on $B(\cdot)$ follows from Gaussian approximation (which gives an upper bound $1/\sqrt{2\pi k q(1-q)}$) 
and Berry-Esseen theorem (which gives an approximation guarantee of $2\times0.4748/\sqrt{kq(1-q)}$) \cite{Ber41},  
for $k=(t_1-3)/2$.
Marginalizing out $T_1\in\{3,5,\ldots, 2\lfloor (n-1)/2\rfloor+ 1\}$ and applying 
an upper bound  $\sum_{i=1}^k 1/\sqrt{i} \leq 2\sqrt{k+1} - 2 \leq 2 \sqrt{k-1} + \sqrt{1/(2(k-1))} -2 \leq \sqrt{4(k-1)}$, we get 
\begin{eqnarray}
	\prob  \big(\, D= r, V^*=\hv_{\rm ML} , T_1\text{ is odd}   \big| Q= \q \big) \leq \nonumber \\ \quad \frac{(1-q)\sqrt{2}}{2\,n\,\sqrt{q(1-q)}}\sqrt{ 8 \,\Big\lfloor \frac{n-1}{2}\Big\rfloor }+ \frac{1-q}{2\,n}\;.
\end{eqnarray}
Similarly, we can show that 
\begin{flalign}
	\prob \big(\,  D=\ell, V^*=\hv_{\rm ML} , T_1\text{ is odd}  \big|  Q=\q  \big) \leq \nonumber \\ 
\frac{\sqrt{2}}{2\,n\,\sqrt{q(1-q)}}\sqrt{ 8 \,\Big\lfloor \frac{n-1}{2}\Big\rfloor }+ \frac{1}{n}\;, 
\end{flalign}
\begin{flalign}
		\prob \big(\, V^*=\hv_{\rm ML} , T_1\text{ is even}  \big|  Q=\q  \big)  \leq \nonumber \\
\frac{q \sqrt{2}}{2\,n\,\sqrt{q(1-q)}}\sqrt{ 8 \,\Big\lfloor \frac{n}{2}\Big\rfloor }+ \frac{1+q}{2\,n}\;, 
\end{flalign}
Summing up, 
\begin{eqnarray}
 \prob(V^* = \hv_{\rm ML}|Q=q) &\leq& \sqrt{\frac{8}{n\,q\,(1-q)}} + \frac{2}{n} \;.
\end{eqnarray}
Recall $Q$ is uniformly drawn from $[0,1]$. Taking expectation over $Q$  gives 
\begin{eqnarray}
 \prob(V^* = \hv_{\rm ML}) &\leq& \pi \sqrt{\frac{8}{n}} + \frac{2}{n} \;,
\end{eqnarray}
where we used $\int_0^1 1/\sqrt{x(1-x)} dx = \arcsin(1)-\arcsin(-1) = \pi$.

\section*{Acknowledgment}

The authors thank Paul Cuff for helpful discussions and for pointing out the Bayesian interpretation of the P\'olya's urn process. 
This work has been supported by NSF CISE
awards CCF-1422278, CCF-1553452, and CCF-1409135, SaTC award
CNS-1527754, ARO W911NF-14-1-0220, and AFOSR 556016.

%
%
%
%
%
\bibliographystyle{IEEEtran}
\bibliography{privacy}
\end{document}